\documentclass{article}

\usepackage{arxiv}

\usepackage{stix}

\usepackage{mismath}
\usepackage{derivative}
\usepackage[most]{tcolorbox}
\usepackage{natbib}
\usepackage{url}
\usepackage[hidelinks]{hyperref}
\usepackage{enumitem}

\let\cite\citep

\newlist{Oitemize}{enumerate}{1}
\setlist[Oitemize,1]{
  label={\textbf{Objective O\arabic*}},
  ref=O\arabic*,
  leftmargin=*,
  labelsep=0.5em
}

\renewcommand{\headeright}{\today}
\renewcommand{\shorttitle}{Reducing spurious mixing in ALE models through variational minimizers}
\renewcommand{\shortauthors}{A.Alexandris and G.Papadakis}

\boldvect

\newcommand{\jump}[1]{\ensuremath{ \left[\!\left[#1\right]\!\right]} }

\author{ \href{https://orcid.org/0000-0002-2124-6670}{\includegraphics[scale=0.06]{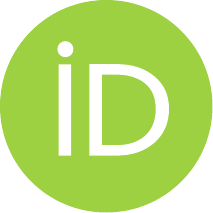}\hspace{1mm}Andreas~Alexandris-Galanopoulos} \\
	School of Naval Architecture and Marine Engineering\\
	National Technical University of Athens\\
	Zografos, 15710, Athens \\
	\texttt{andreas\_alexandris@mail.ntua.gr}
	\And
	\href{https://orcid.org/0000-0002-2742-5258}{\includegraphics[scale=0.06]{orcid.pdf}\hspace{1mm}George~Papadakis} \\
    School of Naval Architecture and Marine Engineering\\
	National Technical University of Athens\\
	Zografos, 15710, Athens \\
	\texttt{papis@fluid.mech.ntua.gr}
}

\title{An ALE approach to reduce spurious numerical mixing through variational minimizers: application to internal waves}

\hypersetup{
pdftitle={An ALE approach to reduce spurious numerical mixing through variational minimizers: application to internal waves},
pdfsubject={sls2preprint},
pdfauthor={Andreas Alexandris-Galanopoulos, George Papadakis},
pdfkeywords={Arbitrary Lagrangian-Eulerian (ALE) , spurious mixing , adaptive mesh , non-hydrostatic , internal solitary waves , ocean model},
}

\RequirePackage{draftwatermark}
\DraftwatermarkOptions{
  text={Under Review at Ocean Modelling},
  color=gray!30,
  angle=90,
  scale=0.2,
  hpos = 0.95\paperwidth,
  vpos=0.5\paperheight
}

\begin{document}

\maketitle

\begin{abstract}
Spurious numerical mixing is a frequent phenomenon in ocean models. In this paper, we present an efficient and robust methodology that defines the vertical grid motion so that this mixing is reduced. This motion is defined as the solution of an optimization problem that --using the ideas of the calculus of variations-- results in an elliptic partial differential equation, which is straightforward to analyze and discretize. This framework is generally applicable to any ocean model that uses an Arbitrary Lagrangian-Eulerian (ALE) vertical coordinate and can be tuned to fit the modeler's specific needs based on the guidelines presented herein. The method is applied to the nonhydrostatic solver presented by the authors in \cite{alexandris2024semi}.

While the majority of spurious numerical mixing studies focus on large-scale processes, herein the proposed method is applied and tested in small-scale nonhydrostatic phenomena. Specifically, the effectiveness of the method in capturing fully nonlinear internal waves is investigated for the test cases of wave propagation, breaking and overturning. Overturning serves as a demanding test for the proposed scheme as it induces rapid vertical accelerations and thus the mesh-moving algorithm must incorporate this motion with the goal of reducing numerical mixing, while not suppressing physically relevant vertical mass transfer. These numerical benchmarks show the ability of the method to reduce spurious mixing, while attaining the physical relevance of the results.
\end{abstract}


\section{Introduction}\label{sec:intro}

Since the mid-20\textsuperscript{th} century, numerical models have become indispensable tools for analyzing and predicting oceanic systems and processes. As a result, considerable research has focused on developing discretization methods that faithfully simulate physical phenomena while minimizing numerical artifacts. Among these artifacts, spurious numerical mixing/diffusion remains a well-known and persistent issue.

Spurious mixing is a frequent phenomenon in ocean models that occurs due to the numerical diffusion introduced by the spatial discretization of advection terms. This is a prominent problem in the advection of tracers, where these diffusive tendencies can introduce significant nonphysical mixing \cite{klingbeil2019reducing}. The most common occurrence of this is in the case of Spurious Diapycnal Mixing (SDM), in which advection introduces mixing across the density layers, thus severely altering the stratification. Consequently, various methods to track and remedy SDM have been proposed. 

Since the first studies on SDM \cite{griffies2000spurious,lee2002spurious}, considerable effort has been devoted to identifying, quantifying, and reducing these artifacts. Various diagnostics have been proposed, including energetics-based approaches, effective diffusivity estimates \cite{lee2002spurious}, and analysis through discrete variance decay \cite{burchard2008comparative,klingbeil2014quantification}.

Following these studies, two primary directions have emerged for mitigating SDM \cite{fox2019challenges}. The first focuses on the development of low-dissipation transport schemes and high-order remapping procedures \cite{klingbeil2019reducing,white2009high}. Although these approaches can substantially reduce numerical diffusion, their secondary properties must be carefully analyzed, especially with respect to monotonicity and mixing behavior (see also \S\ref{sec:sdm_num}).

The second direction involves exploiting vertical coordinate systems and designing the (possibly moving) vertical mesh so that the underlying flow dynamics are better captured and SDM is reduced. The present work contributes primarily toward this second direction through the development of a variational ALE framework for adaptively defining the vertical coordinate.

The choice of vertical coordinate is a crucial aspect in the design of numerical ocean models, and consequently, a substantial body of research has focused on this topic \cite{halliwell2004evaluation,petersen2015evaluation}.
Following \cite{griffies2000developments}, the three vertical coordinate systems that are traditionally used within the oceanographic community are:
\begin{itemize}
    \item $z$-models (Eulerian frame): Popularized by the work of \cite{bryan1997numerical}, these models use a horizontally parallel partition of the domain. The bottom and free-surface boundaries are incorporated through a Marker-and-cell technique.
    \item $\rho$-models (Lagrangian frame): In these isopycnal models, the domain is split into horizontal immiscible layers, so that the fluid density is constant within each of them. Notable contributions to isopycnal models include the works of \cite{hallberg1996buoyancy} and \cite{bleck1998ocean}.
    \item $\sigma$-models (Terrain-following): Pioneered by \cite{phillips1957coordinate}, the $\sigma$-coordinate vertically partitions the water column (from bottom to free-surface) into iso-spaced layers.
\end{itemize}

Each coordinate system has its own advantages and drawbacks and thus the choice among them is not always an obvious one \cite{griffies2000developments}. For example, while $z$-models have straightforward discretization, the inclusion of bottom topography leads to complications. On the other hand, $\sigma$-models naturally include the bottom, but suffer from spurious numerical behavior of terms like the horizontal pressure gradient. Lastly, $\rho$-models are quite good at capturing neutral directions in stratified regions (unlike $z/\sigma$ models), but are not suitable for non-stratified dynamics and in many cases the layers tend to vanish and/or become unstable.

This led to the introduction of a hybrid $\rho$/$z$-model by Bleck \cite{bleck2002oceanic}, in which the vertical coordinate dynamically transitions between isopycnic and fixed-depth coordinates in order to prevent layer collapse and maintain adequate vertical resolution. This Vertically Lagrangian-Remap (VLR) technique has been used with success in the HYCOM \cite{chassignet2007hycom} model, and has also been incorporated in the MOM6 \cite{adcroft2019gfdl} model.

In order to facilitate the design of such schemes, the framework of Generalized Vertical Coordinates (GVC) is employed, where the physical vertical coordinate $z$ is represented through a mapping with a parametric coordinate $\xi$ (in the present notation). Within this framework, the layers are defined by the iso-$\xi$ surfaces, and the choice of the mapping determines the vertical structure and thickness distribution of the layers. In particular, different choices recover different coordinate systems, such as $z$, $\sigma$, $\rho$, or hybrid coordinates.

The versatility of such GVC approaches gives rise to a large number of ocean models with vertical coordinates that are designed within the Arbitrary Lagrangian-Eulerian (ALE) framework. For an extensive review of recent developments in VLR/ALE schemes in ocean models, the reader is referred to \cite{griffies2020primer}. One of the basic advantages of this ALE framework is the flexibility it gives to exploit the vertical mesh movement to achieve specific requirements. For example, in \cite{leclair2011z} and \cite{petersen2015evaluation}, vertical ALE coordinates were added to NEMO and MPAS respectively to reduce SDM through a superposition of imposed motions (barotropic, high/low frequency baroclinic modes, additional smoothing terms etc.). Similar approaches appear also in \cite{burchard2004non,burchard2008comparative,hofmeister2010non}.

Having all the above as inspiration, in this work we aim to present a comprehensive compact mesh-moving ALE method with the goal of reducing SDM. The proposed methodology is introduced to the Semi-Lagrangian Splitting (SLS) model, a nonhydrostatic ALE solver introduced by the authors in \cite{alexandris2024semi}. SLS is built upon a fully versatile GVC framework that allows arbitrary vertical grid motion through a suitable tuning of the dia-surface velocity. The equations are discretized using a second-order Finite Volume (FV) method, while the nonhydrostatic pressure is calculated using a Poisson equation in an operator splitting fashion.

In the present paper, we propose a novel ALE framework that defines the vertical mesh movement through compactly formulated optimality criteria. Although the scheme is implemented and assessed in SLS, it is generally applicable to any GVC ocean model. This mesh movement is prescribed through a multi-objective optimization that is compactly described by a cost functional. The functional is minimized in the spirit of calculus of variations, giving rise to an elliptic partial differential equation that can be efficiently solved numerically. To bolster the utility of the proposed method, the various weighting factors are analyzed and specific values are laid out. While this approach is not restricted to specific objectives, herein we aim to effectively reduce SDM while maintaining the physical fidelity of the results.

Although the majority of the literature on SDM tackles large-scale oceanic processes (see \S\ref{sec:sdm}), in the present paper, we focus on explicitly resolved small-scale nonhydrostatic phenomena. Furthermore, no parametrizations of any kind are considered and a direct simulation approach is adopted. As a result, the proposed mesh movement is benchmarked in scenarios where any instability or artifact of any kind (like SDM) directly pollutes the fidelity of the simulation. This aspires to augment the literature on SDM and ALE schemes at scales and simulation setups that are not usually considered, but might provide valuable insights into the features and limitations of such approaches.

The numerical test cases center around Internal Solitary Waves (ISWs), since they are highly nonlinear and inherently nonhydrostatic, making them a suitable benchmark for the proposed ALE scheme. In particular, ISW-topography interactions are highly transient and induce rapid, strong deformations of the isopycnals, which can provide a challenging test for mesh-moving formulations with a Lagrangian bias. This, combined with the practical relevance of ISWs in regional dynamics (see \cite{stastna2024simulations}), offers an appealing benchmark to assess whether the proposed variational mesh movement can robustly reduce SDM even in such demanding cases.

\section{Governing Equations}\label{sec:formulation}

\begin{figure}
    \centering
    \includegraphics[width=0.7\linewidth]{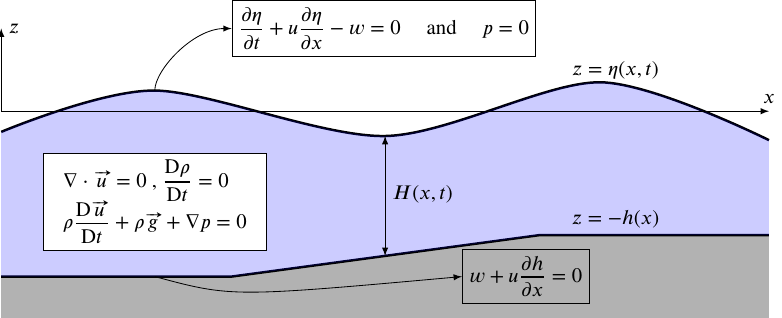}
    \caption{The free-surface Euler system \cite{alexandris2024semi}.}
    \label{fig:domain}
\end{figure}

Let us consider a domain vertically bounded by the bottom $z=-h(x)$ and the free-surface $z=\eta(x,t)$ that are considered to be single-valued functions of the horizontal coordinate $x$ (see Fig.\ref{fig:domain}). Inside the domain the motion is governed by the Euler equations for an incompressible fluid of variable density $\rho$. The bottom and the free-surface are considered as material surfaces and a zero pressure condition is enforced on the latter. On the lateral boundaries, we consider either Neumann (inviscid wall) or Dirichlet boundary conditions based on the case at hand.

\subsection{The Generalized Vertical Coordinate system}\label{sec:gvc}

To express the equations in GVC, we introduce a parametric vertical coordinate $\xi \in [0,1]$ and assume that the physical vertical coordinate is a function of it: $z(x,\xi,t)$ with $x,t$ remaining unchanged \cite{alexandris2024semi}. To mathematically express it, the following key metric is introduced:
\begin{align}
    L \eqdef \pdv{z}{\xi} \then \di z = L \di\xi
\end{align}
Since the equations are discretized directly along the $\xi$-axis, $L$ determines the local vertical grid spacing and therefore plays the role of the layer thickness commonly used in ocean models.

Vertically integrating the above equation, we arrive at the following relation:
\begin{align}\label{coord}
    z & = -h(x) + \int_0^\xi L(x,\xi',t) \di\xi'
\end{align}
By definition, $\xi=0$ corresponds to the bottom boundary and we enforce that $\xi=1$ maps to the free-surface through $\int_0^1 L \di\xi = H \eqdef \eta + h$. The function $L(x,\xi,t)$ is the determinant of the transformation's Jacobian matrix and represents the amount of physical space that is contained within a control volume in the parametric space: $\di V = L \di x \di \xi$ (see Fig.\ref{fig:fv_mesh}).

Note that for a fixed $\xi$, $z(x,\xi,t)$ represents a surface that evolves in time. Thus, discretization along the $\xi$-axis yields a set of surfaces $z(x,\xi_j,t)$, which, following the common practice in ocean modelling, we refer to as layers.

\subsubsection{Notational conventions}

Before proceeding, some remarks concerning the notation must be made. First of all, derivatives of functions on the parametric $(x,\xi,t)$ space are denoted as $\pdv{}{x} , \pdv{}{\xi} , \pdv{}{t}$ and the corresponding vector operators are denoted as $\nabla f \eqdef \left[ \pdv{f}{x} \ , \ \pdv{f}{\xi} \right]^T $ and $ \nabla \cdot \left[ f_1 , f_2 \right]^T \eqdef \pdv{f_1}{x} + \pdv{f_2}{\xi} $. On the other hand, derivatives of functions on the physical space $(x,z,t)$ are denoted as $\mathrm{grad}_x , \mathrm{grad}_z , \mathrm{grad}_t$ respectively. Based on those, the operators $\grad f \eqdef \left[ \mathrm{grad}_x f \ , \ \mathrm{grad}_z f \right]^T $ and $\divg \left[ f_1 , f_2 \right]^T \eqdef \mathrm{grad}_x f_1 + \mathrm{grad}_z f_2 $ are defined in a similar fashion.

The relations between these two sets of derivatives (see also Appendix A of \cite{alexandris2024semi}) can be expressed through:
\begin{subequations}\label{eq:grad}
\begin{align}
    \mathrm{grad}_x f &= \pdv{f}{x} - \pdv{z}{x} L^{-1} \pdv{f}{\xi} \\
    \mathrm{grad}_z f &=  L^{-1} \pdv{f}{\xi} \\ 
    \mathrm{grad}_t f &= \pdv{f}{t} - \pdv{z}{t} L^{-1} \pdv{f}{\xi}
\end{align}
\end{subequations}

\subsection{The Generalized Continuity Equation}\label{sec:gce}

An important aspect of the GVC formulation is how the vertical motion of the grid (i.e., the evolution of layers $z(x,\xi,t)$ in time) is handled. To properly formulate this, we utilize the incompressibility constraint and, through standard algebra (see \cite{alexandris2024semi}), we can derive the following principle:
\begin{align}\label{gcl}
    \pdv{L}{t} + \pdv{}{x}(Lu) + \pdv{\vartheta}{\xi} = 0
\end{align}
where the dia-surface velocity $\vartheta$ is expressed through:
\begin{align}
    \vartheta \eqdef w - u\pdv{z}{x} - \pdv{z}{t}
\end{align}
This is essentially the \textit{continuity equation} that appears in the formulation of layered ocean models and represents the conservation of volume in the parametric $(x,\xi,t)$ system. Throughout the present paper it will be referred to as the Generalized Continuity Equation (GCE) in order to highlight its role in the present ALE formulation. Specifically, $\pdv{L}{t} \di x \di\xi $ expresses the rate of change of volume, with $Lu\di\xi$ and $\vartheta\di x$ being the volume fluxes in the horizontal and vertical directions respectively (see Fig.\ref{fig:fv_mesh}).

In the ALE formulation of SLS, we directly discretize the GCE.
As a result, volume is automatically conserved and the layers' positions $z(x,\xi,t)$ are defined through eq.(\ref{coord}) by tracking $L(x,\xi,t)$ as an independent variable.
The velocity $\vartheta$ is explicitly prescribed in order to achieve the desired type of layer motion (see also \cite{alexandris2024semi}). As such, the design of the proposed scheme essentially reduces to a suitable definition of $\vartheta$ so that the target goals are achieved. This will be described thoroughly in \S\ref{sec:ale}.

\subsection{The incompressible Euler equations in GVC}

Through eq.(\ref{eq:grad}) and by splitting the pressure into the hydrostatic $p_h$ and the nonhydrostatic component $q$, such that $ p = p_h + q = \int_z^{\eta} \rho g \di z + q $, the incompressible inviscid Euler system is transformed into GVC coordinates \cite{alexandris2024semi}:\begin{subequations}
\begin{align}
    & \pdv{L}{t} + \pdv{}{x}(Lu) + \pdv{\vartheta}{\xi} = 0 \label{s1} \\
    & \pdv{}{t}(L\rho) + \pdv{}{x}(Lu\rho ) + \pdv{}{\xi}(\vartheta\rho) = 0 \label{s2} \\
    & \pdv{}{t}(L\vect{V}) + \pdv{}{x}(Lu\vect{V}) + \pdv{}{\xi}( \vartheta \vect{V} ) + \dfrac{L}{\rho} \left( \rho \vect{g} + \grad p_h + \grad q \right) = 0 \label{s3} \\
    & \divg \vect{V} = 0 \label{divu}
\end{align}
\end{subequations}
with $\vect{V} = \left[ u , w \right]^T $ being the velocity field.

Concerning the nature of each equation, eq.(\ref{s1}) represents the conservation of volume, eq.(\ref{s2}) corresponds to the conservation of mass and eq.(\ref{s3}) represents the conservation of momentum under the influence of the hydrostatic and nonhydrostatic pressure fields. Note that since all flows are considered in the $x-z$ plane, no Coriolis term is present in the momentum equation. Lastly, in eq.(\ref{divu}), the divergence-free constraint of incompressibility is enforced.

Note that this formulation corresponds to an idealized thermodynamic closure in which density is materially conserved, $\mdv{\rho}{t}=0$. This assumption neglects diabatic processes and molecular diffusion. In addition, no mixing or viscous terms are included in the equations. These choices, although simplistic, are sufficient for modelling the small-scale phenomena considered in \S\ref{sec:results} since the propagation of ISWs is primarily a hydrodynamic process. Possible extensions of this approach, including more realistic closures and equations of state, are discussed in \S\ref{sec:conclusions}.

Even though we can solve these fully nonlinear equations, there is also the option to solve the system under the Boussinesq approximation where the pressure gradients are divided by a constant reference density: $\frac{1}{\rho} \grad p_h \to \frac{1}{\rho_0} \grad p_h $ and $ \frac{1}{\rho} \grad q \to \frac{1}{\rho_0} \grad q $. When such an approximation is used, it will be noted.

\section{Numerical Framework}\label{sec:numerics}

For the most part, timestepping, variable placement, and discretization follow the scheme presented in \cite{alexandris2024semi}. Improvements were introduced in \cite{alexandris2025development}, including the adoption of a FEM approach for the nonhydrostatic pressure Poisson system and the linear layer reconstruction algorithm. The only additions not covered in these earlier works are the entropy stable fluxes that are presented in Appendix \ref{app:fluxes}.

Although these numerical techniques are quite standard and are described in full in these papers (and the numerics literature in general), a brief overview is provided in the next sections so that the results presented herein can be better interpreted in the context of the production and mitigation of spurious mixing.

\subsection{Timestepping}

In order to treat the divergence-free condition, timestepping in SLS is performed in a split manner. If we consider finite time intervals of $[t^n , t^{n+1}]$ with $ \Delta t^{n} \eqdef t^{n+1} - t^n $ and denote the values of variables at $t^n$ as $(\cdot)^n$, the timestepping is \cite{alexandris2024semi}:

\textbf{Step 1:} From $ L^n , (L\rho)^n , (L\vect{V})^n $, equations (\ref{s1})-(\ref{s3}) are marched using a standard fourth-order Runge-Kutta in order to get the end-of-step values $  L^{n+1} , (L\rho)^{n+1}$ and the intermediate velocity field $\vect{V}^*$.
In this step, the nonhydrostatic pressure is kept fixed at $q^n$.

\textbf{Step 2:} A pressure correction step \cite{van1986second} is applied to calculate $q^{n+1}$ and $\vect{V}^{n+1}$ by enforcing the divergence-free condition:
\begin{subequations}\label{eq:pco}
\begin{align}
   & \dfrac{\vect{V}^{n+1}-\vect{V}^*}{\Delta t^n} + \dfrac{1}{\rho^{n+1}} \grad (q^{n+1}-q^n) = 0 \label{eq:pco1} \\
    & \divg \vect{V}^{n+1} = 0 \label{eq:pco2}
\end{align}
\end{subequations}
By combining eq.(\ref{eq:pco1}) and eq.(\ref{eq:pco2}), one arrives at a standard pressure-Poisson equation.

The acoustic field is suppressed in Step 2 by the implicit treatment of the divergence-free condition. As a result, the only timestep restriction comes from the eigenvalues of Step 1. Since this is performed explicitly, a CFL condition must be enforced to maintain stability \cite{alexandris2024semi}, with the critical eigenvalues being the barotropic ones $\lambda_H = |u|+\sqrt{gH}$ on the horizontal and the advective ones $\lambda_V = \vartheta$  in the vertical directions.

\subsection{Discrete spatial variables}

We discretize the parametric space into $ n_x \times n_l $ orthogonal cells $ \Omega_{ij} \eqdef \left[ x_i , x_{i+1} \right] \times \left[ \xi_{j} , \xi_{j+1} \right] $. The velocities $u,w$ and the variables $\rho,L$ are considered to be cell-centered, while $\vartheta$ is placed on the layer interfaces. This variable placement corresponds to a horizontal Arakawa A-grid with a Lorenz grid in the vertical direction. As such, the fluxes are carefully designed in order to avoid spurious pressure modes (see Appendix \ref{app:fluxes}). The space-discrete variables are noted with the double indexing $(\cdot)_{ij}$.

For the nonhydrostatic pressure, a nodal placement is used to facilitate the FEM treatment of the pressure correction step. Specifically, we utilize linear Q4 elements on the parametric space, giving rise to the representation:
\begin{align}\label{eq:q_fe}
    q(x,\xi) \approx \sum_{i=1}^{n_x} \sum_{j=1}^{n_l+1} q_{ij} N_{ij}(x,\xi)
\end{align}
The nodal Q4 basis functions are presented in local coordinates in eq.(\ref{shapeN}).

\subsection{Explicit terms}\label{sec:fv}

\begin{figure}
    \centering
        \includegraphics[width = 0.8\linewidth]{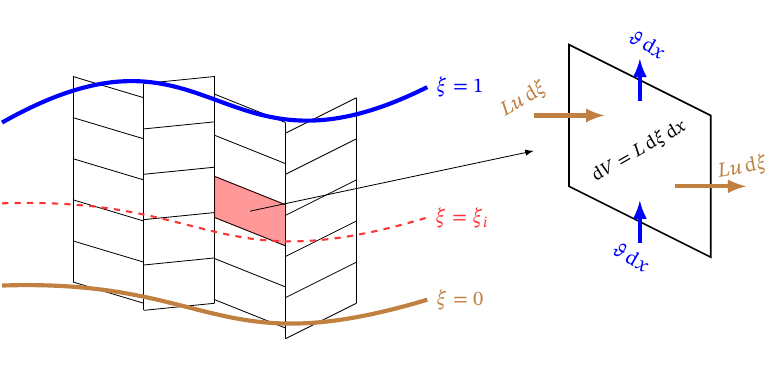}    \caption{ Schematic representation of the GCE on the computational mesh. The GCE fluxes in the horizontal and vertical directions are denoted as $ Lu $ and $\vartheta$ respectively.}
    \label{fig:fv_mesh}
\end{figure}

Within a cell of parametric area $|\Omega_{ij}|=\Delta x_i \times \Delta \xi_j$, eqs.(\ref{s1})--(\ref{s3}) are discretized as follows:\begin{subequations}\label{eq:fv}
\begin{align}
    |\Omega_{ij}|\pdv*{L_{ij}}{t} &+ \Delta \xi_j \left[ F^1_{i+\frac{1}{2},j} - F^1_{i-\frac{1}{2},j} \right] + \Delta x_i \left[ \vartheta_{i,j+1/2} - \vartheta_{i,j-1/2} \right] = 0 \\
    |\Omega_{ij}|\pdv{}{t}(L\rho)_{ij} &+ \Delta \xi_j \left[ {F}^{\rho}_{i+\frac{1}{2},j} - {F}^{\rho}_{i-\frac{1}{2},j} \right] + \Delta x_i \left[ G^{\rho}_{i,j+1/2} - G^{\rho}_{i,j-1/2} \right] = 0 \\
    |\Omega_{ij}|\pdv{}{t}(L\vect{V})_{ij} &+ \Delta \xi_j \left[ \vect{F}^{\vect{V}}_{i+\frac{1}{2},j} - \vect{F}^{\vect{V}}_{i-\frac{1}{2},j} \right] + \Delta x_i \left[ \vect{G}^{\vect{V}}_{i,j+1/2} - \vect{G}^{\vect{V}}_{i,j-1/2} \right] \nonumber \\
    & \hspace{150pt} + \dfrac{|\Omega_{ij}|L_{ij}}{\rho_{ij}} \left[ \rho_{ij} \vect{g} + (\grad p_h)_{ij} + (\grad q)_{ij} \right] = 0
\end{align}    
\end{subequations}, where $F^{(\cdot)}_{i+1/2,j},G^{(\cdot)}_{i,j+1/2}$ are approximations of the fluxes $Lu(\cdot)$ and $\vartheta (\cdot)$.

In SLS, the fluxes are calculated through a second-order 1D MUSCL\footnote{Monotonic Upstream-centered Scheme for Conservation Laws}-type TVD procedure \cite{van1979towards} applied in the $x$ and $\xi$ directions independently. Specifically, a piecewise linear reconstruction coupled with the Superbee limiter is employed \cite{roe1985some}.

In order to maintain stability and negate spurious pressure modes that arise in A-grids, special care is taken in the definition of the fluxes. To achieve this, SLS utilizes Riemann solvers in the spirit of collocated FV schemes \cite{toro2013riemann}. In the present paper, a slight variation of this is adopted. Following closely the analysis of \cite{ersing2025entropy}, the fluxes $F^1_{i+1/2,j},\vect{F}^{\vect{V}}_{i+1/2,j}$ are designed based on the entropy stable framework \cite{tadmor1987numerical}. A brief description of the approach alongside the analytic relations for these fluxes is presented in Appendix \ref{app:fluxes}. These fluxes ensure stability, while respecting the hydrostatic equilibrium, which is crucial for multilayer-type systems.

The horizontal tracer fluxes $F^\rho_{i+1/2,j}$ and all vertical fluxes $G^{(\cdot)}_{i,j+1/2}$ are calculated using an upwinding procedure based on the GCE fluxes $F^1_{i+1/2,j}$ and $\vartheta_{i,j+1/2}$ respectively.

Additionally, the layer interfaces are reconstructed using a second-order scheme. Specifically, we apply the layer reconstruction described in \cite{alexandris2025development}, where the approach of \cite{bollermann2013well} is extended to the multilayer case. This results in a generally nonconforming mesh consisting of skewed quads (see Fig.\ref{fig:fv_mesh}).

For $ (\mathrm{grad}_x p_h)_{ij}$, we adopt the approach of \cite{ersing2025entropy} that is essentially the multilayer extension of the nonconservative product approach that is commonly used in the shallow water equations \cite{audusse2004fast,kurganov2018finite}. The vertical gradient $ \mathrm{grad}_z p_h $ automatically cancels out the buoyancy term $\rho_{ij} g$.

The nonhydrostatic gradient is calculated by directly differentiating eq.(\ref{eq:q_fe}), in accordance with the FEM framework.

\subsection{Nonhydrostatic pressure correction}\label{sec:pco}

For the discretization of eq.(\ref{eq:pco}), the Finite Element framework is employed based on the expansion of eq.(\ref{eq:q_fe}). The complete procedure is presented in Appendix \ref{app:pco}. If we define $\phi_{ij} \eqdef \Delta t^n ( q_{ij}^{n+1} - q_{ij}^n ) $, where $q_{ij}$ are values of the dynamic pressure at the nodes $(x_i,\xi_j)$ of the quadrilaterals, we arrive at the following linear system:
\begin{align}\label{linpco}
    \tensor{K} \vect{\phi} = \tensor{G}^x\vect{u}^* + \tensor{G}^z\vect{w}^* - \tensor{B}\vect{v}_{bc}
\end{align}
with $ \tensor{K} $ being the classic stiffness matrix, whereas $ \tensor{G}^x , \tensor{G}^z , \tensor{B} $ correspond to suitable integrals of the Q4 shape functions over the domain and its boundary (see eq.(\ref{kggb})).

After solving the linear system for $\vect{\phi}$, we can then update the velocity and pressure field using eq.(\ref{eq:pco1}):
\begin{subequations}\label{eq:pco_discrete}
\begin{align}
    \vect{u}^{n+1} = \vect{u}^* - \dfrac{ (\tensor{G}^x)^T \vect{\phi} }{ \vect{ L \Delta x \Delta \xi } } \\
    \vect{w}^{n+1} = \vect{w}^* - \dfrac{ (\tensor{G}^z)^T \vect{\phi} }{ \vect{ L \Delta x \Delta \xi } } \\
    \vect{q}^{n+1} = \vect{q}^n + \dfrac{\vect{\phi}}{ \Delta t^n } 
\end{align}
\end{subequations}
where the division is considered to be element-wise.
For all simulations herein, the \textit{Pardiso} direct sparse solver \cite{schenk2001pardiso} is used for the solution of eq.(\ref{linpco}).

\subsection{Wetting-drying and positivity preservation}

The last thing to specify is the treatment of the wetting-drying of cells. If the cells tend to dry ($L\to 0$), then one cannot get the primitive variables $u,w,\rho$ from their conservative counterparts $ Lu , Lw , L\rho $ without sacrificing numerical stability. Following standard practices, we define two tolerances $\epsilon_{vel} , \epsilon_\rho$ so that if $L_{ij}$ is below these, we set $u=w=0$ and the density is retained at its pre-drying value. For all the present simulations, we set $\epsilon_{vel} = 10^{-3}  \ m $ and $\epsilon_\rho = 10^{-6} \ m $. These values are chosen empirically and the simulation results are insensitive with respect to them.

Additionally, in the FEM solver, cells with $L_{ij}<\epsilon_{vel}$ are considered dried and the condition imposed on them is that the pressure values on the top and bottom sides are equal: $q_{i,j+1} = q_{i,j}$ and $q_{i+1,j+1} = q_{i+1,j}$, thus avoiding any singularities when constructing the stiffness matrix.

Another aspect that must be addressed is that the advection scheme must keep the cell volume positive: $L \geq 0$. In order to directly enforce that, we use the approach presented in \cite{liu2018well}, where the fluxes are directly limited based on a priori indicators. This ensures positivity without imposing time-step or any other restrictions on the method. The exact same procedure is also used on the vertical fluxes.

\section{Spurious Mixing Overview}\label{sec:sdm}

In this section the topic of SDM is expanded in order to clearly place the foundation upon which the present ALE scheme will be formulated.

As stated in \S\ref{sec:intro}, SDM is a numerical artifact that can severely impact the fidelity of ocean models. As such, it has gathered a lot of attention within the ocean modelling literature.
Early studies \cite{griffies2000spurious,lee2002spurious} focused on identifying SDM in $z$-models and proposed methods to quantify and diagnose it. In these works, it is argued that the advection scheme's truncation errors can introduce artificial dissipation that leads to SDM.

Following this, many studies have reported such artifacts in ocean models \cite{urakawa2014effect,garinet2024spurious} and highlighted their negative effect on the fidelity of the simulations. That being said, despite the proposition of various approaches to identify and treat SDM, there exist many relevant open questions to this day concerning the subject \cite{klingbeil2019reducing}. A minimal overview of such topics is presented below.

\subsection{Global and local indicators}

Since the first studies of SDM, the main objective of identifying and quantifying it has been addressed in numerous papers. In \cite{griffies2000spurious}, the framework of \cite{winters1995available} is utilized, where mixing is quantified by tracking the {reference potential energy} (RPE) of the domain. The RPE approach has gained a lot of popularity in quantifying SDM and has appeared in other studies \cite{ilicak2012spurious,gibson2017attribution}. Since RPE is expressed through an integral over the whole domain, it constitutes a \textit{global} indicator of the production of SDM. In order to get the spatial distribution of SDM, in \cite{ilicak2016quantifying} RPE-based diagnostics were extended to obtain spatial indicators.

Another approach appears in \cite{lee2002spurious} and is based on estimating the effective diffusivity through water-mass transformation diagnostics and has also been applied in a more recent analysis \cite{megann2018estimating}. The estimated diffusivity is local and can be directly compared with the physical one, thus automatically resulting in a spatial estimation of SDM.

Even though these indicators have appeared in many studies, their derivation is primarily based on the physical energetics of mixing rather than on the structure of the numerical scheme, which is the primary generator of SDM. This is not the case with the method of {Discrete Variance Decay} (DVD) \cite{burchard2008comparative, klingbeil2014quantification}, where local indicators are calculated based on the discrete form of the equations. This framework will be used herein and will therefore be presented in some detail.

\subsubsection{Discrete Variance Decay}\label{sec:mixing}

The DVD framework relies on quantifying terms that result in the decrease of the variance $\iiint (\cdot)^2 \di V$ of a tracer. For example, in the absence of tracer diffusion (this is the case for $\rho$ throughout this paper), we have $\mdv{\rho}{t}=0$, which implies that in the absence of boundary fluxes we have $\odv{}{t} \iiint \rho^2 \di V = 0 $. Thus, any decrease of the tracer variance will indicate diffusion introduced by the numerical scheme \cite{banerjee2024discrete}.

To further illustrate this using the notation of the present paper, the space-discrete DVD equation for the vertical advection is presented in Appendix \ref{app:dvd}. A notable drawback of this approach is that closed expressions of DVD for arbitrary FV reconstructions are difficult to derive analytically. Thus, in order to ease the analysis, the DVD is calculated under the hypothesis that the variables are piecewise constant per cell. This allows us to derive the following:
\begin{align}\label{eq:dvd}
    \odv{}{t} \sum_{ij} \rho_{ij}^2 ( \Delta x_i L_{ij} \Delta \xi_j) = - \underbrace{\sum_{ij} \Delta x_i |\vartheta_{i,j+1/2}| ( \rho_{i,j+1} - \rho_{ij} )^2}_{S}
\end{align}
The $S$-term is sign-definite and is a discrete analog of the decay due to diffusion. To better understand this, we write:
\begin{align}\label{eq:dec}
    S = \sum_{ij} \Delta x_i \Delta z_{i,j+1/2} \times  |\vartheta_{i,j+1/2}| \times \Delta z_{i,j+1/2}  \left(\dfrac{\rho_{i,j+1} - \rho_{ij}}{\Delta z_{i,j+1/2}}\right)^2 \approx \iiint |\vartheta| \times \Delta z \times \left( \mathrm{grad}_z \rho\right)^2 \di V
\end{align}
where $\Delta z_{i,j+1/2} \eqdef \frac{1}{2} \left( \Delta\xi_{j+1} L_{i,j+1} + \Delta\xi_j L_{ij}  \right)$ is the local vertical resolution of the mesh.

The key takeaway of this is that the numerical diffusion associated with a low-order FV discretization is the product of the strength of vertical advection $|\vartheta|$, the local vertical spacing $\Delta z$ and the magnitude of the vertical gradient of the density $\mathrm{grad}_z \rho$.

From these three terms, the first two can be regulated through the ALE scheme, while the last one stands as an indicator of the local strength of mixing and depends on the stratification itself. Having this low-order estimate as a baseline, we will briefly present the two main approaches that have been proposed for the reduction of SDM.

\subsection{High-order advection schemes}\label{sec:sdm_num}

One of the most direct approaches to reduce SDM is to utilize a higher-order transport scheme. Indeed, the first-order scheme introduces excessive diffusion and thus it is natural to seek high-order alternatives.

A next step might be to adopt a second-order MUSCL approach, such as the one used herein (see \S\ref{sec:fv}). The second-order reconstruction reduces the jumps at the faces, thus resulting in smaller numerical diffusion. In order to get stability and convergence properties, the use of a limiter is needed, which enforces the Total Variation Diminishing (TVD) property, i.e. that $TV(\cdot) \eqdef \sum_i \left| (\cdot)_{i+1} - (\cdot)_i \right|$ decays in time. These limiters directly reduce the reconstructed slopes thus impacting the accuracy and diffusivity of the scheme, especially around extrema.

These MUSCL-TVD approaches serve as a common baseline for transport schemes and attempts of further mitigating SDM have been made. Those include carefully designed dissipation operators \cite{marchesiello2009spurious,garinet2024spurious}, high-order remap schemes \cite{white2009high} and ENO/WENO reconstructions \cite{klingbeil2019reducing}.

Although these schemes result in high-order convergence and reduced SDM, their secondary properties must be carefully studied. This is especially true for mixing, since even MUSCL-TVD schemes have been found to produce negative mixing when using specific limiters \cite{mohammadi2015impact}. This occurs because high-order schemes --unlike the first-order upwind one-- do not necessarily have DVD that is monotonic in time and thus can introduce anti-diffusive tendencies.

Concluding, the design of high-order schemes seems to be a significant avenue to reduce SDM, but attention must be given to the specific properties of the scheme. In the present paper, we work entirely within the confines of MUSCL-TVD procedures. This, however, does not discourage the usage of the presented ALE algorithm alongside other (and possibly high-order) transport schemes.

\subsection{GVC and Vertical ALE treatment}

Parallel to the design of high-order schemes to combat SDM, another approach involves working within the GVC framework and designing the vertical coordinate accordingly. For example, a purely isopycnal model does not permit vertical mass transfer and keeps the density constant per layer. This greatly reduces SDM, but retains the drawbacks of a purely Lagrangian approach (see \S\ref{sec:intro}).

Despite the drawbacks, this gives evidence that the design of vertical coordinates could play an important role in reducing SDM. This is indeed the case and attempts in this direction have been made within the framework of hybrid ALE schemes \cite{klingbeil2019reducing}. The two main hybrid ALE approaches include the $\widetilde{z}$ coordinate \cite{leclair2011z} and the adaptive mesh movement of \cite{hofmeister2010non}. These works showed that the ability of the mesh to follow the underlying dynamics can significantly improve their numerical resolution. The first one was further tested in \cite{petersen2015evaluation}, where evidence that $\widetilde{z}$ can reduce SDM was reported.

In the present work, this avenue is further explored with an emphasis on nonhydrostatic phenomena of smaller spatiotemporal scales with the goal of reducing SDM.

\section{ALE movement through variational minimization}\label{sec:ale}

In the current section the ALE strategy is detailed, alongside the derivation and analysis of the proposed variational scheme.

As was mentioned above, adaptive mesh algorithms already exist in the literature on ocean models \cite{burchard2004non}. The distinctive characteristic of our proposed variational mesh movement is that it is derived through the minimization of a cost functional, thus satisfying an optimality criterion. That being said, the ALE strategies considered herein (Lagrangian bias, smoothing and refinement) have been used in previous works \cite{hofmeister2010non,delandmeter2018fully} and their effectiveness has been demonstrated.

The proposed method aims to gather all these ideas into a compact and consistent mathematical formulation. By doing so, we are able to perform the analysis demonstrated in \S\ref{sec:freqfilter},\S\ref{sec:monitor} and propose how the various target objectives should blend.

\subsection{General strategy}

To properly formulate the ALE scheme, suppose that the layers at the start of the timestep $t^n$ are positioned at $z^n(x,\xi)$ and at the end of the step $t^{n+1}$ will be at $z^{n+1}(x,\xi)$. In the present section, we will focus on expressions that are discrete only in time so that the derivation of the ALE scheme is independent of the spatial discretization.

The evolution of the layers' positions is dictated by the GCE, which in semi-discrete form reads:
\begin{align}\label{eq:dzdt}
  (\ref{s1}) \then  &\dfrac{ L_{}^{n+1} - L_{}^n }{ \Delta t } + \pdv{}{x}(Lu)^* + \pdv{\vartheta}{\xi} = 0 \then
    z^{n+1} = \underbrace{ z^{n} - \Delta t \int_0^\xi \pdv{}{x} (Lu)^* \di \xi }_{z^*_{lag}} - \Delta t \cdot \vartheta
\end{align}
The notation $(\cdot)^*$ signifies calculation in an intermediate state that depends on the timestepping scheme.

Thus, we see that the dia-surface velocity $\vartheta$ (see also \S\ref{sec:gce}) acts as a correction on the Lagrangian forcing $R_{lag} \eqdef  -\int_0^\xi \pdv{}{x} (Lu)^* \di \xi  $. As a result, given the initial positions $z^n$ and $R_{lag}$, we will construct an optimization problem in order to calculate $\vartheta$ and get the end-of-step layer positions $z^{n+1}$.

In order to formally write the optimization criteria, we turn to eq.(\ref{eq:dvd}) with the goal to reduce the $S$ term, which quantifies the production of SDM in the DVD equation. Working term-by-term we see that a suitable strategy could be to limit $|\vartheta|$ and/or make the vertical spacing $\Delta z$ finer in areas with large gradients $|\mathrm{grad}_z \rho|$. The following derivation will be based on this dual approach.

\subsubsection{Remark on the horizontally induced SDM}\label{sdm_hor}

The dual objectives of reducing $|\vartheta|$ and the product $ \Delta z \times \left( \mathrm{grad}_z \rho \right)^2 $ are based on the DVD estimate of eq.(\ref{eq:dvd}) that ignores the horizontal advection terms. This however does not imply that horizontally induced SDM is by any means negligible. If analysis is performed on the horizontal advection too, a dissipation term of similar nature $ \textstyle \Delta x_i \times \left| F^1_{i+1/2,j} \right| \times \pdv{\rho}{x}^2$ would appear in the DVD equation.

From these three terms, only the last one can be reduced by the vertical ALE scheme. Essentially, reducing $\pdv{\rho}{x}^2$ means minimizing density variations along the iso-$\xi$ surfaces (layers), which corresponds to an isopycnal tendency. Although this is not explicitly stated as an optimization target, the proposed variational ALE movement seems to adequately track the density interfaces, thus implicitly regulating $\pdv{\rho}{x}^2$. A more careful study of the influence of each component of SDM alongside other types of isopycnal tendencies is left for future work.

\subsection{Optimization process and variational formulation}

Our approach is to gather all the different target objectives and then construct an appropriate optimization problem in order to minimize the product $ |\vartheta| \times \Delta z \times (\mathrm{grad}_z\rho)^2$. Specifically, we want to:
\begin{tcolorbox}
\begin{Oitemize}
    \item\label{o1} Minimize $\vartheta^2$, so that the vertical mass transfer is limited: this introduces a Lagrangian tendency.
    \item\label{o2} Minimize the gradients $\pdv*{(z-z_{ref})}{x}^2$ and $\pdv*{(z-z_{ref})}{\xi}^2$: this smoothens out the mesh and skews it towards a reference configuration $z_{ref}$. This can be of any desired type (like a $z/\rho$ model). Herein, a $\sigma$ reference grid will be used for all examples: $z_{ref} = -h + \xi H$
    \item\label{o3} Minimize $\left( M \pdv{z}{\xi} \right)^2$: this distributes more mesh points where the function $M(x,\xi)\geq 0$ is larger. If $M \propto \left| \mathrm{grad}_z \rho  \right| $, this distributes the mesh so that the right hand side (RHS) of eq.(\ref{eq:dvd}) is reduced.
\end{Oitemize}
\end{tcolorbox}

The function $M$ essentially offers the option to design the spacing of the mesh according to user-defined criteria. These terms are called \textit{monitor functions} and are heavily used in variational mesh adaptation \cite{cao1999study,huang2003variational}. Having these techniques as inspiration, we propose an optimization procedure that aims to reduce SDM within the framework of ALE ocean models.
\par
In order to enforce the optimality criteria \ref{o1}-\ref{o3} at the end-of-step layer positions $z^{n+1}(x,\xi)$, the following cost functional is proposed:\begin{align}\label{eq:functional0}
    \mathcal{F}(\vartheta) = \iiint {\bigg [} \underbrace{ \vphantom{ \bigg | } T_{ref} a_\vartheta \vartheta^2}_{\text{Lagrangian}} + \underbrace{ a_x \dfrac{\Delta x^2}{\Delta t} \pdv*{(z^{n+1}-z_{ref})}{x}^2 + a_\xi\dfrac{\Delta \xi^2}{\Delta t} \pdv*{(z^{n+1}-z_{ref})}{\xi}^2 }_{\text{Smoothing}} + \underbrace{ a_M \dfrac{\Delta \xi^2}{\Delta t} \left( M\pdv{z^{n+1}}{\xi} \right)^2}_{\text{Monitor function}} {\bigg ]} \di x \di \xi
\end{align}
or in tensorial notation:
\begin{subequations}\label{eq:functional}
\begin{align}
    \mathcal{F}(\vartheta) &= \iiint \left[ T_{ref} a_\vartheta \vartheta^2 + \dfrac{1}{\Delta t} \norm{ \tensor{A}_S \nabla(z^{n+1}-z_{ref})}^2 + \dfrac{1}{\Delta t} \norm{ \tensor{A}_M \nabla z^{n+1} }^2 \right] \di x \di \xi \\
    \text{where:} \quad \tensor{A}_S &= \begin{bmatrix} \sqrt{a_x} \Delta x & 0 \\ 0 & \sqrt{a_\xi} \Delta \xi \end{bmatrix} \quad \text{and} \quad \tensor{A}_M = \begin{bmatrix} 0 & 0 \\ 0 & \sqrt{a_M} \Delta \xi M \end{bmatrix}
\end{align}
\end{subequations}

The coefficients $a_\vartheta,a_x , a_\xi $ are non-dimensional, whereas $a_M$ must be carefully chosen based on the form of $M$.

The coefficients determine the relative influence of the objectives \ref{o1}--\ref{o3}: $a_\vartheta$ controls the strength of the Lagrangian bias, $a_x$ and $a_\xi$ regulate grid smoothing in the $x$ and $\xi$ directions, respectively, while $a_M$ controls the degree of refinement induced by the monitor function $M$.
The modeler has the freedom to specify these four parameters based on the particular configuration under study.
Below, it will be shown how these four values should blend.

$T_{ref}$ represents a characteristic time scale. Since this study concerns the evolution of waves in stratified flows, we chose $T_{ref} = \sqrt{{h}{/}{g'}}$ where the reduced gravity is given by $g' \eqdef g ( { \rho_{max} - \rho_{min} } ) { / } { \rho_0 } $.

Gathering all these, the optimization problem states:

\begin{tcolorbox}[ams align*]
    \text{find} \qquad \vartheta \in \mathcal{V}
    \qquad \text{so that} \qquad
    \mathcal{F}(\vartheta) =
    \min_{\vartheta_* \in \mathcal{V}}
    \left\{ \mathcal{F}(\vartheta_*) \right\}
\end{tcolorbox}
where $\mathcal{V}$ is a suitable function space. Through the framework of the calculus of variations, in Appendix \ref{app:var} we conclude that $\vartheta$ obeys the following:
\begin{align}\label{eq:elliptic}
     a_\vartheta \dfrac{T_{ref}}{\Delta t} \vartheta - \nabla \cdot \left[ \tensor{A}^2_S \nabla(\vartheta - v_{lag}) + \tensor{A}^2_M \nabla \left( \vartheta - \dfrac{z_{lag}^*}{\Delta t} \right) \right] = 0
\end{align}
where $v_{lag} \eqdef \dfrac{ z^*_{lag} - z_{ref} }{\Delta t}$.
The expression in eq.(\ref{eq:elliptic}) is a Helmholtz equation on the $x-\xi$ plane, hereafter referred to as the elliptic equation. This is significant because the optimization process yields the optimality criterion in equation form through a well-studied elliptic partial differential equation that can be solved efficiently using standard numerical methods, such as Finite Difference, Finite Element, or Finite Volume schemes.

To better understand eq.(\ref{eq:elliptic}), we can reformulate it by substituting $z^{n+1} = z^{n} + \Delta t ( R_{lag} - \vartheta)$, which results in:
\begin{align}\label{eq:elliptic2}
    a_\vartheta T_{ref}
    \dfrac{z^{n+1}-z^n}{\Delta t} =
    a_\vartheta T_{ref} R_{lag} +
    \nabla \cdot \left[
    \tensor{A}^2_S \nabla \left( z^{n+1} - z_{ref} \right) +
    \tensor{A}^2_M \nabla z^{n+1} \right]
\end{align}

In order to balance these 4 coefficients ($a_\vartheta , a_x , a_\xi , a_M$), the analysis is divided into two parts. First, we will describe the interaction of $a_\vartheta$ with $a_x,a_\xi$ followed by the balance between $a_M$ and $a_\xi$.

\subsection{The variational minimizer as a frequency filter}\label{sec:freqfilter}

Firstly, we aim to balance the Lagrangian and smoothing terms. To do so, let $a_M=0$ and $a_x = a_\xi = a_{x\xi}$. Then, all coefficients in eq.(\ref{eq:elliptic2}) are constant and thus Fourier analysis can be performed based on a space/time expansion of the type:
\begin{align}
    z = \sum_{\{ \omega , k_x , k_\xi \}} \left. \widehat{z} \right._{\{ \omega , k_x , k_\xi \}} \exp\left( {i \omega t + i  k_x  x + i  k_\xi  \xi} \right)
\end{align}
with $i\eqdef\sqrt{-1}$. Inserting this into eq.(\ref{eq:elliptic2}) and supposing that $\Delta t$ is small enough so that $\frac{z^{n+1}-z^n}{\Delta t} \to \pdv{z}{t}$ we get:
\begin{align}\label{eq:fourier0}
    i a_\vartheta \left( \omega T_{ref} \right) \widehat{z} = a_\vartheta T_{ref} \widehat{R}_{lag}  -  a_{x\xi} (k_*) ^2(\widehat{z}-\widehat{z}_{ref})
\end{align}
where $k_* \eqdef \sqrt{(k_x\Delta x)^2 + (k_\xi \Delta \xi )^2 } $ is the combined non-dimensional wavenumber.

If, in addition, we define $\omega_* \eqdef \omega T_{ref}$ and $\widehat{z}_{lag} \eqdef \widehat{R}_{lag} / (i\omega) $, we get:
\begin{align}\label{eq:fourier}
    \widehat{z} = \dfrac{ \left( a_\vartheta i \omega_*\right)  \widehat{z}_{lag} \ + \ \left( a_{x\xi} k_*^2 \right)  \widehat{z}_{ref} }{ \left( a_\vartheta i\omega_* \right) \ + \ \left( a_{x \xi} k_*^2 \right) }
\end{align}
According to eq.(\ref{eq:fourier}), the ratio between $a_\vartheta$ and $a_{x\xi}$ dictates how aggressively the high-wavenumber spatial Lagrangian modes are damped. Also note that with respect to the time frequency, higher $\omega_*$ favor the Lagrangian term $\widehat{z}_{lag}$. Consequently, the mesh follows the fast Lagrangian motions that are prescribed by the horizontal fluxes, while smoothing slow motions.

In order to balance $a_\vartheta$ and $a_{x\xi}$, we study the ratio between the Lagrangian and smoothing terms for the case of $\omega_*=2\pi$, i.e. motions that have a period equal to the reference one.
Since all the functions are defined on a grid with spacings $\Delta x , \Delta \xi$ , the highest wavenumbers that can be expected are the Nyquist ones: $ k_x \leq \pi / \Delta x $ and $ k_\xi \leq \pi / \Delta \xi $. This corresponds to $k_* \leq \pi = k_{\text{Nyquist}}$.
Based on this, the ratio for $T=T_{ref}$ is:
\begin{align}\label{eq:ratio}
    (\ref{eq:fourier}) \then \dfrac{\text{Lagrangian amplitude}}{\text{Smoothing amplitude}} = \dfrac{a_\vartheta\omega_*}{a_{x\xi}k_*^2} = \dfrac{a_\vartheta}{a_{x\xi}} \times \dfrac{2}{\pi} \times \left( \dfrac{k_{\text{Nyquist}}}{k_*} \right)^2
\end{align}

\begin{figure}
    \centering
    \includegraphics[width=0.6\textwidth]{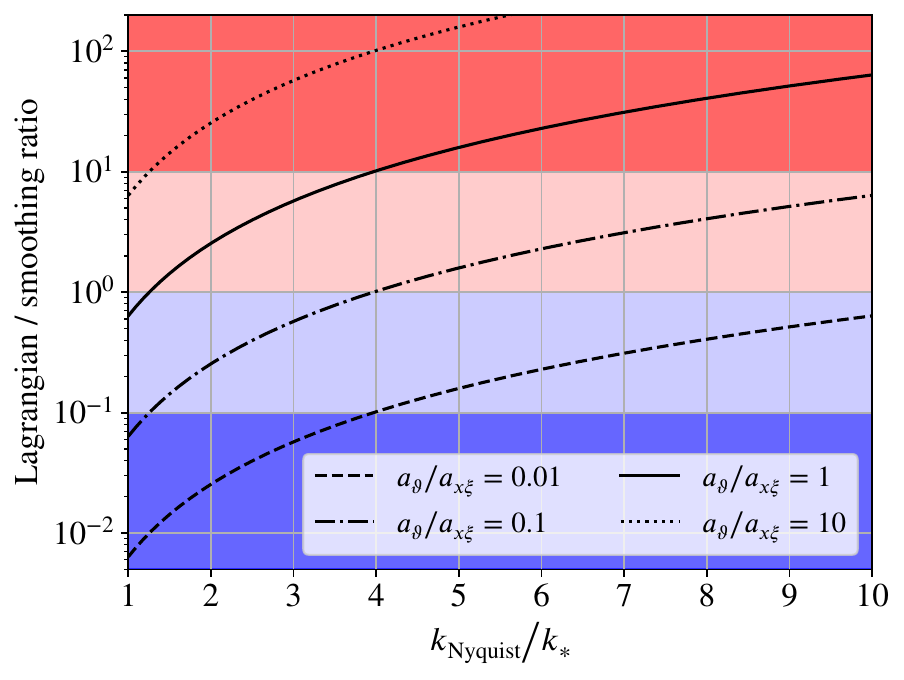}
    \caption{Diagrams of the Lagrangian-Smoothing ratio of eq.(\ref{eq:ratio}) based on the wavenumber for different values of $a_\vartheta/a_{x\xi}$ at $\omega_*=2\pi$. The y axis is in logarithmic scale. Lagrangian dominated areas are colored with red, while Smoothing dominated ones with blue.}
    \label{fig:filter}
\end{figure}

In Fig.\ref{fig:filter} this ratio is plotted for different values of $a_\vartheta/a_{x\xi}$. For $a_\vartheta / a_{x\xi}=0.1$, at the Nyquist limit we have a 1 to 10 Lagrangian-Smoothing ratio. At $k_{\text{Nyquist}}/k_*=4$ the split is even, while at $k_{\text{Nyquist}}/k_*>10$ the curve enters the Lagrangian dominated regime. This means that the noise-like grid scale oscillations will be strongly dampened, while medium wavenumber Lagrangian motions will remain unchanged.

As the $a_\vartheta/a_{x\xi}$ ratio goes up, we see that this damping effect is reduced. With $a_\vartheta/a_{x\xi}=1$, the split at the Nyquist wavenumber is even and the ratio quickly enters the Lagrangian regime. This is more apparent at $a_\vartheta/a_{x\xi}=10$ where at all wavenumbers the curve is essentially within the Lagrangian regime, while on the other hand the $a_\vartheta/a_{x\xi}=0.01$ curve is entirely below the Lagrangian regime for $1<k_{\text{Nyquist}}/k_*<10$.

This analysis suggests that the ratio $a_\vartheta/a_{x\xi}=0.1$ produces a balanced filter that smoothens grid-scale noise while also resolving the medium-sized motions. Values larger than this bias the solver toward Lagrangian dynamics and should be chosen carefully depending on the application.
Ratios below $ 0.1 $ further suppress Lagrangian motion and are advisable in the case where rapid vertical mass transfer is essential, such as during wave breaking and overturning (see \S\ref{sec:results}).

In summary, for cases with strong isopycnal deformation, $ a_\vartheta/a_{x\xi} \sim 0.01 $ is advised, whereas $ a_\vartheta/a_{x\xi} \sim 0.1 $ is preferable if a moderate and balanced smoothing is desired.

Of course, these values should not be viewed as universal choices, but rather as indicative balance points between Lagrangian forcing and mesh smoothing. In applications where an almost fully Lagrangian configuration is desired, larger values of $a_\vartheta/a_{x\xi}$ can be preferable. This may well be the case in stably stratified scenarios, where isopycnal-type configurations can be suitable. In such cases, the above analysis and Fig.\ref{fig:filter} can be used to choose the ratio in order to remain within the Lagrangian regime as much as possible.

\subsection{Effect of the monitor function}\label{sec:monitor}

Unlike the above analysis, when dealing with the case of $a_M \neq 0$, the elliptic equation has space-varying coefficients and thus we cannot utilize Fourier analysis. In order to better understand the monitor function term, we choose to study $a_M$ alongside $a_\xi$ with $a_x=a_\vartheta=0$. Then, eq.(\ref{eq:elliptic}) gives:
\begin{align}
    \pdv{}{\xi} \left[ \left( 1 + \dfrac{a_M}{a_\xi} M^2 \right) \pdv{z}{\xi}  \right] = \pdv[order=2]{z_{ref}}{\xi}
\end{align}

Integrating the above we get:
\begin{align}
        \pdv{z}{\xi} = \dfrac{ \pdv{z_{ref}}{\xi} + C }{ 1 + A }
\end{align}
with $A \eqdef  \dfrac{a_M}{a_\xi} M^2$ and $C$ a suitable integration constant.

If now we consider the reference configuration to almost be iso-spaced\footnote{in the present paper, this assumption is exact, since the reference configuration is a $\sigma$-grid}, i.e. $ \pdv{z_{ref}}{\xi} \approx H(x,t) $ we get:
\begin{align}
    \pdv{z}{\xi} = \dfrac{H}{1+A} \left( \int_0^1 \dfrac{1}{1+A}\di \xi \right)^{-1}
\end{align}

The above equation expresses how the grid is vertically distributed based on the values of the monitor function $M$ since $\pdv{z}{\xi}$ is essentially the vertical size of the cells with $\Delta z_j \approx \pdv{z}{\xi}_j \times \Delta \xi_j$ (see also \S\ref{sec:gvc}). We, then, see that areas where $M$ is larger result in smaller values for $\pdv{z}{\xi}$, thus increasing the local resolution.

To further investigate the balance between $a_\xi$ and $a_M$, we can reformulate the above equation to show the ratio between the maximum and the minimum value of $\pdv{z}{\xi}$, which essentially corresponds to the ratio of the tallest and shortest cells in the vertical column:
\begin{align}\label{eq:Lratio1}
         \dfrac{ \pdv{z}{\xi}_{\max} }{ \pdv{z}{\xi}_{\min} }
    = \dfrac{ 1 + \dfrac{a_M}{a_\xi} M_{\max}^2 }{ 1 + \dfrac{a_M}{a_\xi} M_{\min}^2 }
\end{align}

This relation shows that  $a_M/a_\xi$ dictates how aggressively the mesh is refined in areas where $M$ is maximized. Even though the choice of the monitor function $M$ can be of any desired type (see \S\ref{sec:remarks}), to conclude the present analysis, we now specify its form in the context of reducing SDM.
In all numerical examples that follow, in order to minimize the spurious numerical mixing (see \S\ref{sec:mixing}) we chose the following monitor function, so that the RHS of eq.(\ref{eq:dvd}) is regulated:

\begin{align}\label{eq:monitor}
    M = \left. { \left| \pdv{\rho}{\xi}\right| }  \right/ { \left| \pdv{\rho}{\xi}\right|_{\max} }
\end{align}
By doing this, the ratio $a_M/a_\xi$ regulates the vertical cell size in areas with large density gradients. For example, according to eq.(\ref{eq:Lratio1}), if $a_M/a_\xi=1$, areas where the density gradient is maximized will result in almost 2 times smaller cells than those where density is almost uniform, whereas the choice $a_M/a_\xi = 2 $ triples the mesh resolution in these areas.

That being said, high $a_M/a_\xi$ ratios result in low numerical mixing, although excessively large values will deteriorate the overall quality since areas of constant density would become severely under-resolved. For the test cases considered herein, it was found through experimentation that $a_M/a_{\xi}=10$ results in a balanced mesh resolution that greatly reduces SDM, as will be demonstrated in \S\ref{sec:results}.

\subsection{Remarks on the scheme}\label{sec:remarks}

Before proceeding to the implementation of the variational ALE algorithm, some remarks must be made.
Although the present analysis is presented through the lens of FV schemes, terms like eq.(\ref{eq:dec}) serve only as error indicators and can be replaced if the underlying scheme is inherently different. For example, in the case of FEM methods, a plethora of literature is dedicated to the construction of such indicators both in \textit{a priori} and in \textit{a posteriori} form \cite{john2000numerical}. Those can be naturally included in the present formulation by suitably tuning the monitor function $M(x,\xi)$ without changing the rest of the algorithm.

Moreover, while the present paper focuses on the reduction of SDM, other target objectives can be naturally included in the variational framework. Specifically, $M(x,\xi)$ can be augmented in order to distribute more mesh points in areas of interest other than those with large density gradients. For example, in a more realistic ocean study, one might tune $M(x,\xi)$ to refine areas near the bottom and free surface in order to better capture boundary layer and mixed-layer dynamics, respectively.

The same is true for the reference grid $z_{ref}(x,\xi)$. This can be freely chosen based on experience and common practices in the ocean modelling literature on vertical coordinates (see \S\ref{sec:intro}) to fit specific needs.

Another thing that must be examined is the applicability of the proposed scheme in more realistic ocean models that operate at larger scales than the ones considered herein. Even though the cases in \S\ref{sec:results} focus primarily on ISWs in small-scale, idealized scenarios, effort was made in the derivation of the scheme to keep the framework as general and scale-independent as possible. To be specific, in the smoothing and monitor function tensors $\tensor{A}_S$, $\tensor{A}_M$, the characteristic lengths are the grid spacings $\Delta x$, $\Delta\xi$. As a result, the horizontal scales are treated implicitly since the analysis of grid smoothing must be performed using grid-scaled local distances so that highly skewed cells are avoided.

On the other hand, in the Lagrangian term, the ad hoc definition of the time scale parameter $T_{ref}$ is needed. While in the case of internal waves the choice of $T_{ref}$ is quite straightforward, there may exist situations (for example in realistic large-scale ocean simulations), where the definition of a single characteristic $T_{ref}$ is infeasible. This poses a limitation of the present approach that is left to be further studied in future works.

Lastly, we should note that the proposed variational mesh movement framework is not necessarily restricted to the target objectives considered herein, and can be extended to include extra optimality criteria.
To be specific, an additional term could be added to the functional $\mathcal{F}$, multiplied by a constant. Then, a variational derivative of this term should be calculated based on the process of Appendix \ref{app:var}, thus resulting in an extra term in eq.(\ref{eq:elliptic}). This procedure is well-defined and results in mathematically consistent schemes that can be studied accordingly.

This property is particularly important as it broadens the applicability of the proposed framework, allowing additional physical or numerical constraints to be incorporated. This strengthens the flexibility of the approach when the method is employed in different ocean models or in future applications of SLS to large-scale configurations.

\subsection{Numerical implementation of the variational minimizer}

Having described the variational mesh movement, we now specify its implementation in the solver:
\begin{enumerate}
    \item Calculate the horizontal GCE fluxes $F^1_{i+1/2,j}$ through eq.(\ref{eq:flux_L}).
    \item Based on this, use eq.(\ref{eq:dzdt}) to get $z_{lag}^*$ and specify the monitor function $M(x,\xi)$.
    \item Based on the chosen values for ($a_\vartheta,a_x,a_\xi,a_M$), solve the elliptic equation eq.(\ref{eq:elliptic}) numerically on the $x-\xi$ plane to calculate the optimal values for $\vartheta_{i,j+1/2}$.
    \item Use these $\vartheta_{i,j+1/2}$ to calculate the vertical numerical fluxes $G^{(\cdot)}_{i,j+1/2}$.
\end{enumerate}

An important feature of this approach is that it can be applied to any ocean model that is based on a GCE-type formulation. This is true for a large portion of ocean models, since the layers' heights are usually specified using the continuity equation. To adopt the proposed method, one can specify the Lagrangian impulse, the desired monitor function $M$ and solve an elliptic problem without altering the structure of the solver.

With regard to the numerical cost, we note that the proposed procedure is an optional optimization. Consequently, the elliptic equation can be solved with moderate accuracy without significantly affecting the solver’s overall performance. In the present study, we use the Jacobi iterative method with a finite-difference space discretization on a classic 5-point orthogonal stencil. In this case, only a small number of iterations were found to be sufficient. More specifically, when using $a_\vartheta=0.1$ , $a_x=a_\xi=1$, $a_M=10$, only 3-7 iterations are needed when using the convergence threshold $ || { \vect{\vartheta}^\text{previous}-\vect{\vartheta}^\text{next} } ||_\infty < 10^{-5} $.

\section{Results}\label{sec:results}

In the following subsections, the ability of the proposed ALE scheme to efficiently simulate inviscid stratified flows is demonstrated through test cases of varying complexity. Specifically, in order to validate the ability of SLS to maintain the form of a traveling ISW, a convergence test towards an analytical solution is presented in \S\ref{sec:djl}. \par 
Next, in \S\ref{sec:isw_tanh} the ability of the ALE scheme in non-breaking ISW propagation and shoaling is presented and benchmarked with published results.
In \S\ref{sec:isw_wedge}, the ability of the mesh movement to handle cases with rapid and considerable isopycnal deformation with stability and accuracy is tested and validated using experimental data. Lastly, in \S\ref{sec:agh} the breaking of an ISW over a sloping beach is investigated and comments are made with regard to the effectiveness of the variational mesh movement strategy to mitigate SDM, while preserving relevant flow characteristics.

In all simulations, the timestep is calculated through the CFL condition found in \cite{alexandris2024semi} with a CFL constant of 0.45.

Additionally, the Boussinesq approximation (see \S\ref{sec:formulation}) is utilized in all cases except that of \S\ref{sec:isw_wedge}.
Regarding the mesh sizes, about 100 cells per characteristic horizontal length are used, which in \S\ref{sec:isw_wedge},\S\ref{sec:agh} corresponds to the length of the soliton, while in \S\ref{sec:isw_tanh} to the smallest harmonic we wish to resolve. In the vertical direction, $n_l \approx 30$ layers were found to exhibit good linear dispersion properties (see also \cite{alexandris2024semi}), which is necessary for the resolution of the phenomena presented herein. In \S\ref{sec:agh}, a larger number of layers $n_l=75$ is used in order to properly capture the vertical advection during the overturning. In each test case, the results presented are mesh-independent and produce stable solutions.

\subsection{ISW generation}\label{sec:isw_gen}

In the following, the Internal Solitary Waves (ISWs) are initialized using solutions of the Dubreil-Jacotin-Long (DJL) equation \cite{long1953some}. Specifically, we utilize PyDJL\footnote{\url{https://github.com/golnazir/PyDJL}}, an open-source solver that employs the methodology described in \cite{dunphy2011} to numerically solve the DJL using Fourier expansions. In all cases, 4096 modes are used in each direction and the density profile is:
\begin{align}
    & {\rho(z)} = \dfrac{\rho_1+\rho_2}{2} - \dfrac{\rho_2-\rho_1}{2} \tanh \left( \dfrac{ z - z_{pyc} }{h_{pyc}} \right)  
\end{align}
with $z_{pyc}$ and $h_{pyc}$ being the location and thickness of the pycnocline. The soliton's amplitude is chosen by calibrating the Available Potential Energy (APE) that the code uses as input. Details about the ISWs for all benchmark cases considered herein are displayed in Table \ref{tab:isw}.

\begin{table}
    \centering
    \begin{tabular}{c|c|c|c|c|c|c|c}
        & $\rho_1[kg/m^3]$ & $\rho_2[kg/m^3]$ & depth$[m]$ & $z_{pyc}[m]$ & $h_{pyc}[mm]$ & $APE[J]$ & Amplitude$[m]$ \\ \hline
        \S \ref{sec:djl} \& \S\ref{sec:agh} &  1000 & 1040 & 0.15 & -0.02 & 2.50 & 3.8000$\cdot 10^{-5}$ & 0.033  \\
        \S \ref{sec:isw_tanh} &  990 & 1010 &  1.00 & -0.25 & 8.75 & 4.9335$\cdot 10^{-3}$ & 0.048  \\
        \S \ref{sec:isw_wedge} &  996 & 1030 &  0.50 & -0.10 & 40.00 & 3.2700$\cdot 10^{-4}$ & 0.056
    \end{tabular}
    \caption{Details of the ISWs used in the numerical test cases of \S\ref{sec:results}.}
    \label{tab:isw}
\end{table}

\subsection{Convergence test on a traveling soliton}\label{sec:djl}

For the purpose of validating the solver's accuracy in capturing a traveling DJL soliton, a convergence test is conducted. Specifically, an ISW of amplitude $0.033 m$ (see Tab.\ref{tab:isw}) on the domain $(x,z) \in \left[ 0 , 1.5 m \right] \times \left[ -0.15 m , 0 \right] $ is used. The analytical solution at the domain boundaries is enforced through Dirichlet boundary conditions. The pressure obeys Neumann BC.

As shown in Appendix \ref{app:djl},
the solver converges in a robust way, while staying below the relative error of 1\% even at a small number of DOFs. This indicates that SLS can accurately and consistently capture the form of the traveling nonlinear soliton. This is a necessary condition for the test cases that follow, since DJL solitons will be used for the initialization of the simulations and associated errors would severely impact the quality of the results.

\subsection{Shoaling of an ISW on a shelf}\label{sec:isw_tanh}

\begin{figure}[t]
    \centering
    \includegraphics{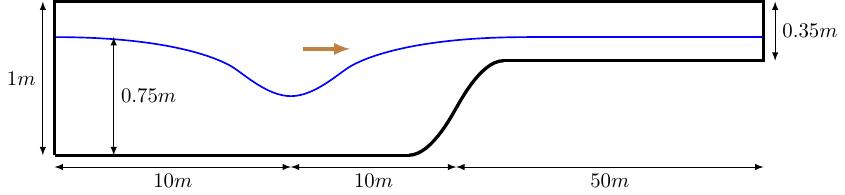}
    \caption{Sketch of the ISW shoaling test case of \S\ref{sec:isw_tanh}. The ISW data can be seen in Tab.\ref{tab:isw}, while the topography is defined by eq.(\ref{eq:tanh_topo}). The ISW propagates rightwards, while all boundaries are considered as impermeable walls.}
    \label{fig:tanh_tikz}
\end{figure}

\begin{figure}[t]

    \centering
    \includegraphics[width=0.32\linewidth]{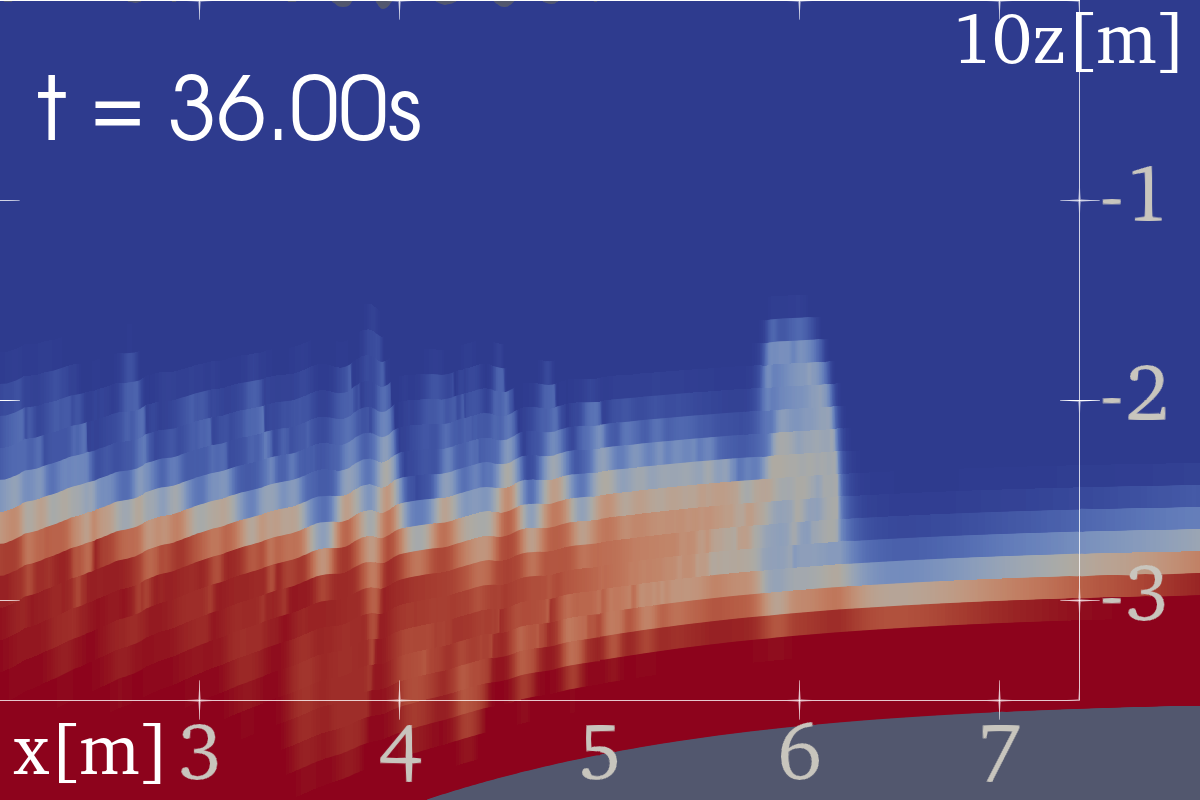}
    \includegraphics[width=0.32\linewidth]{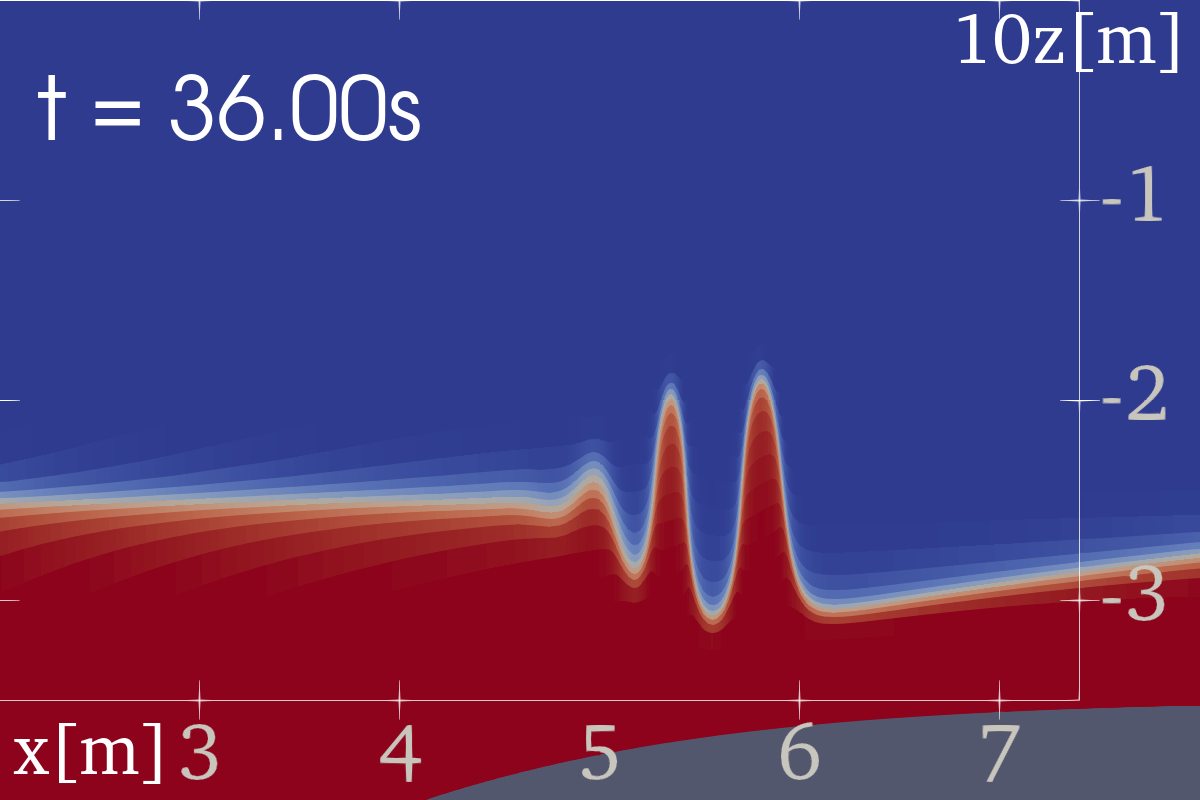}
    \includegraphics[width=0.32\linewidth]{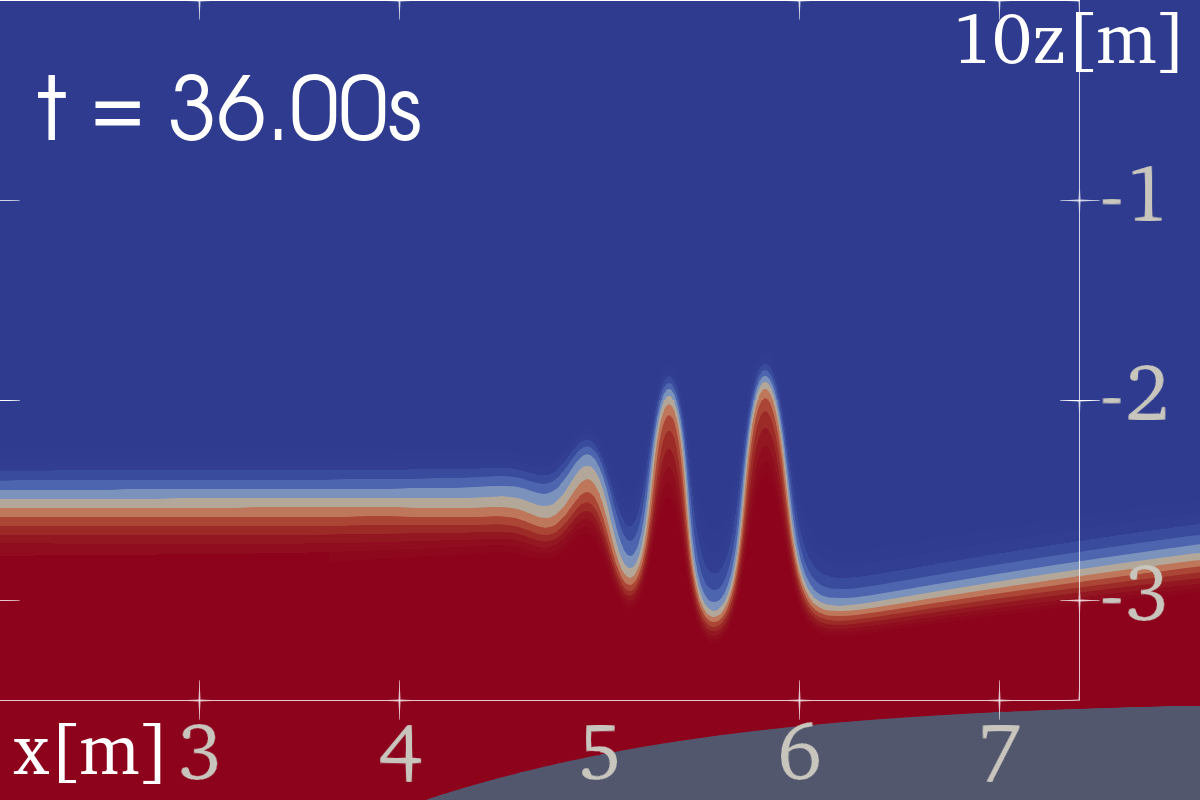}
    \includegraphics[width=0.4\linewidth]{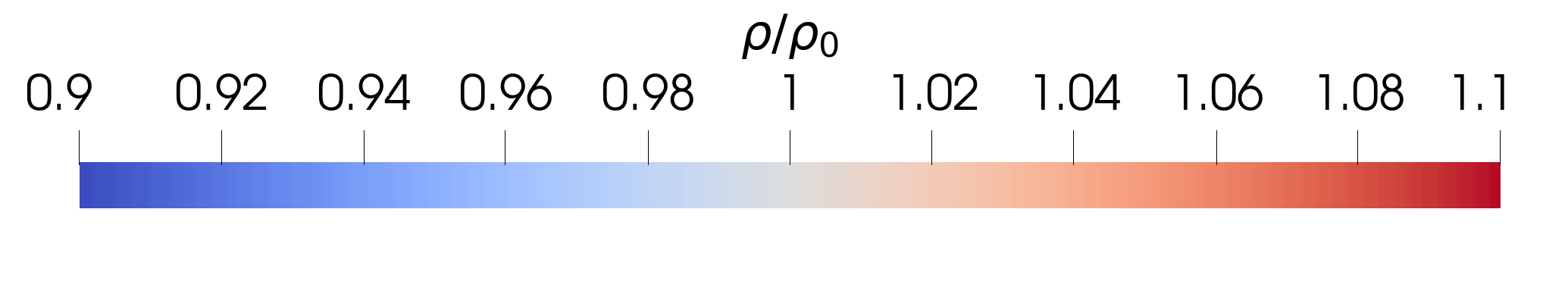}
        \caption{Density contours from the SLS simulations of the shoaling ISW at $t=36s$. \textbf{Left:} SLS with $\sigma$-grid configuration, \textbf{middle:} SLS with variational ALE scheme, \textbf{right:} isopycnal mode with $\vartheta = 0$.}
    \label{fig:tanh_compare}
\end{figure}

\begin{figure}[t]
    \centering
    \begin{minipage}[t]{0.48\linewidth}
        \centering
        \vspace{-1pt}
        \includegraphics[width=\linewidth]{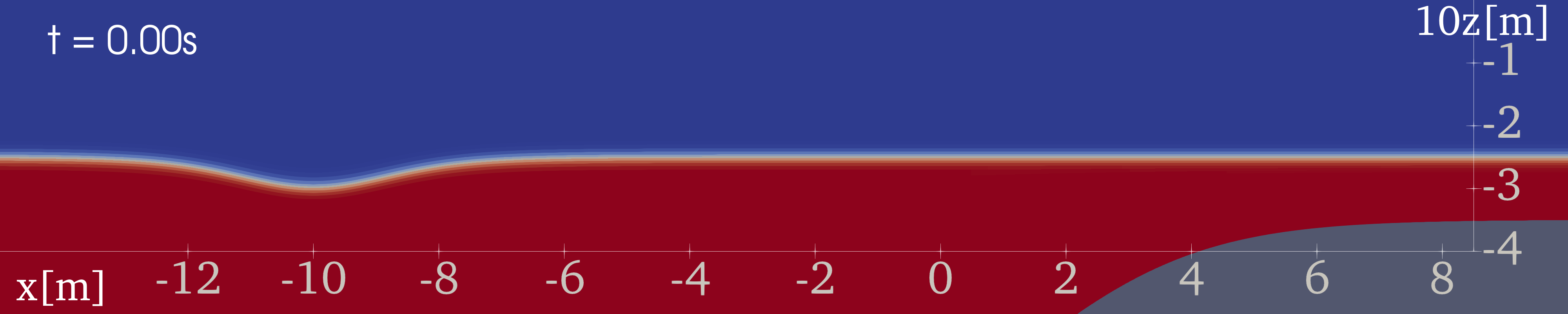}\vspace{4pt}
        \includegraphics[width=\linewidth]{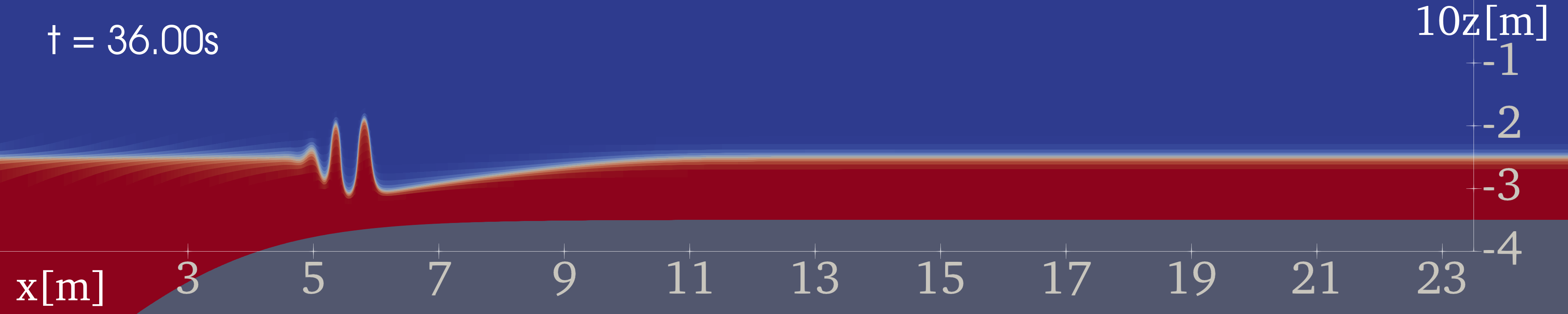}\vspace{4pt}
        \includegraphics[width=\linewidth]{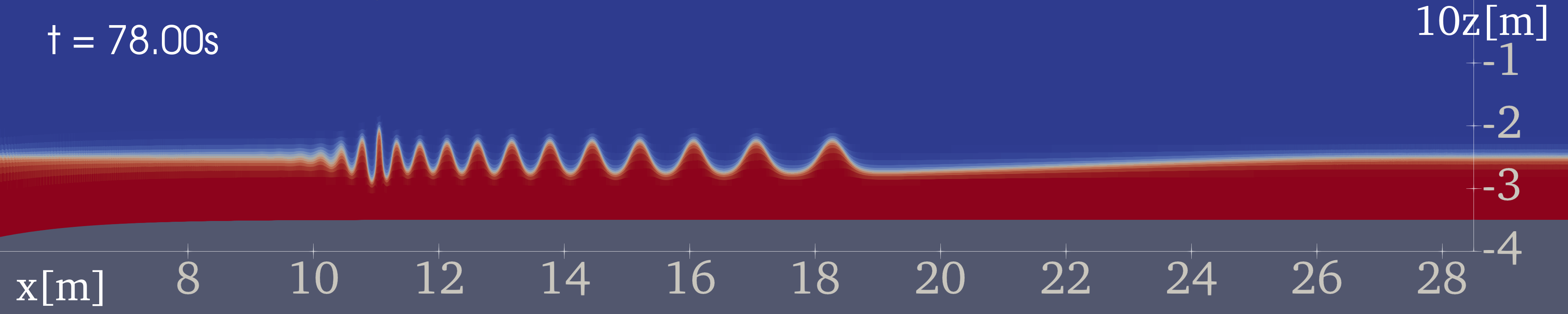}\vspace{4pt}
        \includegraphics[width=\linewidth]{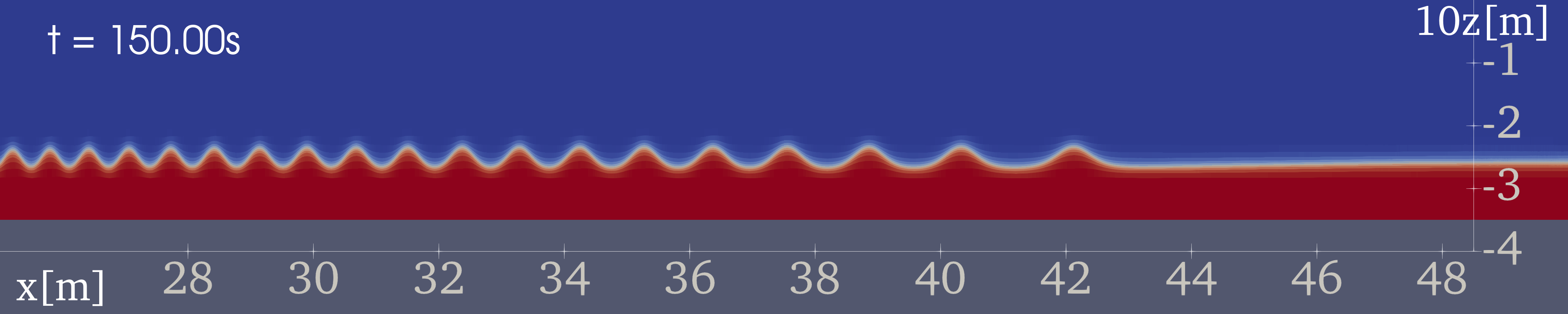}
        \includegraphics[width=0.8\linewidth]{tanh2_bar.png}
    \end{minipage}%
    \hfill
    \begin{minipage}[t]{0.52\linewidth}
        \vspace{0pt}
        \includegraphics[width=\linewidth]{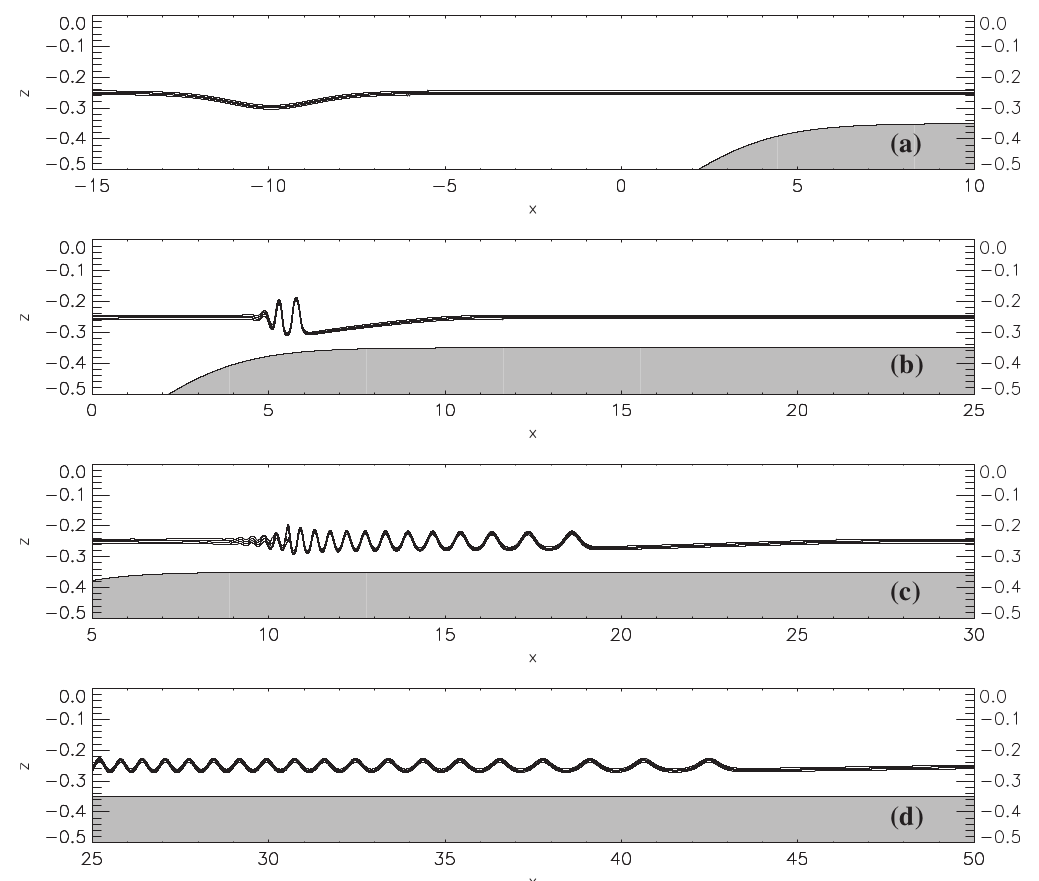}
    \end{minipage}
    \caption{Snapshots of density contours at various timestamps for the shoaling ISW of \S\ref{sec:isw_tanh}. On the left are the SLS results with the variational ALE scheme and on the right, those from \cite{lamb2014internal}. In the SLS contours, the z axis is magnified by a factor of 10 for better readability.}
    \label{fig:tanh}
\end{figure}

Next, we consider the shoaling of an ISW by considering the \verb|s8c1c| test case of \cite{lamb2014internal}. The data of the soliton is again detailed in Tab.\ref{tab:isw}, while the configuration of the domain can be seen in Fig.\ref{fig:tanh_tikz}. The topography changes with a maximum slope of $1/10$ at $x=0$ through the following relation taken from \cite{lamb2014internal}:
\begin{align}\label{eq:tanh_topo}
    & h(x) = 1 - \dfrac{2.5}{20} \left[ \mathrm{inttanh}\left( \dfrac{x+3.25}{2.5} \right) - \mathrm{inttanh}\left( \dfrac{x-3.25}{2.5} \right) \right]
\end{align}
with $ \displaystyle \mathrm{inttanh}(x) \eqdef \int_{-\infty}^x [ 1 + \tanh(s) ] \di s = x + \ln \left( 2\cosh x \right) $

For the simulation, a mesh of $ 19200 \times 32 $ cells is used and $ a_\vartheta = 0.1 $ , $a_x = a_\xi = 1$ and $ a_M = 10 $ are considered. Furthermore, to assess the performance of the ALE scheme, two additional simulations are carried out using a $\sigma$-grid and an isopycnal configuration ($\vartheta = 0$). In all three cases, the mesh resolution and total number of layers are kept identical to ensure a fair comparison. On all boundaries, inviscid wall BCs are imposed. The comparison between the three vertical meshes is displayed in Fig.\ref{fig:tanh_compare}. Snapshots of SLS's results alongside those of \cite{lamb2014internal} are presented in Fig.\ref{fig:tanh}.

From a physical standpoint, the soliton propagates without breaking, but it is subject to strong bottom-induced dispersion that generates a wavetrain containing more than 10 distinct wave crests. Since dispersion is inherently nonhydrostatic, such phenomena necessitate a careful treatment of the nonhydrostatic pressure, such as the one presented herein.

By comparing the vertical mesh configurations, it becomes apparent from Fig.\ref{fig:tanh_compare} that the $\sigma$-grid is inadequate for resolving this phenomenon with only $32$ vertical layers, producing a heavily diffused and under-resolved wave crest. In contrast, SLS equipped with either the proposed ALE framework or the isopycnal configuration captures the shoaling dynamics with substantially improved resolution. This indicates that the ALE scheme performs adequately and is capable of accurately tracking the underlying Lagrangian dynamics.

In Fig.\ref{fig:tanh}, the density field contours from the SLS results are presented side-by-side with those of \cite{lamb2014internal} at various timestamps. The contours have the same scale and bounding boxes as those in \cite{lamb2014internal} and the comparison between them indicates no noticeable differences. This observed agreement further validates SLS in this challenging case, since dispersive phenomena of this nature necessitate a good combined resolution of multiple harmonics at different frequencies.

With respect to the ALE movement, its performance is robust and no serious SDM appears. This indicates a good performance in ISW propagation phenomena, when no excessive isopycnal deformation occurs, while its robustness in more demanding configurations is assessed below.

\subsection{ISW interaction with a submerged wedge}\label{sec:isw_wedge}

\begin{figure}[t]
        \centering
    \includegraphics[width=0.32\linewidth]{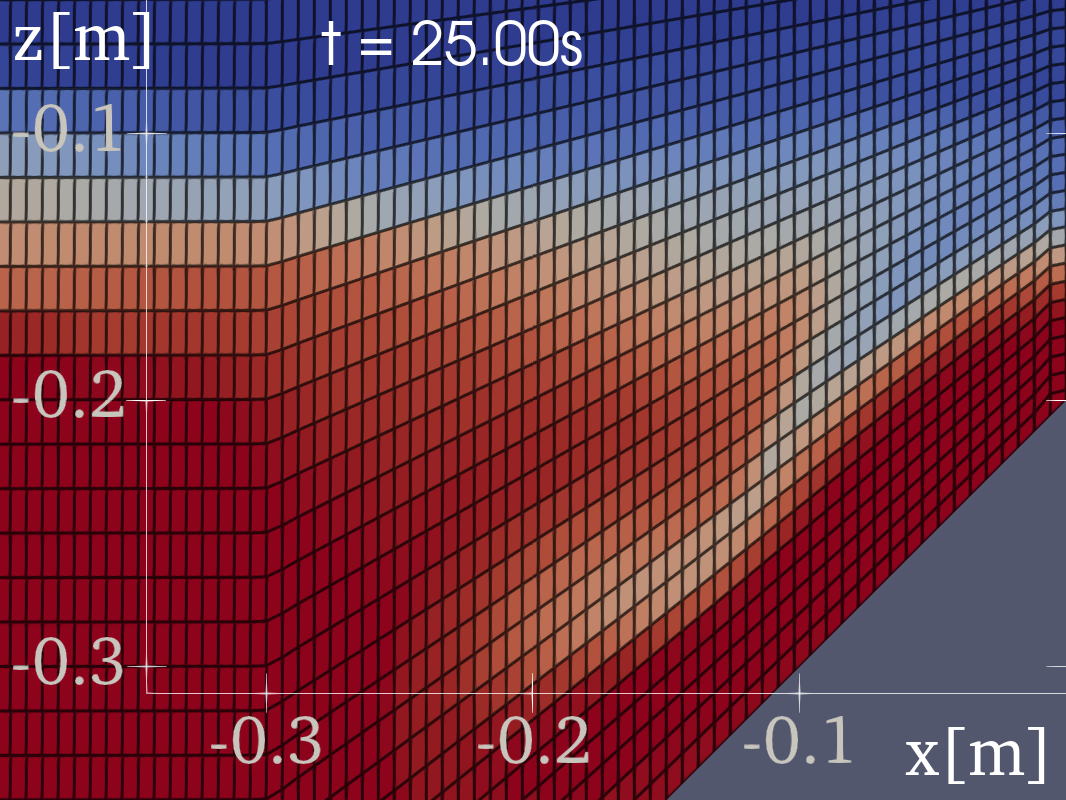}
    \includegraphics[width=0.32\linewidth]{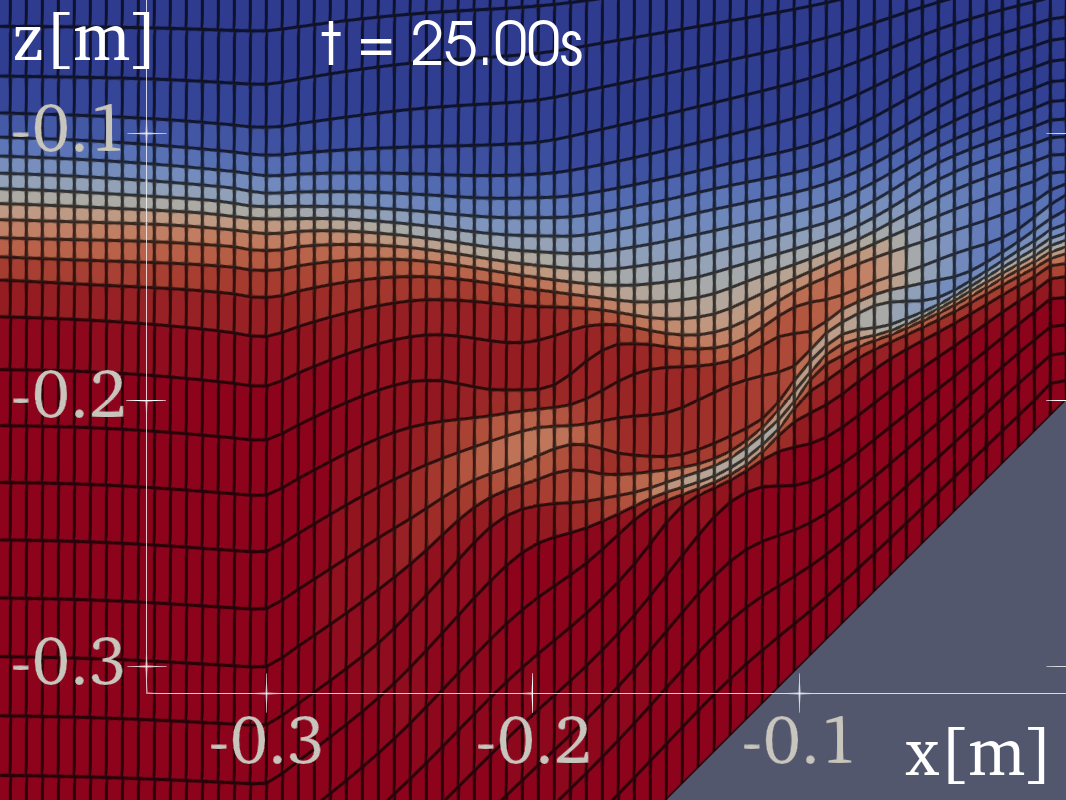}
    \includegraphics[width=0.32\linewidth]{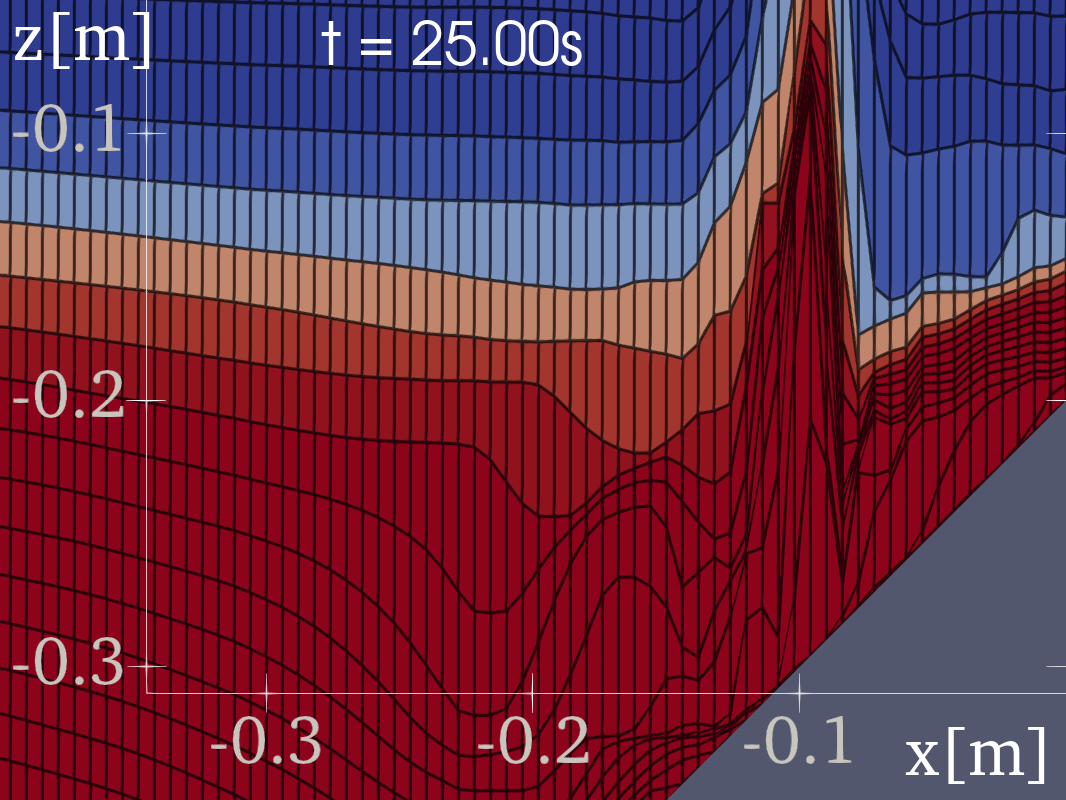}
    \includegraphics[width=0.60\linewidth]{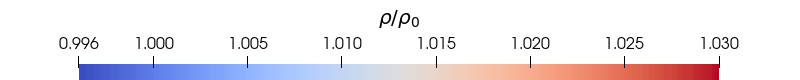}
    \caption{Density contours of the ISW impacting on the submerged wedge at $ t = 25s $ alongside the computational mesh. \textbf{Left:} SLS with $\sigma$-grid configuration, \textbf{middle:} SLS with variational ALE scheme, \textbf{right:} isopycnal mode with $\vartheta = 0$.}
    \label{fig:isw_wedge_comparison}
\end{figure}

\begin{figure}[t]
    \centering
    \includegraphics[width=0.45\linewidth]{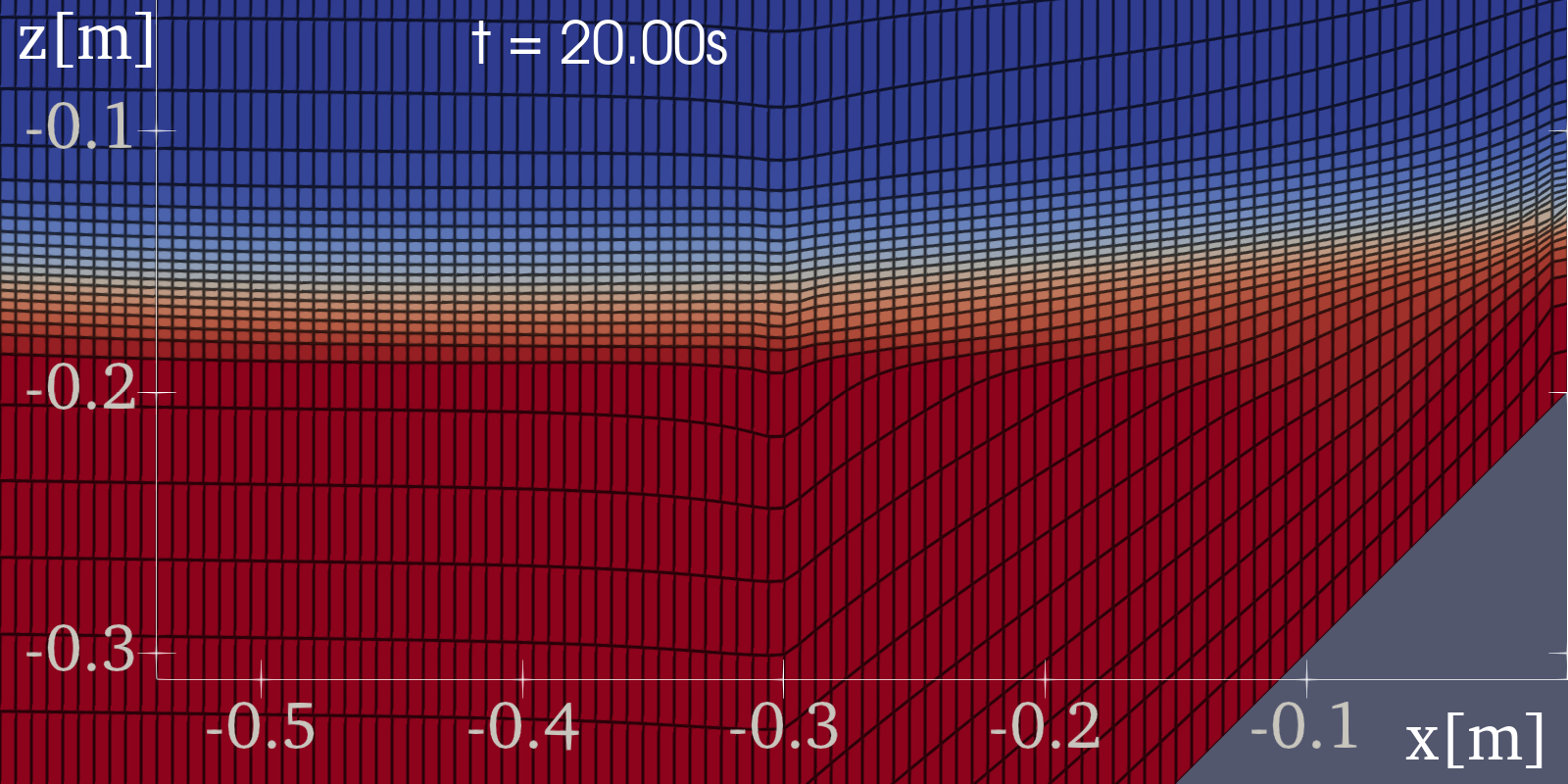}
    \includegraphics[width=0.45\linewidth]{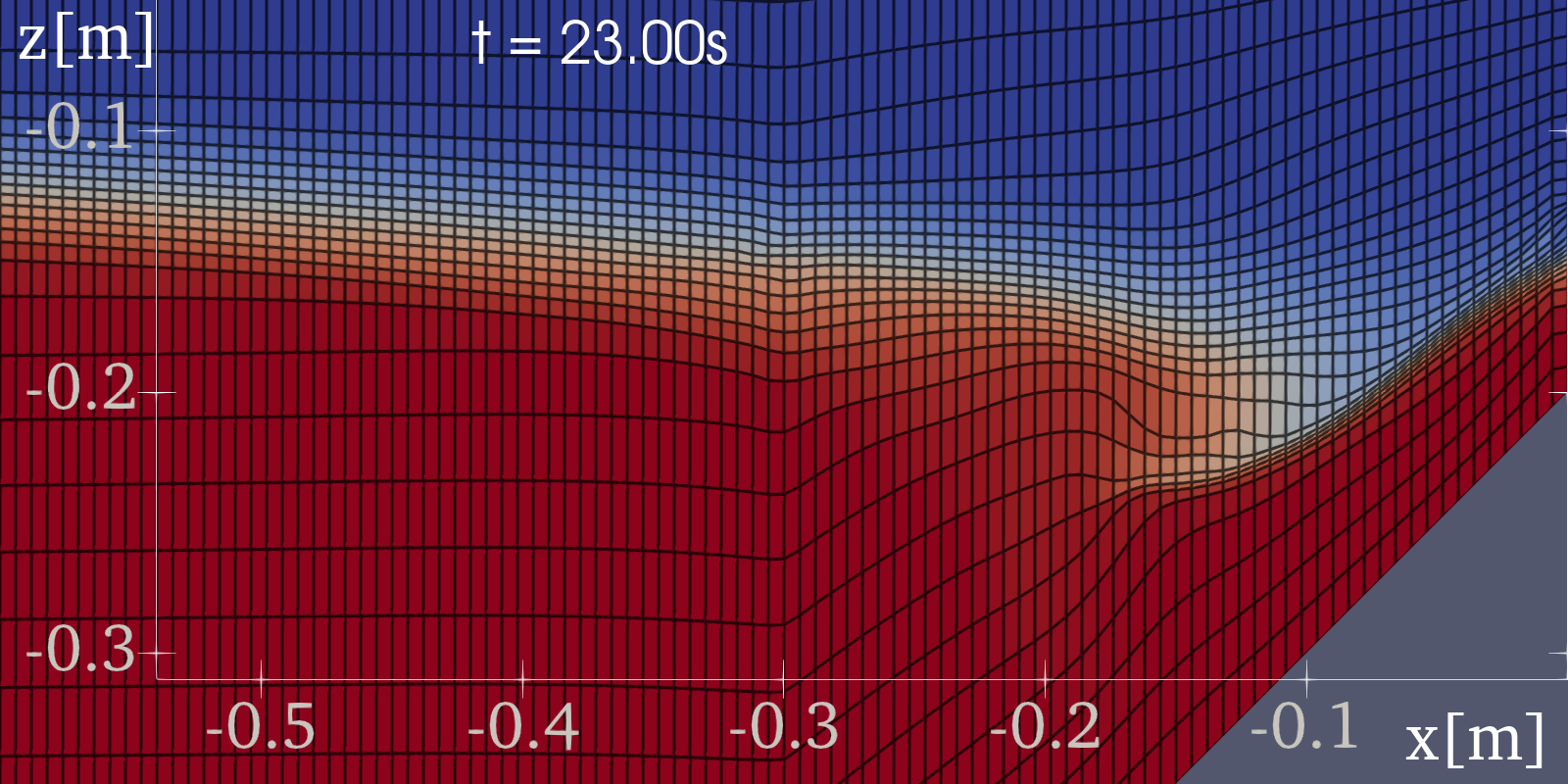}
    \includegraphics[width=0.45\linewidth]{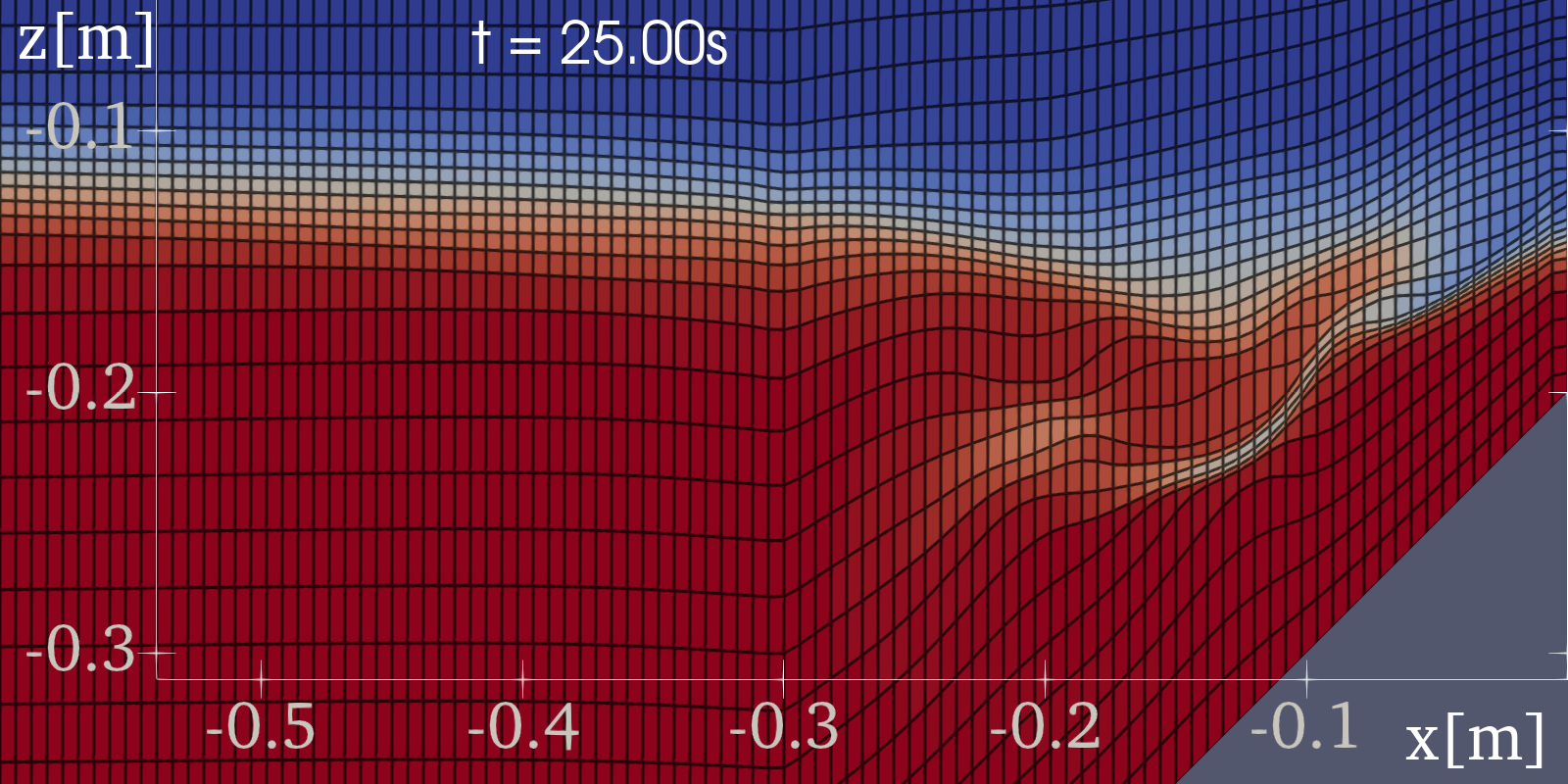}
    \includegraphics[width=0.45\linewidth]{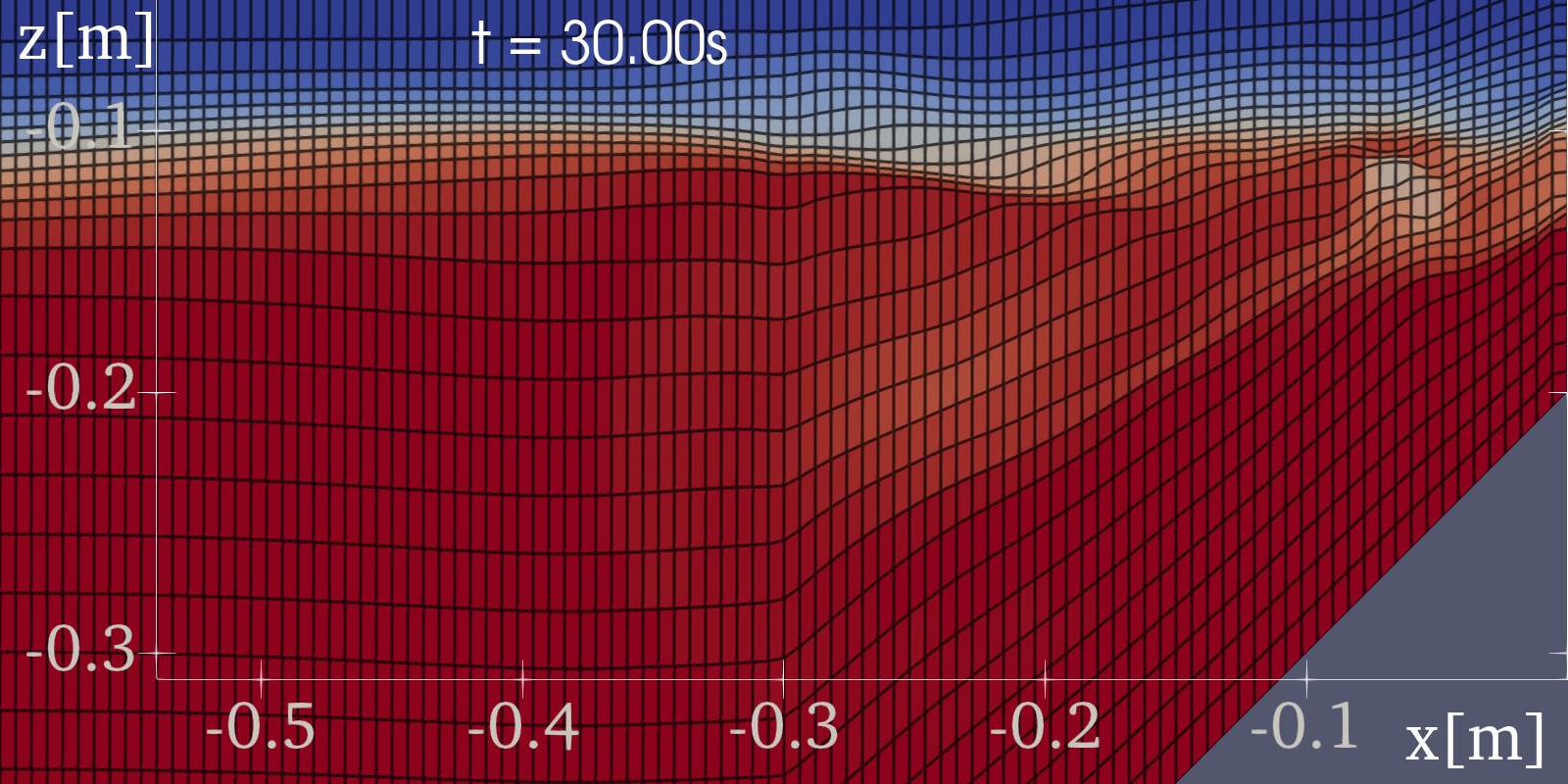}
    \includegraphics[width=0.65\linewidth]{wedge_bar.png}
    \caption{Density contours of the ISW propagating and impacting on the submerged wedge at $t[s] = \{20,23,25,30\}$ alongside the computational mesh. The results are obtained from SLS simulations using the variational ALE scheme with parameters $a_\vartheta = 0.1$, $a_x = a_\xi = 1$, and $a_M = 10$.}
    \label{fig:isw_wedge_contour}
\end{figure}

\begin{figure}[t]
    \centering
    \includegraphics[width=0.6\linewidth]{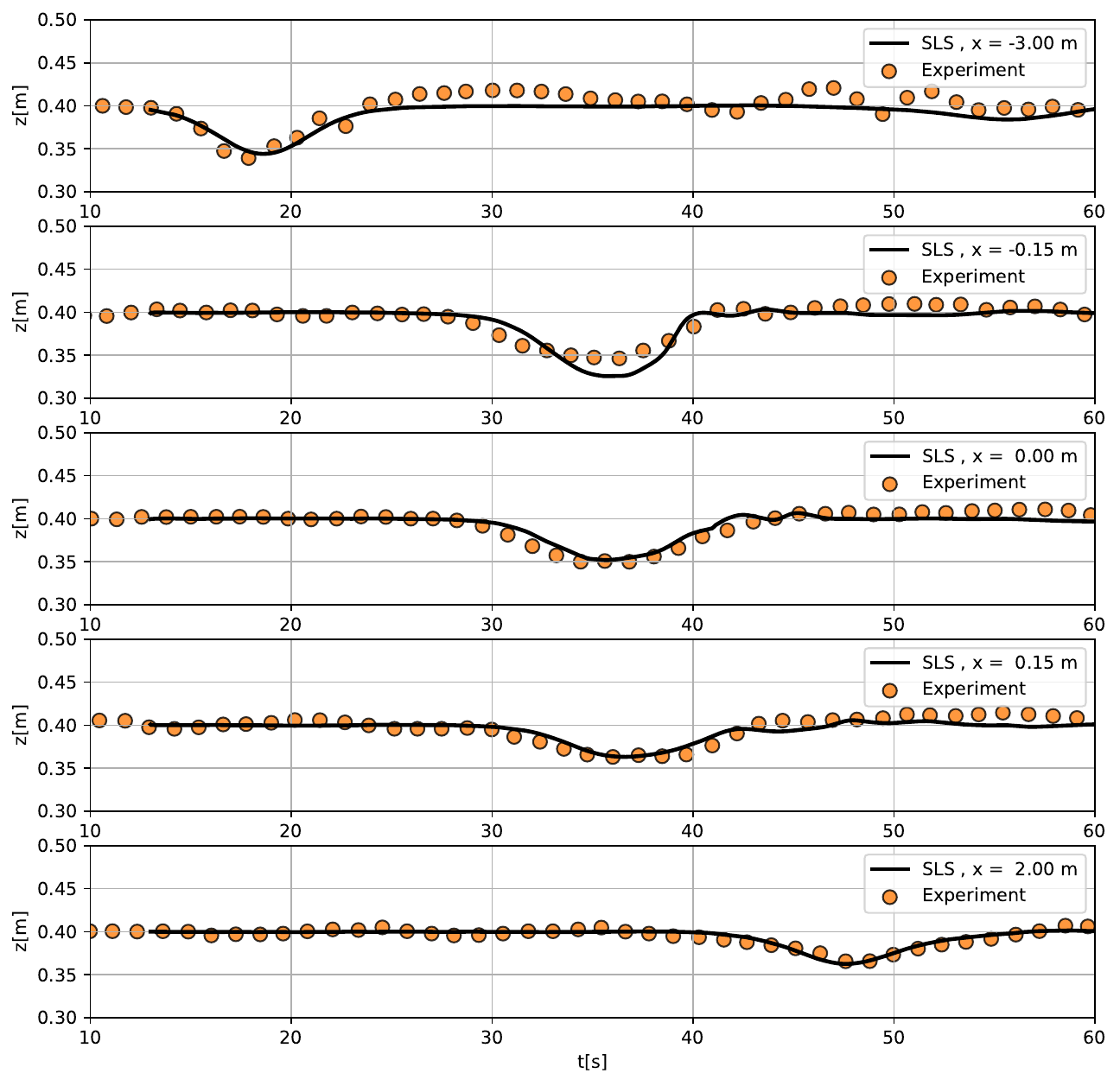}
    \caption{Time series of the pycnocline vertical location at various locations throughout the domain for the case of an ISW propagating above a submerged wedge. SLS results with the use of the proposed ALE algorithm are presented with continuous lines, while experimental measurements from \cite{hsieh2015numerical} are noted with circles.}
    \label{fig:isw_wedge}
\end{figure}

To further investigate the effectiveness of the ALE scheme in soliton-topography interactions, we focus on the interaction of an ISW with a submerged obstacle. Specifically, we consider the experiments of \cite{hsieh2015numerical}, where a submerged wedge with a $45^\circ$ angle and a height of $0.3m$ is placed at $x=0$. The depth of the tank is $0.5m$ and the data of the soliton can be found in Tab.\ref{tab:isw}.

According to the experiments, the ISW is expected to impact the wedge in a violent manner and then to continue and propagate past it. In order to properly capture this numerically, the solver must be able to sharply capture the ISW without severe numerical diffusion/mixing, while retaining a robust mesh quality during the impact upon the obstacle. This poses a challenge that, if properly addressed, will further solidify the validity of the proposed method.

A mesh of $ 2000 \times 30 $ cells in total is employed that covers the domain $ (x,z) \in [-6 m ,4 m ] \times [-0.5 m , 0 ] $. Again, SLS uses the variational mesh movement $ a_x = a_\xi = 1 $ and $ a_M = 10 $. Since the impact on the wedge heavily distorts the isopycnals, the value $a_\vartheta=0.01$ is utilized based on the discussion of \S\ref{sec:freqfilter}. Similar to \S\ref{sec:isw_tanh}, the performance of the proposed ALE scheme is assessed through two additional simulations using a $\sigma$-grid configuration and the solver in isopycnal mode. All boundaries are considered impenetrable walls except the top one, where a zero pressure free-surface condition is enforced.

In Fig.\ref{fig:isw_wedge_comparison}, the results obtained using the three vertical mesh approaches are compared at the instant when the ISW impacts the wedge. Straightaway, we note that the pycnocline undergoes considerable deformation and this, unlike the comparison shown in Fig.\ref{fig:tanh_compare} of the previous test case, severely impacts the performance of vertically Lagrangian approaches. Indeed, in the purely isopycnal configuration, the layers become heavily distorted and eventually unstable, resulting in the solver crashing shortly after $t=25s$. Although this behavior could potentially be stabilized through the ad-hoc addition of diffusive smoothing, the ALE approach provides such stabilization without sacrificing the accuracy of the results.

Additionally, the isopycnal approach fails to capture the overturning phenomenon, in which a mass of light water penetrates into the denser region, whereas both the $\sigma$ and ALE simulations successfully resolve it. Comparing the $\sigma$-grid with the proposed variational ALE approach, we observe that the latter exhibits a higher effective resolution, capturing both the smooth and strongly deformed parts of the pycnocline more sharply. Overall, Fig.\ref{fig:isw_wedge_comparison} demonstrates that the proposed ALE approach reduces the numerical mixing apparent in the $\sigma$-configuration while simultaneously avoiding the instabilities of the isopycnal configuration.

In Fig.\ref{fig:isw_wedge_contour}, density contours over the computational mesh are presented at various time values, just to the left of the obstacle during the impact of the ISW upon it. Regarding the mesh movement algorithm, we observe that indeed with $a_M=10$ and $a_\xi=1$ areas with large density gradients have almost 10 times finer vertical resolution compared to the constant density regions, thus resulting in a sharp pycnocline tracking. We also observe that the algorithm correctly triggers vertical mass transfer when the isopycnals tend to distort and overturn at the point where the ISW impacts the wedge. The result is a smooth mesh that concentrates cells in areas where the production of SDM would otherwise be significant (see Fig.\ref{fig:isw_wedge_comparison}, left). Some numerical mixing is noticed at the point of overturning when vertical advection is triggered, but it remains localized and doesn't severely pollute the solution. When the ISW passes above the wedge, the mesh progressively returns to its original configuration.

To assess the results, in Fig.\ref{fig:isw_wedge}, the time series of the pycnocline's center elevation at various locations throughout the domain is compared with the experimental measurements of \cite{hsieh2015numerical}. The pycnocline center is defined as the location of the point where the density is equal to $({\rho_{\min}+\rho_{\max}}) \big / {2}$. The comparison between the SLS and the experiments is generally good. Some discrepancies exist at the $x=-3 m$ probe, where some wiggles that SLS doesn't capture appear in the experiments. These wiggles are attributed to the ISW generation in the experimental setup, in which a sliding gate is used, which might have created some trailing secondary waves. This is further reinforced by the fact that those wiggles are also absent in \cite{li2022iswfoam}, where, as here, the DJL equation is used to initialize this configuration. That being said, the SLS results seem to be in agreement with the experiments, thus further validating the ability of the proposed mesh movement to handle cases with considerable isopycnal deformation accurately and stably without exhibiting numerical artifacts.

\subsection{ISW breaking on a sloping beach}\label{sec:agh}

The last test case investigates the effectiveness of the proposed variational mesh movement on a complex phenomenon that includes overturning of the isopycnal surfaces, and thus more nuanced approaches must be considered. Specifically, we focus on the breaking of an ISW on a linear and emerging beach-like topography. The breaking of ISWs has been the subject of numerous studies, due to the complexity of the phenomenon and its critical role in the continental shelf cascade of energy. For an extensive review of the subject, the reader is referred to \cite{stastna2024simulations}.

Unlike the previous ones, in this test case a purely isopycnal formulation would totally fail to capture the phenomenon since such a scheme negates vertical mass transfer which is an inherent feature of wave breaking. As such, it is natural to assume that overly limiting $\vartheta$ through a large Lagrangian bias ($a_\vartheta$) would produce erroneous results. Thus, in order to reduce spurious mixing, one should consider the use of the $a_M$ parameter so that the vertical spacing is denser in areas with large vertical density gradients. The results that follow will demonstrate the effectiveness of this approach.

Specifically, we consider \verb|case 60| from \cite{aghsaee2010breaking}, where an ISW with the same characteristics as the one in \S\ref{sec:djl} propagates towards a beach with constant slope equal to $0.3$. The phenomenon consists of a plunging breaker and thus overturning and vertical mass transfer are prevalent.

The numerical domain consists of $600 \times 75$ cells that cover the area $ (x,z) \in [ 0 , 3 m ] \times [ -0.15 m , 0 ] $, while the bottom starts to slope upwards at $x=2.4m$. All boundaries are considered as impenetrable walls. The variational mesh movement is utilized with $a_x=a_\xi=1$, with $a_\vartheta,a_M$ varying between simulations. A simulation on a fixed $\sigma$-grid is also performed by prescribing $\vartheta$ through \cite[eq.(16b)]{alexandris2024semi} in order to provide a non-ALE configuration.

\begin{figure}[t]
    \centering
    \includegraphics[width=0.6\linewidth]{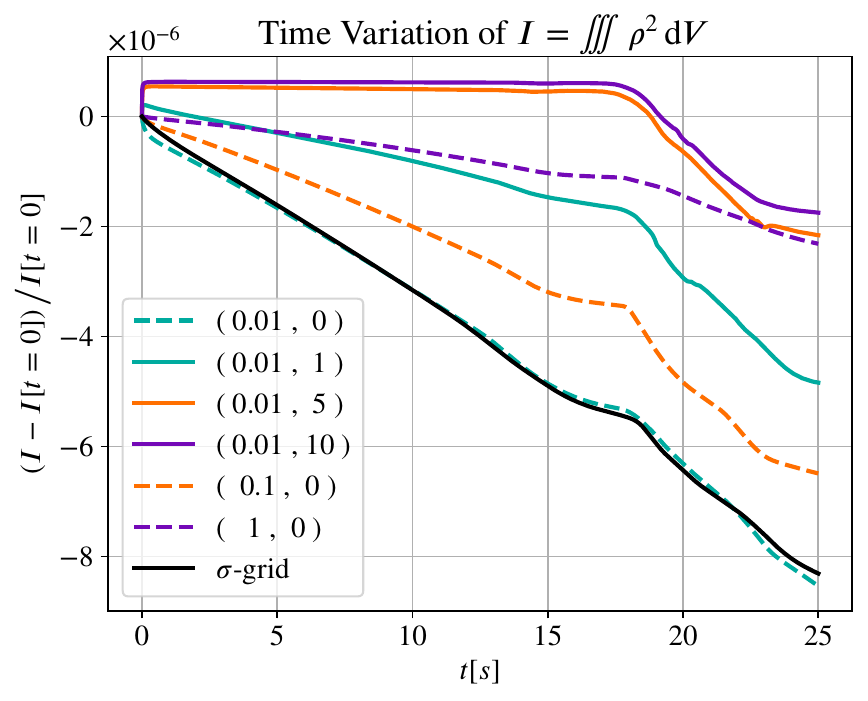}
    \caption{Decay of the relative second-order moment of the density field for the case of a plunging breaker in \S\ref{sec:agh}. In the legend, the two numbers indicate the values of the parameters $(a_\vartheta,a_M)$. }
    \label{fig:aghsaee_dvd}
\end{figure}

To properly quantify spurious numerical mixing of the density, at each timestep we calculate the discrete second-order moment of the density $ \iiint \rho^2 \di V \approx I \eqdef \sum ( \rho_{ij} )^2 L_{ij} \Delta x_i \Delta \xi_j $. In general, the decrease of $I$ quantifies the magnitude of both physically induced and numerical mixing. Since no mixing parameterizations are used in the present formulation, the rate of decay of $I$ will indicate the amount of numerical diffusion that the solver introduces (see the discussion in \S\ref{sec:mixing}).
In Fig.\ref{fig:aghsaee_dvd}, the time series of the relative deviation $ \frac{I(t)-I(t=0)}{I(t=0)} $ are presented for various values of $(a_\vartheta,a_M)$. Specifically, we use $a_\vartheta=0.01$ and $a_M=0$ as a baseline configuration and try to see how SDM behaves by increasing one or both of them.

\begin{figure}[t]
    \centering
    \includegraphics[width=0.32\linewidth]{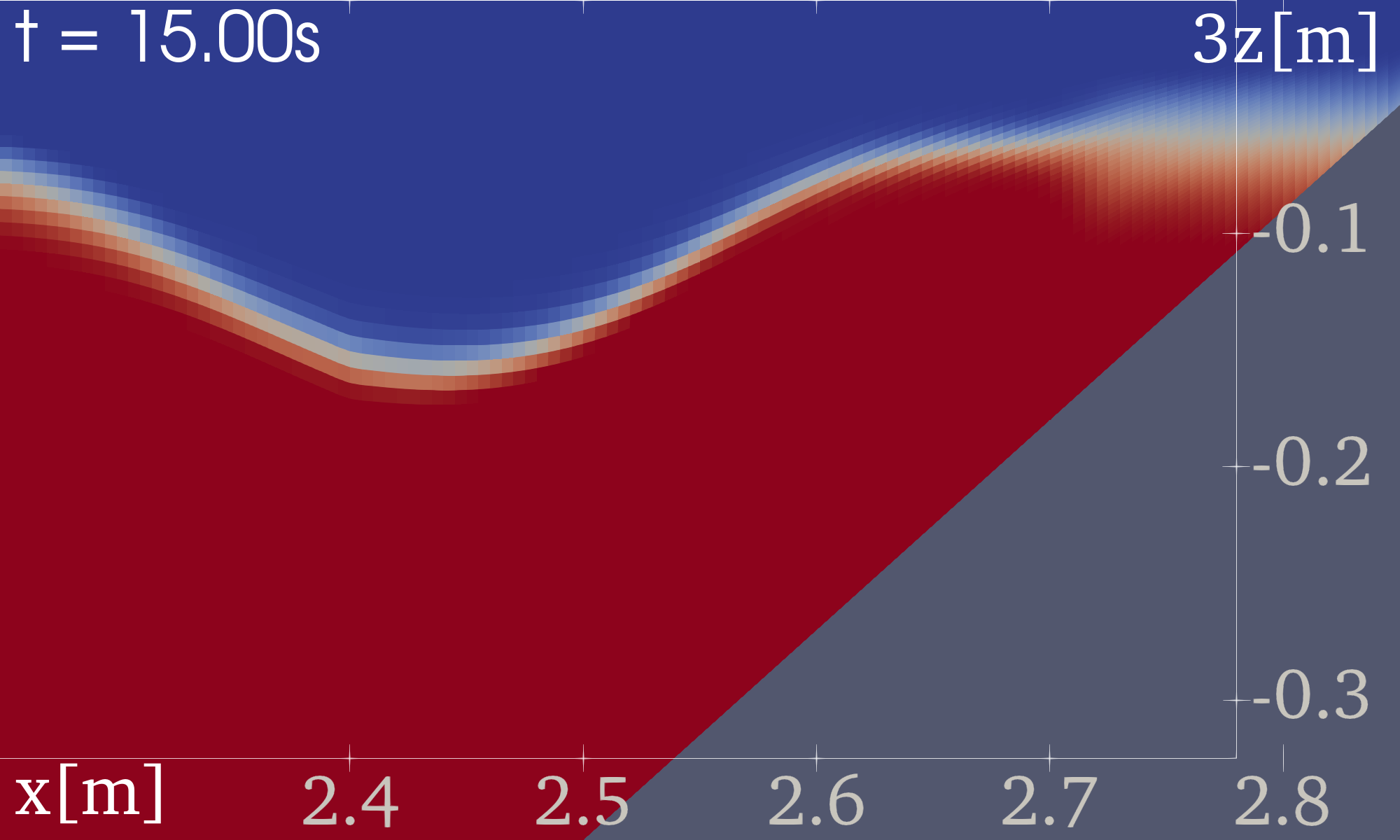}
    \includegraphics[width=0.32\linewidth]{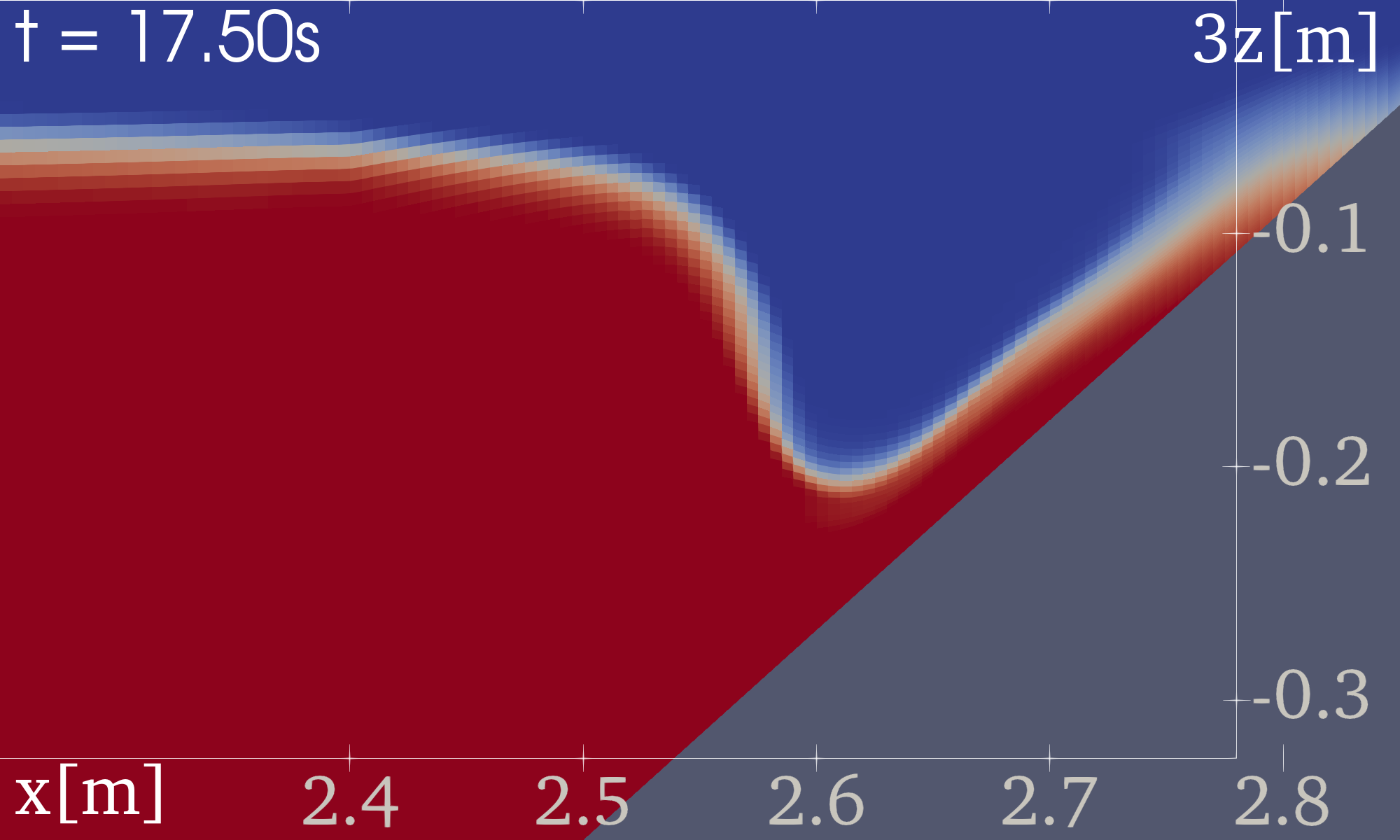}
    \includegraphics[width=0.32\linewidth]{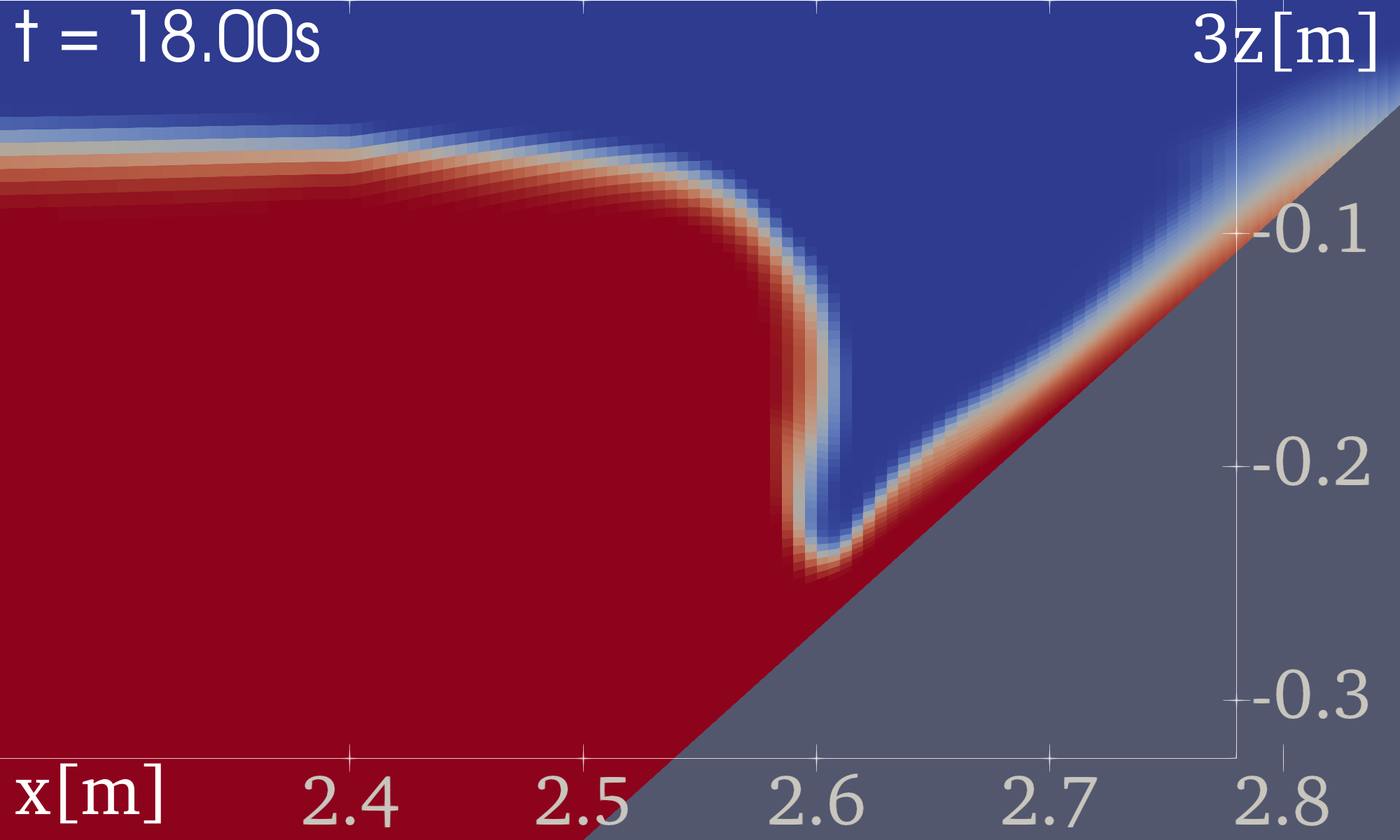}
    \includegraphics[width=0.32\linewidth]{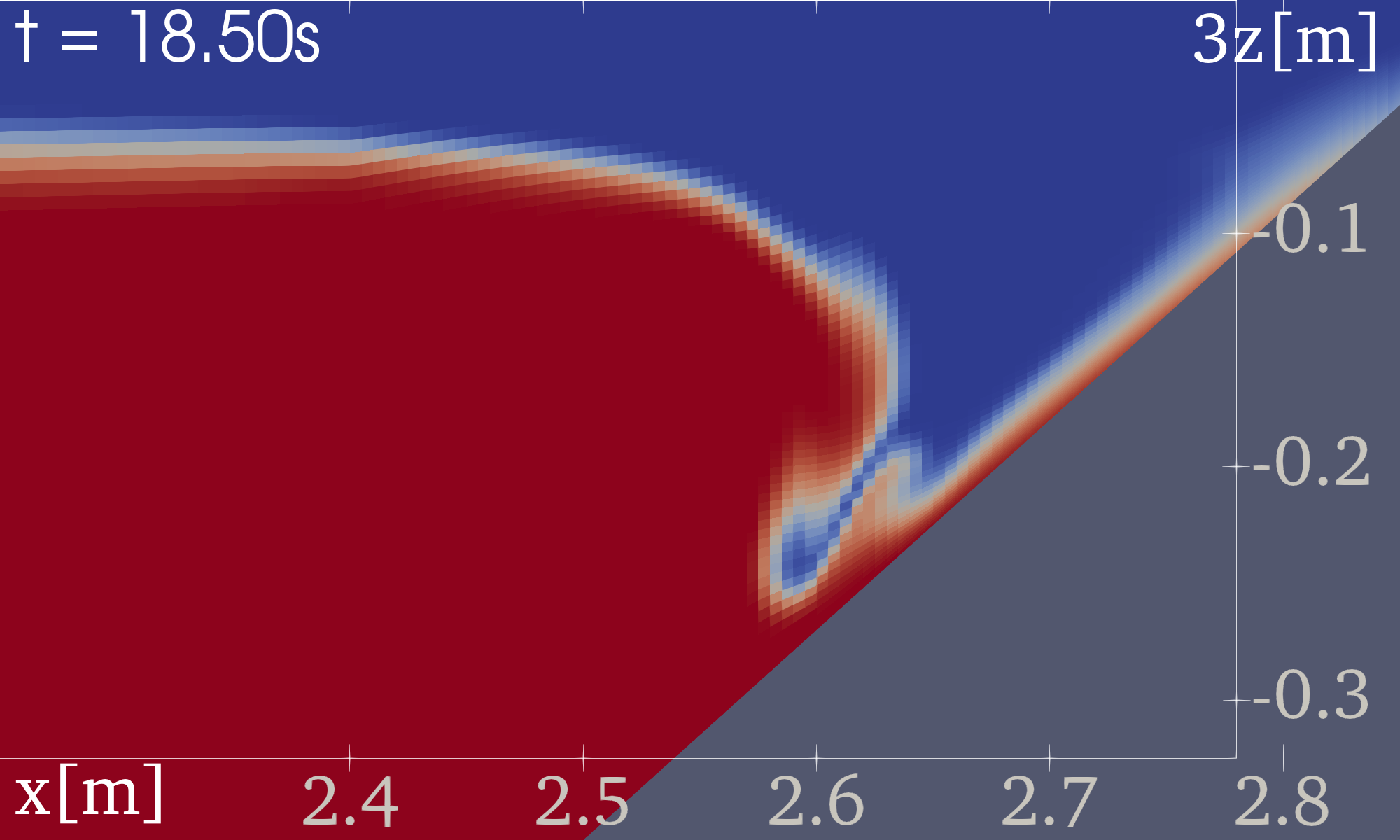}
    \includegraphics[width=0.32\linewidth]{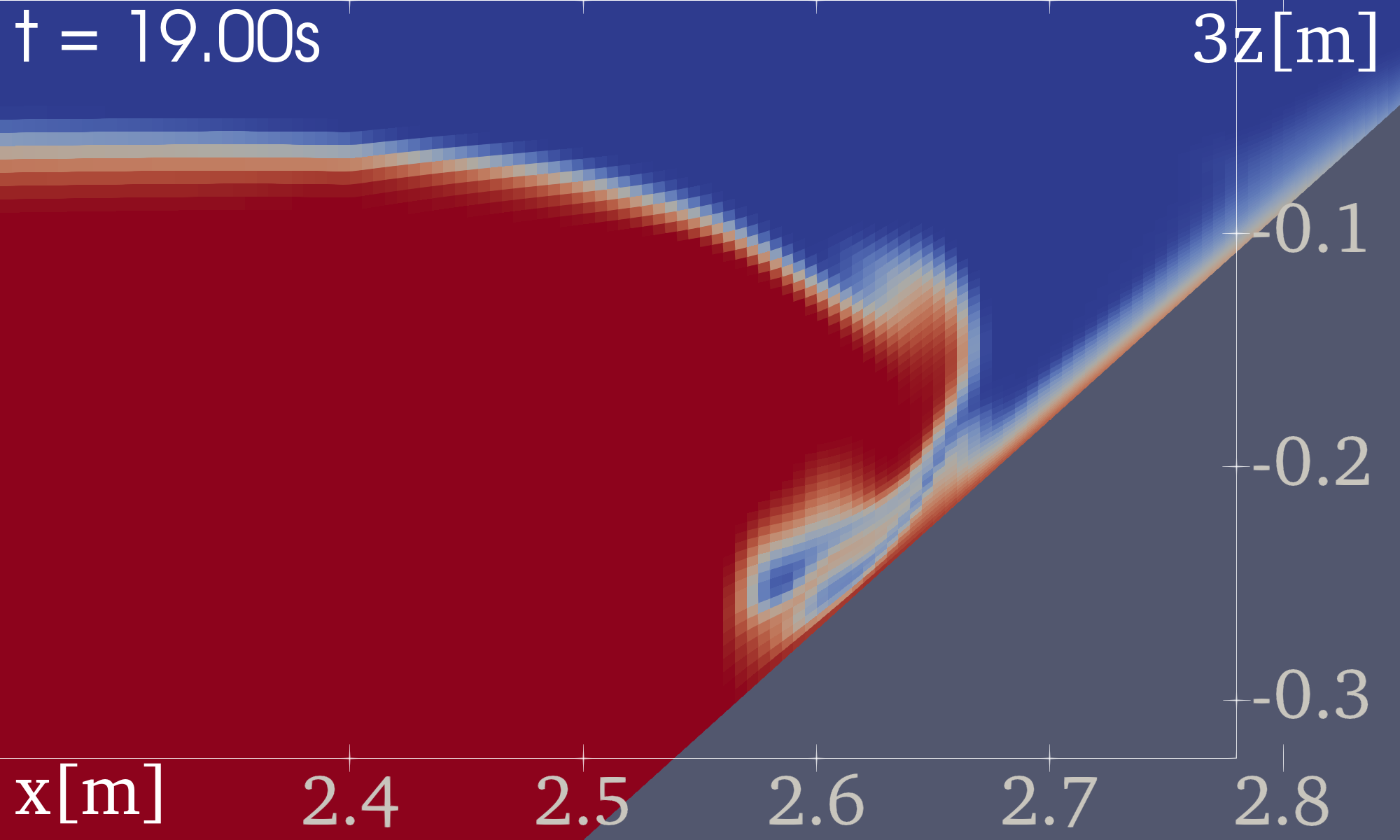}
    \includegraphics[width=0.32\linewidth]{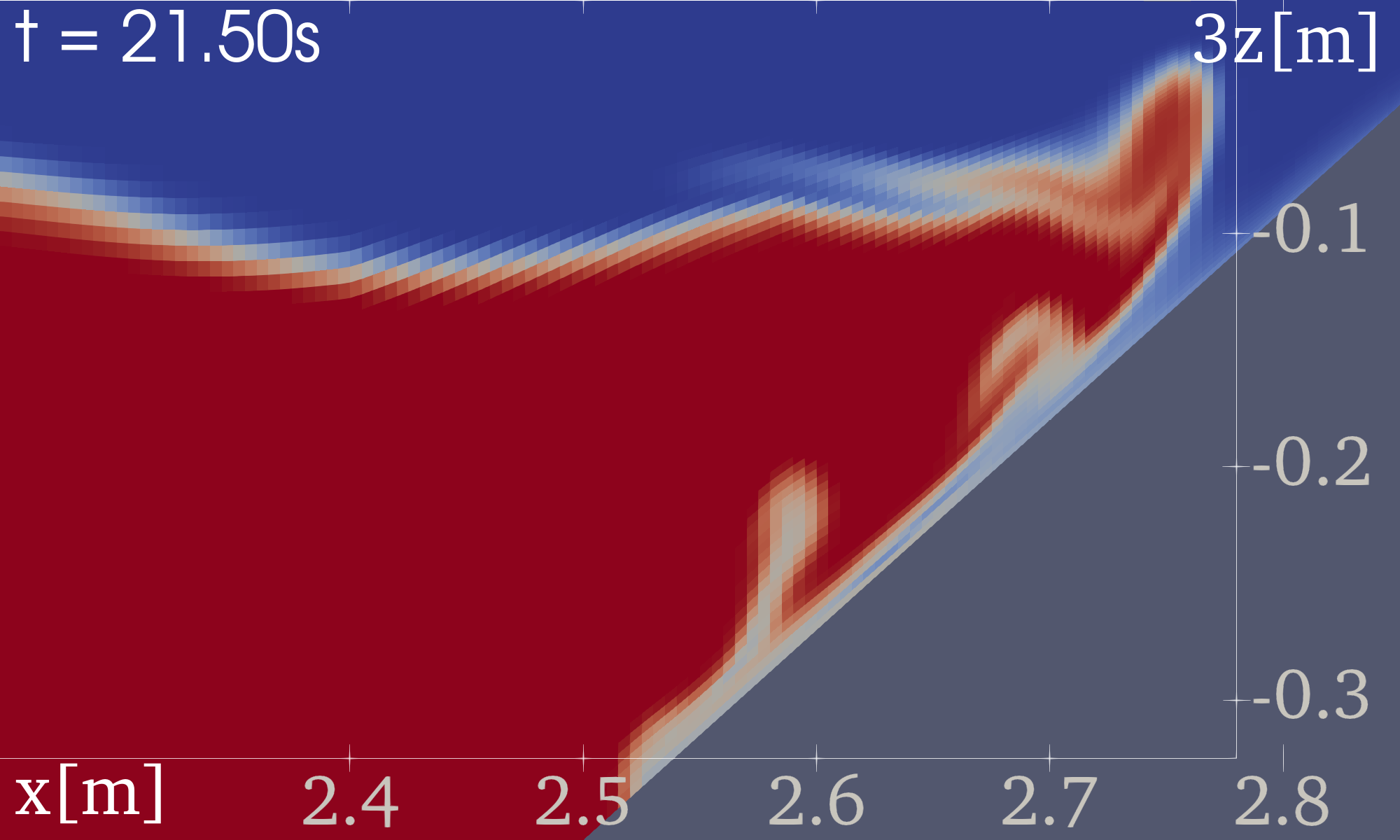}
    \includegraphics[width=0.6\linewidth]{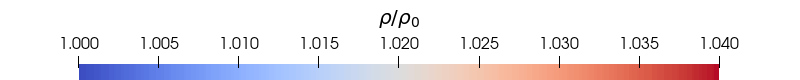}
    \caption{ISW plunging into a sloping beach, normalized density contours from SLS simulation with $(a_\vartheta,a_M) = (0.01,0)$. The vertical axis and its labels are magnified by a factor of 3 and the dimensional time value is noted at each frame.}  
    \label{fig:ag10_0}
\end{figure}

In this baseline configuration, the effect of the numerical mixing in the results is apparent. Even before the ISW reaches the slope, heavy numerical diffusion is observed in Fig.\ref{fig:ag10_0} at the point where the pycnocline intersects the topography. Also, numerical mixing is observed at the ISW at the left side of its crest. This diffusive behavior of the $(a_\vartheta,a_M)=(0.01,0)$ configuration is also displayed in Fig.\ref{fig:aghsaee_dvd}, where the decay of $I$ is almost linear up to the point of breaking ($t\approx 18s$) and almost identical with the $\sigma$-grid configuration.

\begin{figure}[t]
    \centering
    \includegraphics[width=0.32\linewidth]{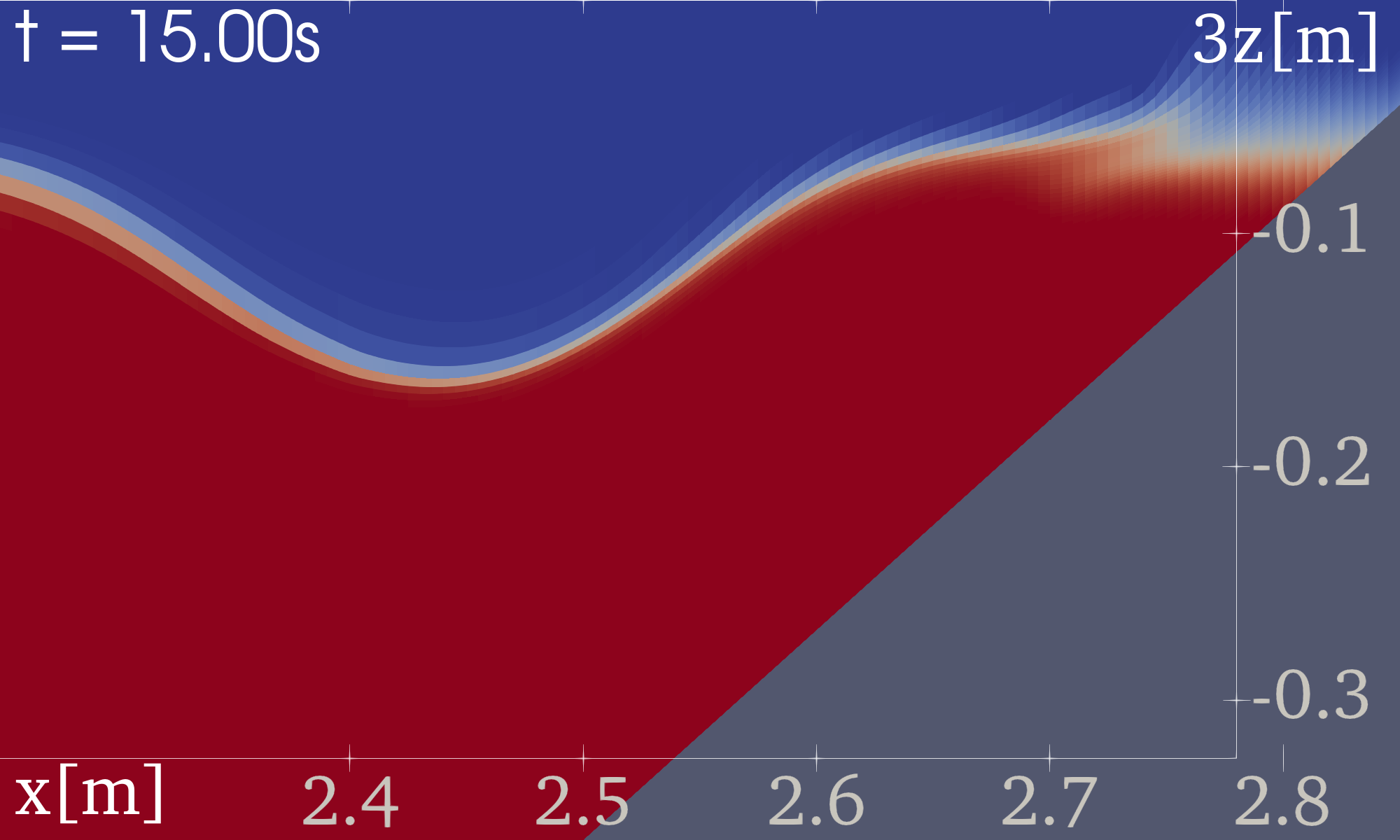}
    \includegraphics[width=0.32\linewidth]{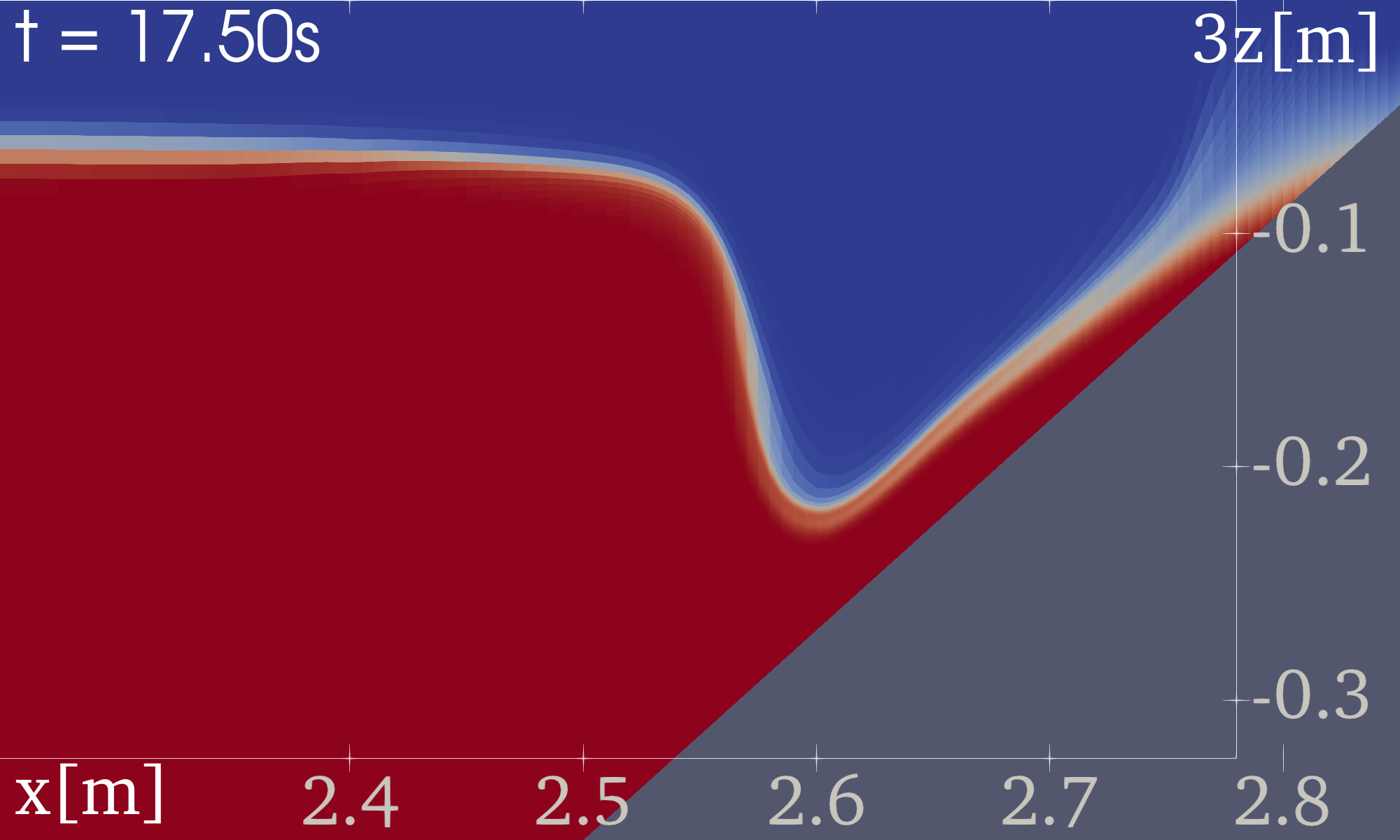}
    \includegraphics[width=0.32\linewidth]{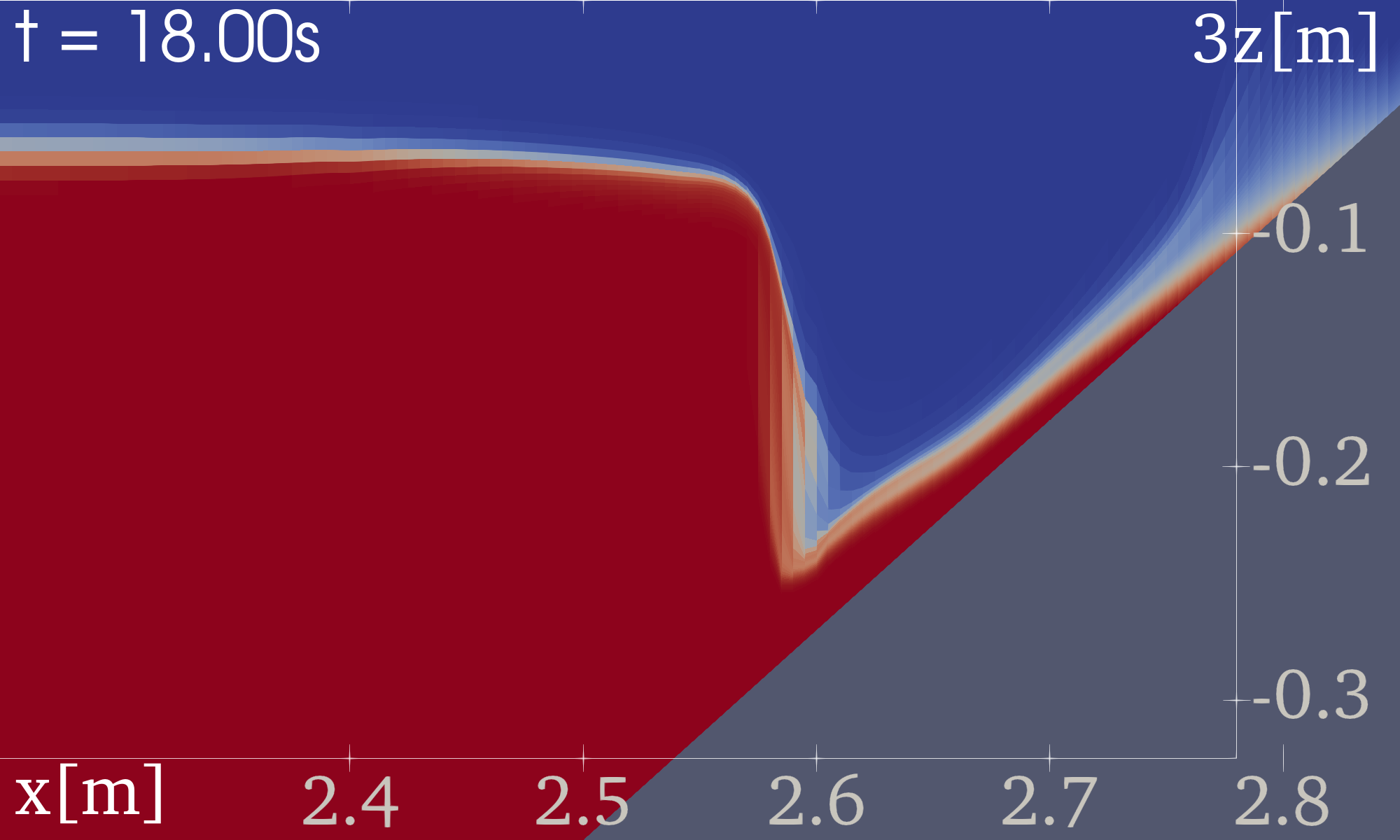}
    \includegraphics[width=0.32\linewidth]{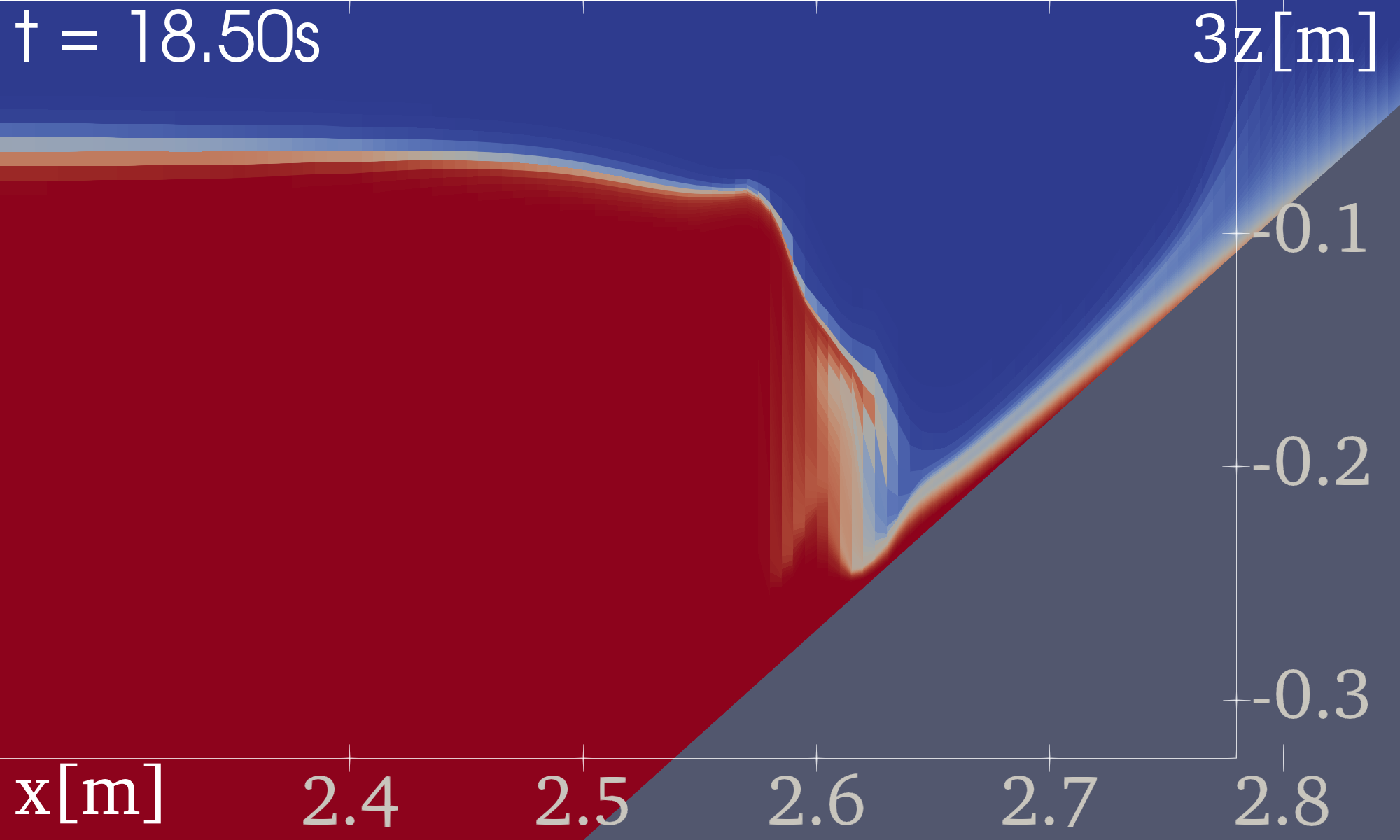}
    \includegraphics[width=0.32\linewidth]{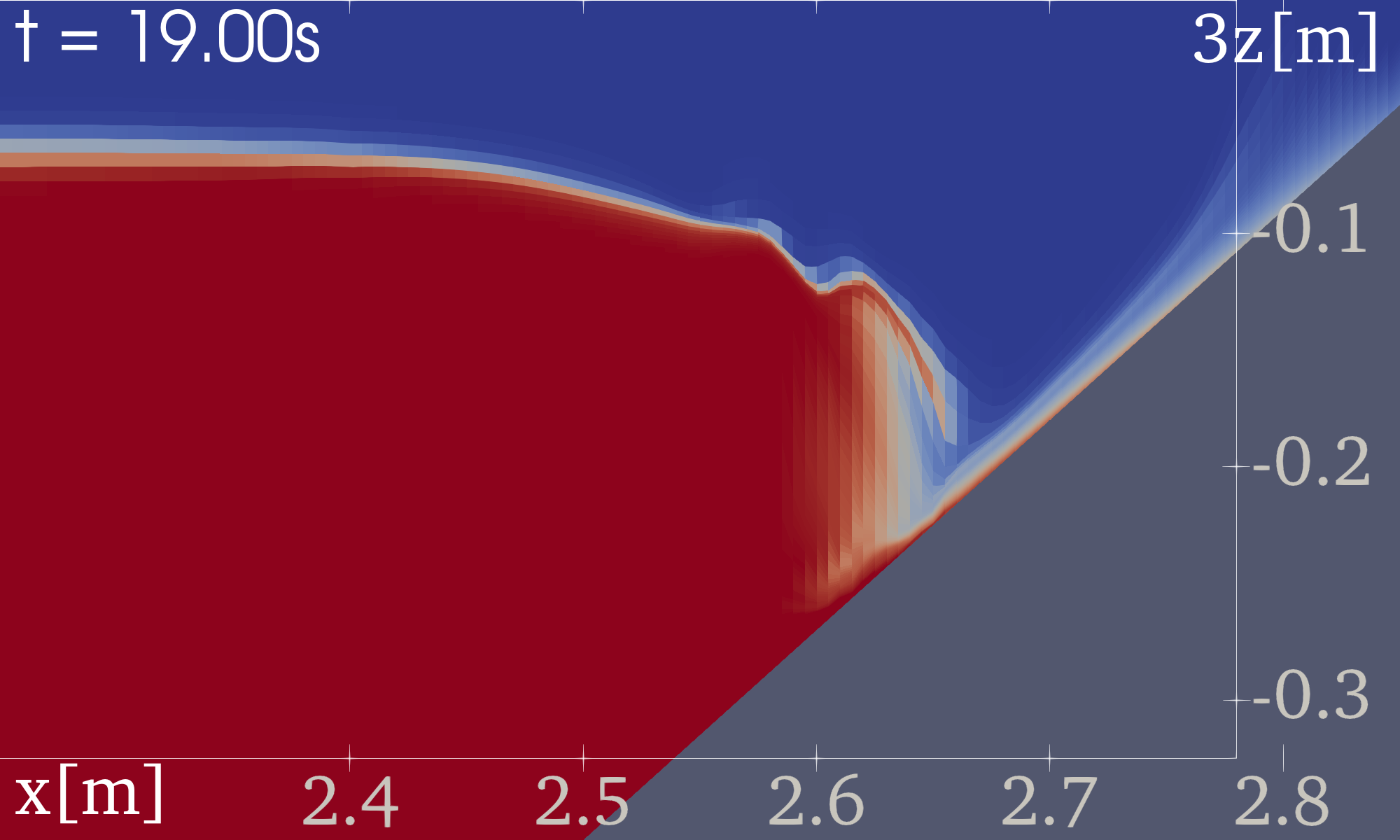}
    \includegraphics[width=0.32\linewidth]{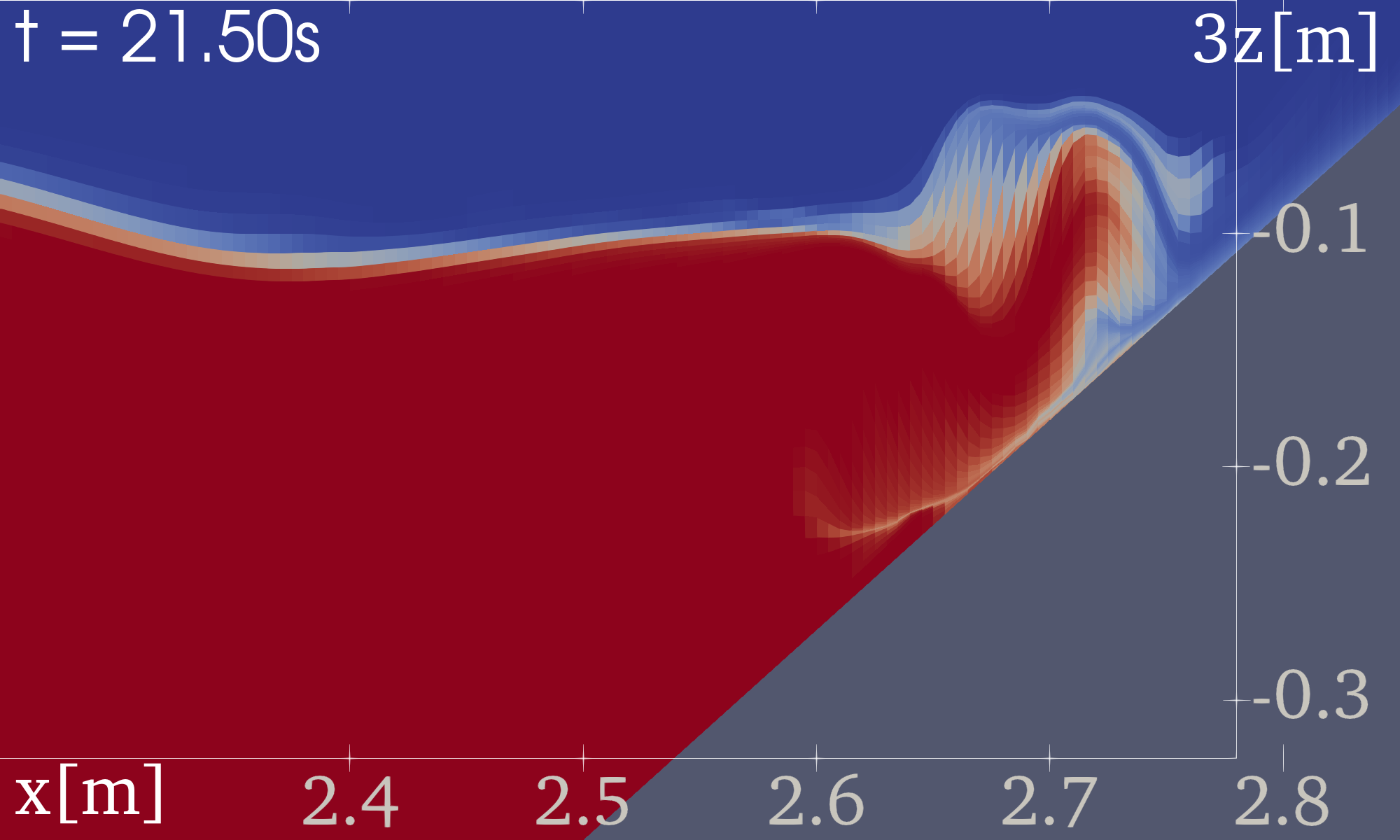}
    \includegraphics[width=0.6\linewidth]{ag_bar.png}
    \caption{ISW plunging into a sloping beach, normalized density contours from SLS simulation with $(a_\vartheta,a_M) = (1,0)$. The vertical axis and its labels are magnified by a factor of 3 and the dimensional time value is noted at each frame.} 
    \label{fig:ag100_0}
\end{figure}

In order to counteract this, we could introduce Lagrangian bias through $a_\vartheta$ and keep $a_M=0$. From Fig.\ref{fig:aghsaee_dvd}, we see that as $a_\vartheta$ increases, the decay of $I$ is reduced. But, as we see in Fig.\ref{fig:ag100_0}, when using large values ($a_\vartheta=1$) vertical mass transfer is suppressed in a highly nonphysical manner and the results do not capture the overturning correctly.

\begin{figure}[t]
    \centering
    \includegraphics[width=0.32\linewidth]{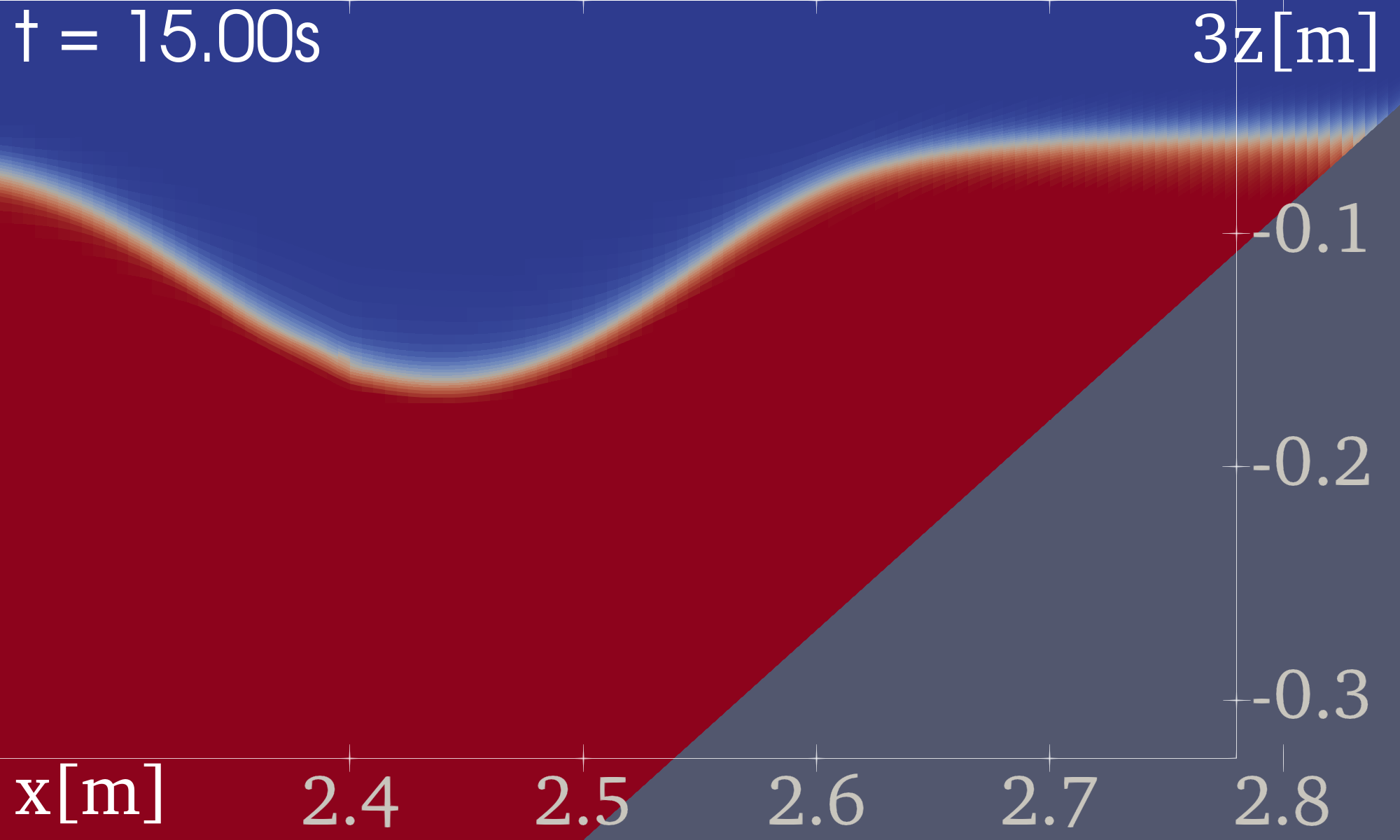}
    \includegraphics[width=0.32\linewidth]{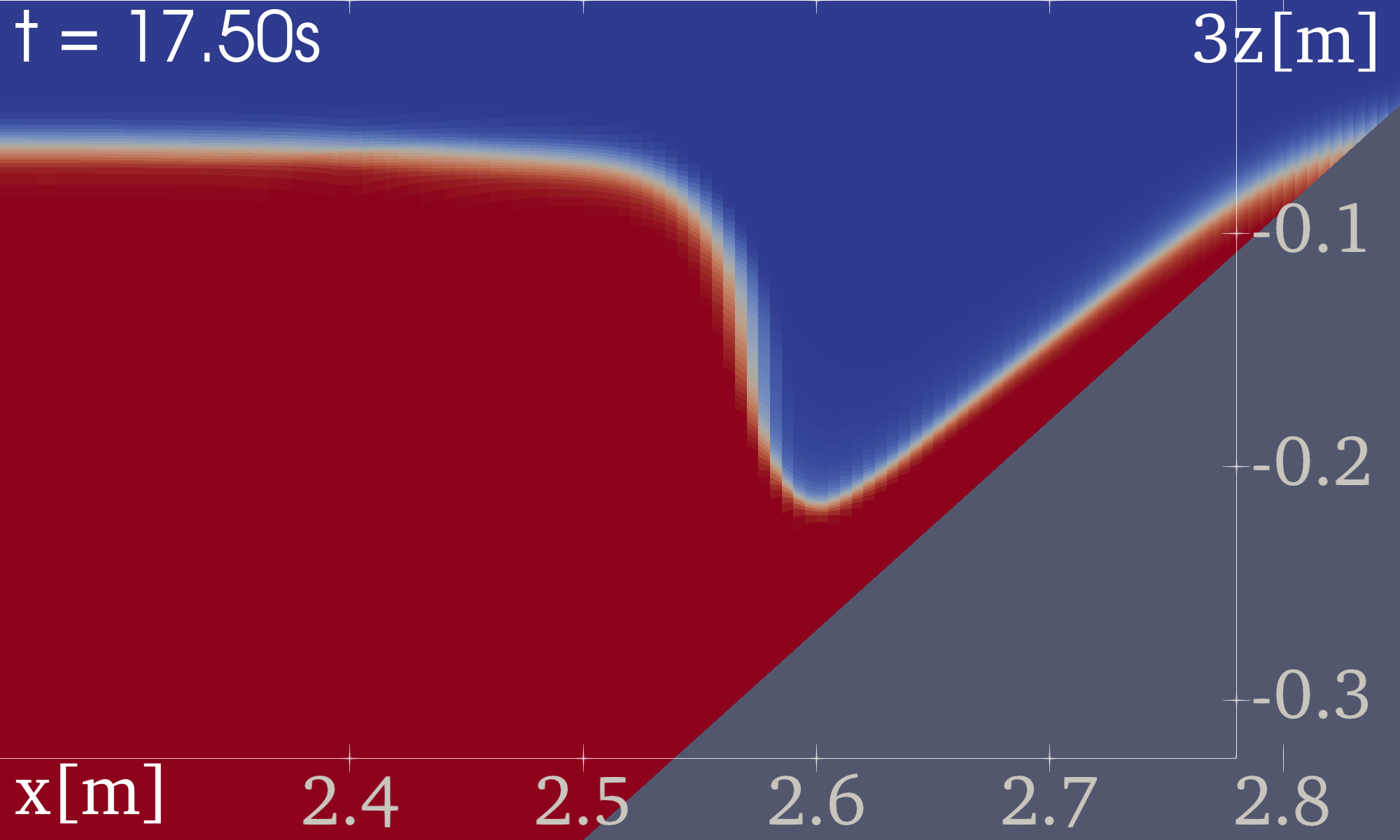}
    \includegraphics[width=0.32\linewidth]{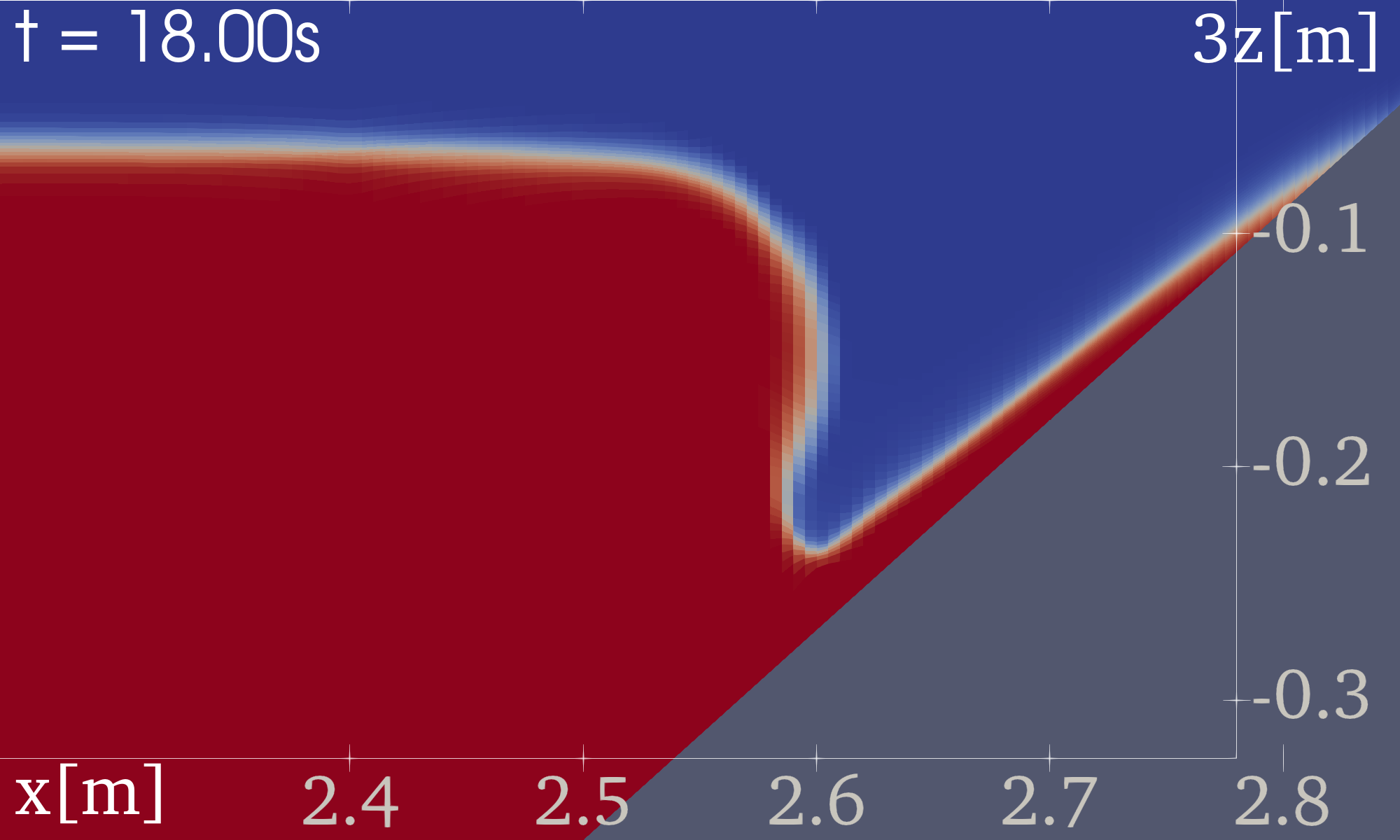}
    \includegraphics[width=0.32\linewidth]{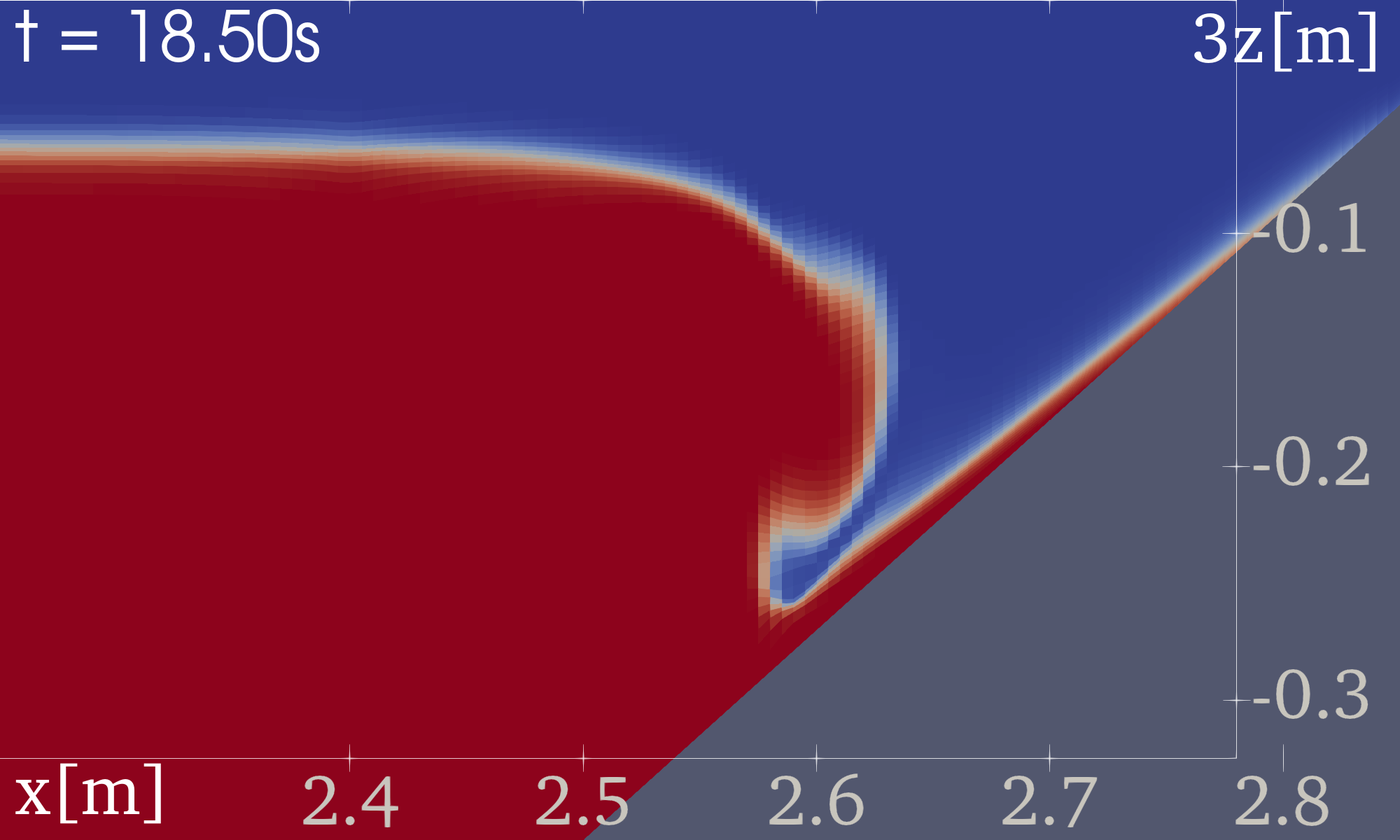}
    \includegraphics[width=0.32\linewidth]{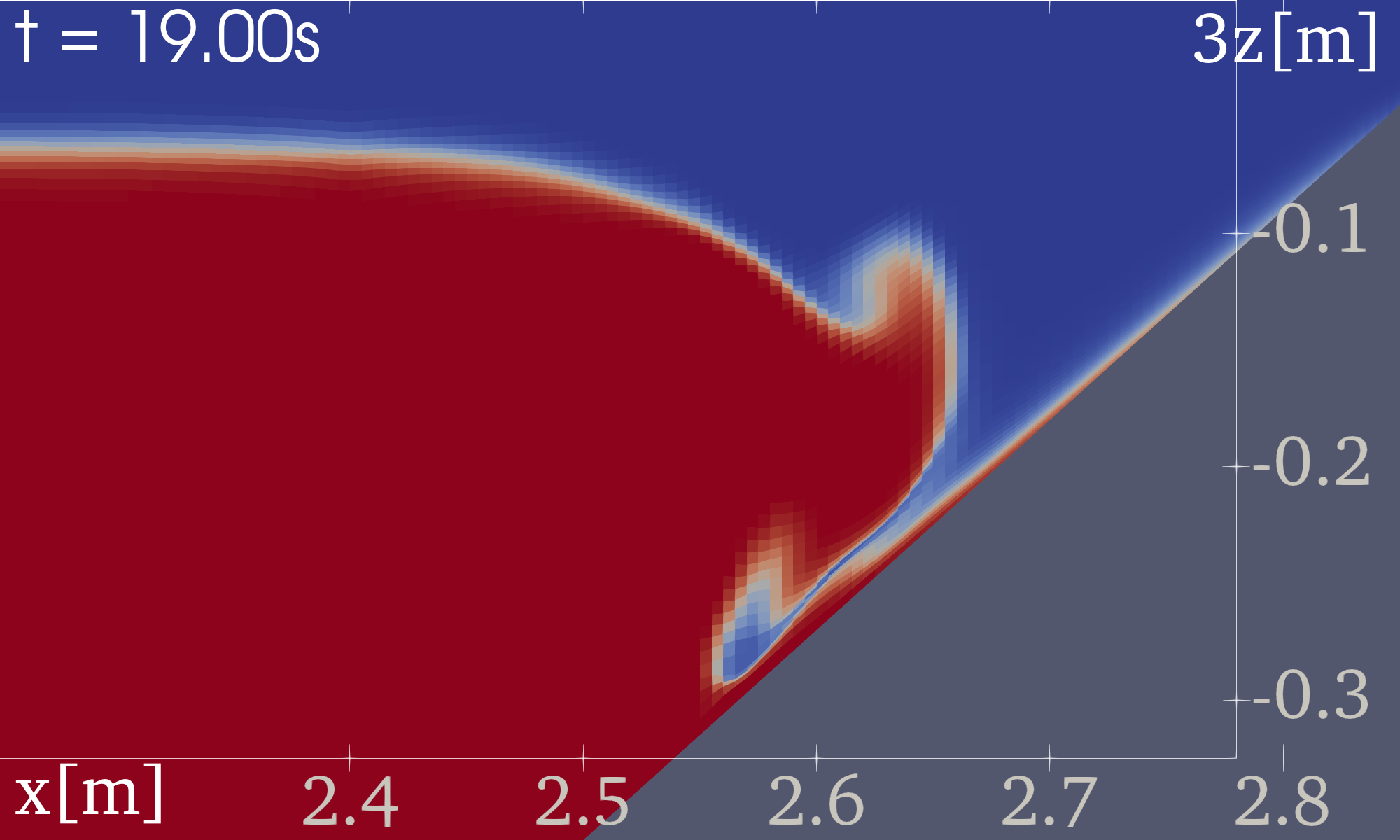}
    \includegraphics[width=0.32\linewidth]{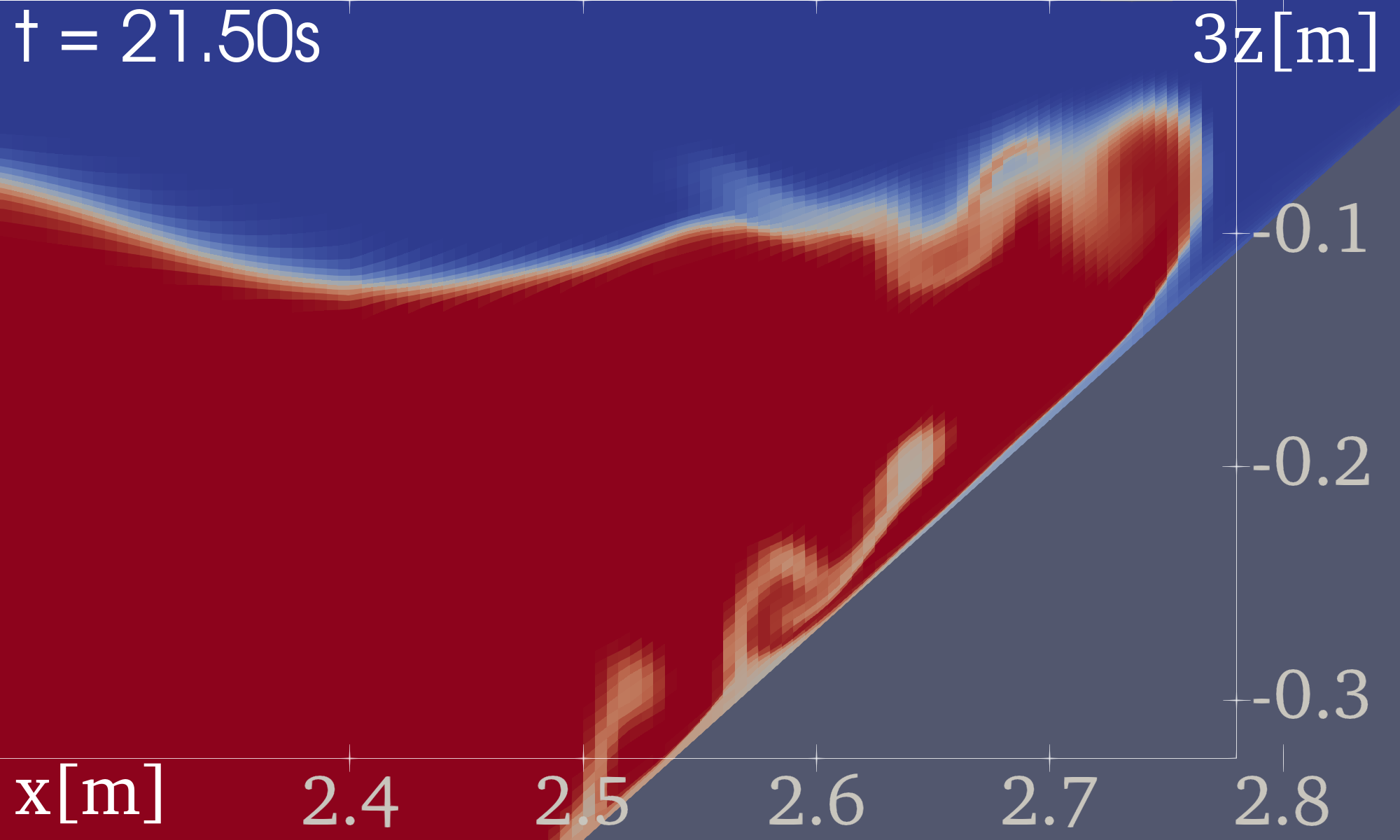}
    \includegraphics[width=0.6\linewidth]{ag_bar.png}
    \caption{ISW plunging into a sloping beach, normalized density contours from SLS simulation with $(a_\vartheta,a_M) = (0.01,10)$. The vertical axis and its labels are magnified by a factor of 3 and the dimensional time value is noted at each frame.} 
    \label{fig:ag1010}
\end{figure}

Because of this, if we want to further diminish numerical mixing, we must also employ the monitor function strategy through $a_M$. Indeed, we see that by using $a_\vartheta=0.01$, values of $a_M$ between $1$ and $10$ seem to greatly reduce numerical mixing up until the point that the ISW starts to break. After the point of overturning ($t>18s$), numerical mixing is mainly the result of the horizontal advection scheme that a vertical ALE method cannot directly reduce.
Indeed, by comparing Fig.\ref{fig:ag10_0} and Fig.\ref{fig:ag1010}, we see that the latter displays a much clearer result that greatly improves the resolution around the density interface.

It should be noted that a small jump of the variance is observed in Fig.\ref{fig:aghsaee_dvd} at the start of the simulation for $a_M>1$. In these cases, the factor $a_M$ dictates a strong refinement of the mesh around the pycnocline (see \S\ref{sec:monitor}). Consequently, since the initial $\sigma$-mesh is iso-spaced, the whole grid moves at the start of the simulation to meet this refinement criterion. Thus, the jump of the variance is attributed to this abrupt movement of the mesh. Furthermore, the solver exhibits this behavior only at the start of the simulation, and no spontaneous "unmixing" is observed thereafter. As such, this is considered a minor issue related to the initialization process, rather than the proposed scheme itself.

Summarizing the above observations, we conclude that when vertical mass transfer is important, it is advisable to keep $a_\vartheta$ at a moderate value $ a_\vartheta \sim 0.01 $ and try to reduce SDM through $a_M$. Indeed, the pair $(a_\vartheta,a_M)=(0.01,10)$, as demonstrated in Fig.\ref{fig:aghsaee_dvd}, almost negates the decay of $ \iiint \rho^2 \di V $ and simultaneously captures the overturning accurately (Fig.\ref{fig:ag1010}).

With regard to the phenomenon, the results of SLS closely resemble those of \cite{aghsaee2010breaking}: When the ISW approaches the slope, its crest steepens, until the critical point of overturning, when a mass of light water enters the dense region, while the dense front propagates to the right. Afterwards, overturning occurs and a portion of the ISW is reflected leftwards. The above results indicate the ability of the scheme to simulate nonhydrostatic flows of complex nature. Furthermore, through the proposed mesh moving formulation, SLS is able to have a low level of numerical dissipation without overly suppressing the physically relevant mass transfer.

\section{Concluding remarks}\label{sec:conclusions}

In the present paper, an ALE vertical mesh movement strategy was applied to a nonhydrostatic numerical model. The mesh movement is defined through the minimization of the functional in eq.(\ref{eq:functional}) that results in the elliptic equation of eq.(\ref{eq:elliptic}). The cost functional includes various target objectives that are weighted by user-defined coefficients. These goals include a Lagrangian tendency and a monitor function that increases the vertical mesh resolution in areas of interest, accompanied by smoothing towards a reference configuration. Those coefficients were analyzed in \S\ref{sec:freqfilter},\S\ref{sec:monitor} and some guidelines were laid out.

The effectiveness of the variational mesh movement in mitigating SDM is thoroughly investigated in \S\ref{sec:agh} for the demanding case of ISW breaking. Various values of the cost functional coefficients were tested and SDM was quantified through the second moment of density $\iiint \rho^2 \di V$. The results indicate that although numerical mixing can be severely reduced by the Lagrangian bias, an excessive value of the corresponding coefficient might result in suppression of physically relevant vertical mass transfer. On the other hand, keeping the Lagrangian bias at the mild value ($a_\vartheta \sim 0.01$) and using a monitor function that increases resolution in areas with large density gradients, one can achieve significant reduction in numerical mixing without producing nonphysical results as indicated in Fig.\ref{fig:aghsaee_dvd}.

Although these findings indicate the effectiveness of the method in the test cases of \S\ref{sec:results}, the implications of using the present variational ALE scheme and the proposed values for $(a_\vartheta,a_x,a_\xi,a_M)$ in larger-scale, general-purpose ocean models may not be straightforward. In particular, applying the method in realistic ocean scenarios requires the inclusion of Coriolis terms and mixing or turbulence parameterizations. These processes can have a nontrivial impact on the production of SDM and, consequently, the tuning of parameters may differ or even require the introduction of additional optimality criteria for the method to operate effectively.

Despite this, the derivation and analysis of the present scheme are largely scale-independent and the incorporation of additional optimality criteria remains a possibility (see \S\ref{sec:remarks}). As such, the present framework provides a flexible foundation upon which further extensions can be built. The application of the proposed ALE method to more realistic ocean simulations is planned as the next step in the development of SLS.

\section*{Acknowledgments}

The research work was supported by the Greek State Scholarship Foundation (IKY) under the Chrysovergis grant. The first author is the recipient of the 2024 scholarship in the domain of Physics that fully funds his PhD studies at NTUA and gratefully acknowledges this financial support. The authors also extend their gratitude to S. Zafeiris and two anonymous reviewers for their constructive comments that helped improve the quality of the manuscript.

\bibliography{sls2_preprint_rev1}

@article{audusse2004fast,
  title={A fast and stable well-balanced scheme with hydrostatic reconstruction for shallow water flows},
  author={Audusse, Emmanuel and Bouchut, Fran{\c{c}}ois and Bristeau, Marie-Odile and Klein, Rupert and Perthame, Benoît},
  journal={SIAM Journal on Scientific Computing},
  volume={25},
  number={6},
  pages={2050--2065},
  year={2004},
  publisher={SIAM}
}

@article{kurganov2018finite,
  title={Finite-volume schemes for shallow-water equations},
  author={Kurganov, Alexander},
  journal={Acta Numerica},
  volume={27},
  pages={289--351},
  year={2018},
  publisher={Cambridge University Press}
}

@article{griffies2020primer,
  title={A primer on the vertical Lagrangian-remap method in ocean models based on finite volume generalized vertical coordinates},
  author={Griffies, Stephen M and Adcroft, Alistair and Hallberg, Robert W},
  journal={Journal of Advances in Modeling Earth Systems},
  volume={12},
  number={10},
  pages={e2019MS001954},
  year={2020},
  publisher={Wiley Online Library}
}

@article{phillips1957coordinate,
  title={A coordinate system having some special advantages for numerical forecasting},
  author={Phillips, Norman A},
  journal={J. Meteor.},
  volume={14},
  pages={184--185},
  year={1957}
}

@article{bleck2002oceanic,
  title={An oceanic general circulation model framed in hybrid isopycnic-Cartesian coordinates},
  author={Bleck, Rainer},
  journal={Ocean modelling},
  volume={4},
  number={1},
  pages={55--88},
  year={2002},
  publisher={Elsevier}
}

@article{chassignet2007hycom,
  title={The HYCOM (hybrid coordinate ocean model) data assimilative system},
  author={Chassignet, Eric P and Hurlburt, Harley E and Smedstad, Ole Martin and Halliwell, George R and Hogan, Patrick J and Wallcraft, Alan J and Baraille, Remy and Bleck, Rainer},
  journal={Journal of Marine Systems},
  volume={65},
  number={1-4},
  pages={60--83},
  year={2007},
  publisher={Elsevier}
}

@article{griffies2000developments,
  title={Developments in ocean climate modelling},
  author={Griffies, Stephen M and B{\"o}ning, Claus and Bryan, Frank O and Chassignet, Eric P and Gerdes, R{\"u}diger and Hasumi, Hiroyasu and Hirst, Anthony and Treguier, Anne-Marie and Webb, David},
  journal={Ocean Modelling},
  volume={2},
  number={3-4},
  pages={123--192},
  year={2000},
  publisher={Elsevier}
}

@article{bleck1998ocean,
  title={Ocean modeling in isopycnic coordinates},
  author={Bleck, Rainer},
  journal={Ocean modeling and parameterization},
  pages={423--448},
  year={1998},
  publisher={Springer}
}

@article{leclair2011z,
  title={z-Coordinate, an Arbitrary Lagrangian--Eulerian coordinate separating high and low frequency motions},
  author={Leclair, Matthieu and Madec, Gurvan},
  journal={Ocean Modelling},
  volume={37},
  number={3-4},
  pages={139--152},
  year={2011},
  publisher={Elsevier}
}

@article{adcroft2019gfdl,
  title={The GFDL global ocean and sea ice model OM4. 0: Model description and simulation features},
  author={Adcroft, Alistair and Anderson, Whit and Balaji, V and Blanton, Chris and Bushuk, Mitchell and Dufour, Carolina O and Dunne, John P and Griffies, Stephen M and Hallberg, Robert and Harrison, Matthew J and others},
  journal={Journal of Advances in Modeling Earth Systems},
  volume={11},
  number={10},
  pages={3167--3211},
  year={2019},
  publisher={Wiley Online Library}
}

@article{petersen2015evaluation,
  title={Evaluation of the arbitrary Lagrangian--Eulerian vertical coordinate method in the MPAS-Ocean model},
  author={Petersen, Mark R and Jacobsen, Douglas W and Ringler, Todd D and Hecht, Matthew W and Maltrud, Mathew E},
  journal={Ocean Modelling},
  volume={86},
  pages={93--113},
  year={2015},
  publisher={Elsevier}
}

@article{van1986second,
  title={A second-order accurate pressure-correction scheme for viscous incompressible flow},
  author={Van Kan, JJIM},
  journal={SIAM journal on scientific and statistical computing},
  volume={7},
  number={3},
  pages={870--891},
  year={1986},
  publisher={SIAM}
}

@article{roe1985some,
  title={Some contributions to the modelling of discontinuous flows},
  author={Roe, Philip L},
  journal={Large-scale computations in fluid mechanics},
  pages={163--193},
  year={1985}
}

@article{alexandris2024semi,
  title={A semi-Lagrangian Splitting framework for the simulation of non-hydrostatic free-surface flows},
  author={Alexandris-Galanopoulos, Andreas and Papadakis, George and Belibassakis, Kostas},
  journal={Ocean Modelling},
  volume={187},
  pages={102290},
  year={2024},
  publisher={Elsevier}
}

@article{bollermann2013well,
  title={A well-balanced reconstruction of wet/dry fronts for the shallow water equations},
  author={Bollermann, Andreas and Chen, Guoxian and Kurganov, Alexander and Noelle, Sebastian},
  journal={Journal of Scientific Computing},
  volume={56},
  pages={267--290},
  year={2013},
  publisher={Springer}
}

@article{banerjee2024discrete,
  title={Discrete variance decay analysis of spurious mixing},
  author={Banerjee, Tridib and Danilov, Sergey and Klingbeil, Knut and Campin, Jean-Michel},
  journal={Ocean Modelling},
  volume={192},
  pages={102460},
  year={2024},
  publisher={Elsevier}
}

@article{burchard2004non,
  title={Non-uniform adaptive vertical grids in one-dimensional numerical ocean models},
  author={Burchard, Hans and Beckers, Jean-Marie},
  journal={Ocean Modelling},
  volume={6},
  number={1},
  pages={51--81},
  year={2004},
  publisher={Elsevier}
}

@article{burchard2008comparative,
  title={Comparative quantification of physically and numerically induced mixing in ocean models},
  author={Burchard, Hans and Rennau, Hannes},
  journal={Ocean modelling},
  volume={20},
  number={3},
  pages={293--311},
  year={2008},
  publisher={Elsevier}
}

@article{liu2018well,
  title={Well-balanced positivity preserving central-upwind scheme with a novel wet/dry reconstruction on triangular grids for the Saint-Venant system},
  author={Liu, Xin and Albright, Jason and Epshteyn, Yekaterina and Kurganov, Alexander},
  journal={Journal of Computational Physics},
  volume={374},
  pages={213--236},
  year={2018},
  publisher={Elsevier}
}

@Article{dunphy2011,
AUTHOR = {Dunphy, M. and Subich, C. and Stastna, M.},
TITLE = {Spectral methods for internal waves: indistinguishable density profiles and double-humped solitary waves},
JOURNAL = {Nonlinear Processes in Geophysics},
VOLUME = {18},
YEAR = {2011},
NUMBER = {3},
PAGES = {351--358},
URL = {https://npg.copernicus.org/articles/18/351/2011/},
DOI = {10.5194/npg-18-351-2011}
}

@article{lamb2014internal,
  title={Internal solitary waves shoaling onto a shelf: Comparisons of weakly-nonlinear and fully nonlinear models for hyperbolic-tangent stratifications},
  author={Lamb, Kevin G and Xiao, Wenting},
  journal={Ocean Modelling},
  volume={78},
  pages={17--34},
  year={2014},
  publisher={Elsevier}
}

@article{li2022iswfoam,
  title={ISWFoam: A numerical model for internal solitary wave simulation in continuously stratified fluids},
  author={Li, Jingyuan and Zhang, Qinghe and Chen, Tongqing},
  journal={Geoscientific Model Development},
  volume={15},
  number={1},
  pages={105--127},
  year={2022},
  publisher={Copernicus Publications G{\"o}ttingen, Germany}
}

@article{hsieh2015numerical,
  title={Numerical modeling of flow evolution for an internal solitary wave propagating over a submerged ridge},
  author={Hsieh, Chih-Min and Hwang, Robert R and Hsu, John R-C and Cheng, Ming-Hung},
  journal={Wave Motion},
  volume={55},
  pages={48--72},
  year={2015},
  publisher={Elsevier}
}

@article{stastna2024simulations,
  title={Simulations of shoaling large-amplitude internal waves: perspectives and outlook},
  author={Stastna, Marek and Legare, Sierra},
  journal={Flow},
  volume={4},
  pages={E11},
  year={2024},
  publisher={Cambridge University Press}
}

@article{aghsaee2010breaking,
  title={Breaking of shoaling internal solitary waves},
  author={Aghsaee, Payam and Boegman, Leon and Lamb, Kevin G},
  journal={Journal of Fluid Mechanics},
  volume={659},
  pages={289--317},
  year={2010},
  publisher={Cambridge University Press}
}

@article{ersing2025entropy,
  title={Entropy stable hydrostatic reconstruction schemes for shallow water systems},
  author={Ersing, Patrick and Goldberg, Sven and Winters, Andrew R},
  journal={Journal of Computational Physics},
  volume={527},
  pages={113802},
  year={2025},
  publisher={Elsevier}
}

@article{tadmor1987numerical,
  title={The numerical viscosity of entropy stable schemes for systems of conservation laws. I},
  author={Tadmor, Eitan},
  journal={Mathematics of Computation},
  volume={49},
  number={179},
  pages={91--103},
  year={1987}
}

@book{gelfand2000calculus,
  title={Calculus of variations},
  author={Gelfand, Izrail Moiseevitch and Silverman, Richard A and others},
  year={2000},
  publisher={Courier Corporation}
}

@article{hofmeister2010non,
  title={Non-uniform adaptive vertical grids for 3D numerical ocean models},
  author={Hofmeister, Richard and Burchard, Hans and Beckers, Jean-Marie},
  journal={Ocean Modelling},
  volume={33},
  number={1-2},
  pages={70--86},
  year={2010},
  publisher={Elsevier}
}

@inproceedings{alexandris2025development,
  title={Development of a Semi-lagrangian-Splitting Ocean Model: Extension to Abrupt Bathymetries and Wave-Structure Interactions},
  author={Alexandris-Galanopoulos, Andreas and Papadakis, George},
  booktitle={International Congress of the International Maritime Association of the Mediterranean},
  pages={338--350},
  year={2025},
  organization={Springer}
}

@article{fox2019challenges,
  title={Challenges and prospects in ocean circulation models},
  author={Fox-Kemper, Baylor and Adcroft, Alistair and B{\"o}ning, Claus W and Chassignet, Eric P and Curchitser, Enrique and Danabasoglu, Gokhan and Eden, Carsten and England, Matthew H and Gerdes, R{\"u}diger and Greatbatch, Richard J and others},
  journal={Frontiers in Marine Science},
  volume={6},
      pages={65},
  year={2019},
  publisher={Frontiers Media SA}
}

@article{delandmeter2018fully,
  title={A fully consistent and conservative vertically adaptive coordinate system for SLIM 3D v0. 4 with an application to the thermocline oscillations of Lake Tanganyika},
  author={Delandmeter, Philippe and Lambrechts, Jonathan and Legat, Vincent and Vallaeys, Valentin and Naithani, Jaya and Thiery, Wim and Remacle, Jean-Fran{\c{c}}ois and Deleersnijder, Eric},
  journal={Geoscientific Model Development},
  volume={11},
  number={3},
  pages={1161--1179},
  year={2018},
  publisher={Copernicus Publications G{\"o}ttingen, Germany}
}

@article{cao1999study,
  title={A study of monitor functions for two-dimensional adaptive mesh generation},
  author={Cao, Weiming and Huang, Weizhang and Russell, Robert D},
  journal={SIAM Journal on Scientific Computing},
  volume={20},
  number={6},
  pages={1978--1994},
  year={1999},
  publisher={SIAM}
}

@article{huang2003variational,
  title={Variational mesh adaptation II: error estimates and monitor functions},
  author={Huang, Weizhang and Sun, Weiwei},
  journal={Journal of computational physics},
  volume={184},
  number={2},
  pages={619--648},
  year={2003},
  publisher={Elsevier}
}

@article{van1979towards,
  title={Towards the ultimate conservative difference scheme. V. A second-order sequel to Godunov's method},
  author={Van Leer, Bram},
  journal={Journal of computational Physics},
  volume={32},
  number={1},
  pages={101--136},
  year={1979},
  publisher={Elsevier}
}

@incollection{klingbeil2019reducing,
  title={Reducing spurious diapycnal mixing in ocean models},
  author={Klingbeil, Knut and Burchard, Hans and Danilov, Sergey and Goetz, Claus and Iske, Armin},
  booktitle={Energy Transfers in Atmosphere and Ocean},
  pages={245--286},
  year={2019},
  publisher={Springer}
}

@article{ilicak2012spurious,
  title={Spurious dianeutral mixing and the role of momentum closure},
  author={Il{\i}cak, Mehmet and Adcroft, Alistair J and Griffies, Stephen M and Hallberg, Robert W},
  journal={Ocean Modelling},
  volume={45},
  pages={37--58},
  year={2012},
  publisher={Elsevier}
}

@article{megann2018estimating,
  title={Estimating the numerical diapycnal mixing in an eddy-permitting ocean model},
  author={Megann, Alex},
  journal={Ocean Modelling},
  volume={121},
  pages={19--33},
  year={2018},
  publisher={Elsevier}
}

@book{toro2013riemann,
  title={Riemann solvers and numerical methods for fluid dynamics: a practical introduction},
  author={Toro, Eleuterio F},
  year={2013},
  publisher={Springer Science \& Business Media}
}

@article{long1953some,
  title={Some aspects of the flow of stratified fluids: I. A theoretical investigation},
  author={Long, Robert R},
  journal={Tellus},
  volume={5},
  number={1},
  pages={42--58},
  year={1953},
  publisher={Taylor \& Francis}
}

@article{lee2002spurious,
  title={Spurious diapycnal mixing of the deep waters in an eddy-permitting global ocean model},
  author={Lee, Mei-Man and Coward, Andrew C and Nurser, AJ George},
  journal={Journal of physical oceanography},
  volume={32},
  number={5},
  pages={1522--1535},
  year={2002}
}

@article{garinet2024spurious,
  title={Spurious numerical mixing under strong tidal forcing: a case study in the south-east Asian seas using the Symphonie model (v3. 1.2)},
  author={Garinet, Adrien and Herrmann, Marine and Marsaleix, Patrick and P{\'e}nicaud, Juliette},
  journal={Geoscientific Model Development},
  volume={17},
  number={18},
  pages={6967--6986},
  year={2024},
  publisher={Copernicus Publications G{\"o}ttingen, Germany}
}

@article{griffies2000spurious,
  title={Spurious diapycnal mixing associated with advection in a z-coordinate ocean model},
  author={Griffies, Stephen M and Pacanowski, Ronald C and Hallberg, Robert W},
  journal={Monthly Weather Review},
  volume={128},
  number={3},
  pages={538--564},
  year={2000}
}

@article{ilicak2016quantifying,
  title={Quantifying spatial distribution of spurious mixing in ocean models},
  author={Il{\i}cak, Mehmet},
  journal={Ocean Modelling},
  volume={108},
  pages={30--38},
  year={2016},
  publisher={Elsevier}
}

@article{marchesiello2009spurious,
  title={Spurious diapycnal mixing in terrain-following coordinate models: The problem and a solution},
  author={Marchesiello, Patrick and Debreu, Laurent and Couvelard, Xavier},
  journal={Ocean Modelling},
  volume={26},
  number={3-4},
  pages={156--169},
  year={2009},
  publisher={Elsevier}
}

@article{white2009high,
  title={High-order regridding--remapping schemes for continuous isopycnal and generalized coordinates in ocean models},
  author={White, Laurent and Adcroft, Alistair and Hallberg, Robert},
  journal={Journal of Computational Physics},
  volume={228},
  number={23},
  pages={8665--8692},
  year={2009},
  publisher={Elsevier}
}

@article{urakawa2014effect,
  title={Effect of numerical diffusion on the water mass transformation in eddy-resolving models},
  author={Urakawa, L Shogo and Hasumi, Hiroyasu},
  journal={Ocean Modelling},
  volume={74},
  pages={22--35},
  year={2014},
  publisher={Elsevier}
}

@article{winters1995available,
  title={Available potential energy and mixing in density-stratified fluids},
  author={Winters, Kraig B and Lombard, Peter N and Riley, James J and D'Asaro, Eric A},
  journal={Journal of Fluid Mechanics},
  volume={289},
  pages={115--128},
  year={1995},
  publisher={Cambridge University Press}
}

@article{gibson2017attribution,
  title={Attribution of horizontal and vertical contributions to spurious mixing in an Arbitrary Lagrangian--Eulerian ocean model},
  author={Gibson, Angus H and Hogg, Andrew McC and Kiss, Andrew E and Shakespeare, Callum J and Adcroft, Alistair},
  journal={Ocean Modelling},
  volume={119},
  pages={45--56},
  year={2017},
  publisher={Elsevier}
}

@article{klingbeil2014quantification,
  title={Quantification of spurious dissipation and mixing--Discrete variance decay in a Finite-Volume framework},
  author={Klingbeil, Knut and Mohammadi-Aragh, Mahdi and Gr{\"a}we, Ulf and Burchard, Hans},
  journal={Ocean Modelling},
  volume={81},
  pages={49--64},
  year={2014},
  publisher={Elsevier}
}

@article{john2000numerical,
  title={A numerical study of a posteriori error estimators for convection--diffusion equations},
  author={John, Volker},
  journal={Computer methods in applied mechanics and engineering},
  volume={190},
  number={5-7},
  pages={757--781},
  year={2000},
  publisher={Elsevier}
}

@article{mohammadi2015impact,
  title={The impact of advection schemes on restratifiction due to lateral shear and baroclinic instabilities},
  author={Mohammadi-Aragh, Mahdi and Klingbeil, Knut and Br{\"u}ggemann, Nils and Eden, Carsten and Burchard, Hans},
  journal={Ocean Modelling},
  volume={94},
  pages={112--127},
  year={2015},
  publisher={Elsevier}
}

@article{bryan1997numerical,
  title={A numerical method for the study of the circulation of the world ocean},
  author={Bryan, Kirk},
  journal={Journal of computational physics},
  volume={135},
  number={2},
  pages={154--169},
  year={1997},
  publisher={Elsevier}
}

@article{hallberg1996buoyancy,
  title={Buoyancy-driven circulation in an ocean basin with isopycnals intersecting the sloping boundary},
  author={Hallberg, Robert and Rhines, Peter},
  journal={Journal of Physical Oceanography},
  volume={26},
  number={6},
  pages={913--940},
  year={1996}
}

@article{schenk2001pardiso,
  title={PARDISO: a high-performance serial and parallel sparse linear solver in semiconductor device simulation},
  author={Schenk, Olaf and G{\"a}rtner, Klaus and Fichtner, Wolfgang and Stricker, Andreas},
  journal={Future Generation Computer Systems},
  volume={18},
  number={1},
  pages={69--78},
  year={2001},
  publisher={Elsevier}
}

@article{halliwell2004evaluation,
  title={Evaluation of vertical coordinate and vertical mixing algorithms in the HYbrid-Coordinate Ocean Model (HYCOM)},
  author={Halliwell, George R},
  journal={Ocean Modelling},
  volume={7},
  number={3-4},
  pages={285--322},
  year={2004},
  publisher={Elsevier}
}

\appendix

\section{Flux solvers}\label{app:fluxes}

For the discretization of $F^1_{i+1/2,j} , \vect{F}^{\vect{V}}_{i+1/2,j}$, we utilize the entropy stable framework (see e.g. \cite{ersing2025entropy}). Thus, the numerical fluxes are:
\begin{subequations}
\begin{align}
     F^1_{i+\frac{1}{2},j} &= \overline{Lu} - \dfrac{c_{bar}}{\overline{\rho}g} \jump{ \rho M_p} - | \overline{u} | \jump{L} \label{eq:flux_L} \\
    \vect{F}^{\vect{V}}_{i+\frac{1}{2},j} &= \overline{Lu} \times \overline{\vect{V}} - \overline{L} \left(|\overline{u}|+c_{bar}\right) \jump{\vect{V}}
\end{align}
\end{subequations}
where $\overline{(\cdot)}\eqdef \frac{1}{2} \left[ (\cdot)_R + (\cdot)_L \right]$ and $\jump{(\cdot)} \eqdef \frac{1}{2} \left[ (\cdot)_R - (\cdot)_L \right]$ where $(\cdot)_{L/R}$ are the reconstructed values on the left and right of the face respectively. Note that in SLS, these values are reconstructed through a second-order MUSCL-TVD procedure and thus --in smooth regions-- the jumps $\jump{\cdot}$ are noticeably smaller than those of a first-order scheme. The barotropic speed is denoted as $c_{bar} \eqdef \sqrt{g\overline{H}}$, while $M_p$ is the Montgomery potential $M_p \eqdef {p_h}{\big /}{\rho} + g z$.

Stability is achieved because the entropy variables of the multilayer system are $ \vect{U}^e = \left[ \ M_p \ , \ u \ , w \ \right]^T $ \cite{ersing2025entropy}, and thus terms like $\tensor{A} \jump{\vect{U}^e}$ result in a reduction of the system's mechanical energy if $\tensor{A}$ is a negative definite matrix \cite{tadmor1987numerical}. The jump of the entropy variables is scaled according to the maximum eigenvalue of the system $|u|+\sqrt{gH}$ (the barotropic one). Aside from these, the upwinding-like term $ |\overline{u}| \jump{L} $ is added to the GCE flux so that wiggles are negated.

\section{Finite Element formulation of pressure correction}\label{app:pco}

Let us consider test functions $N\in \mathcal{H}_{0,fs}^1$ that have zero trace at the free-surface and multiply eq.(\ref{eq:pco2}) by them:
\begin{align}
    \iiint_{\mathcal{D}} N \divg (\vect{V}^{n+1}) \di V = 0 \iff \iiint_{\mathcal{D}} \vect{V}^{n+1} \cdot \grad N \di V= \iiint_{\mathcal{D}} \divg (\vect{V}^{n+1} N) \di V
\end{align}
By applying the Green-Gauss theorem and considering the Neumann boundary conditions $\vect{V}^{n+1}\cdot n = v_{bc} $ on $\partial \mathcal{D}_N = \partial \mathcal{D} \setminus \partial \mathcal{D}_{fs}$ we get:
\begin{align}\label{eq:weak_form1}
    \iiint_{\mathcal{D}} \left[ \vect{V}^{n+1} \cdot \grad N \right] \di V = \iint_{\partial \mathcal{D}_N} v_{bc} N \di S
\end{align}
Inserting eq.(\ref{eq:pco1}) with $\phi \eqdef \Delta t (q^{n+1}-q^n)$ into eq.(\ref{eq:weak_form1}) we get:
\begin{align}\label{eq:weak}
   \iiint_{\mathcal{D}} \dfrac{1}{\rho^{ n+1 }} (\grad \phi) \cdot (\grad N) \di V = \iiint_{\mathcal{D}} \vect{V}^* \cdot (\grad N) \di V - \iint_{\partial \mathcal{D}_N} v_{bc} N \di S
\end{align}
Thus, the weak formulation of eq.(\ref{eq:pco}) is to find $\phi \in \mathcal{H}^1_{0,fs}$ so that eq.(\ref{eq:weak}) holds for every $N \in \mathcal{H}^1_{0,fs}$.

Now, we proceed to the definition of the various matrices. Inside a cell $\Omega_{ij}$ the Q4 elements have the following local parametric shape functions:
\begin{subequations}\label{shapeN}
\begin{align}
    N^{(1)} &= ts \\
    N^{(2)} &= (1-t) s \\
    N^{(3)} &= t(1-s) \\
    N^{(4)} &= (1-t)(1-s)
\end{align}
\end{subequations}
with $s = \dfrac{x-x_{i}}{\Delta x_i}$ and $t = \dfrac{\xi - \xi_j}{ \Delta \xi_j }$.

Using these local shape functions, to discretize eq.(\ref{eq:weak}), we need to calculate their gradients on the physical space. Thus, inside a cell $\Omega_{km}$ we approximate the gradient of $N_{ij}$ using:
\begin{align}
    \grad N_{ij} {\Big|}_{\Omega_{km}}
    \approx
    \begin{bmatrix}
        \displaystyle \pdv{N_{ij}}{x} - \dfrac{\overline{(Z_x)}_{km}}{L_{km}} \pdv{N_{ij}}{\xi} \vspace{10pt}\\
        \displaystyle \dfrac{1}{L_{km}}\pdv{N_{ij}}{\xi}
    \end{bmatrix} \label{gradN}
\end{align}
with $(Z_x)_{ij} = \dfrac{ z_{i+1,j} - z_{i,j} }{ ( \Delta x_{i+1} + \Delta x_{i} ) / 2}$ and $\overline{(Z_x)}_{km} = \dfrac{(Z_x)_{ij}+(Z_x)_{i+1,j}+(Z_x)_{i,j+1}+(Z_x)_{i+1,j+1}}{4}$ being the nodal and cell-averaged gradients respectively.

By doing so, the gradients inside $\Omega_{ij}$ are polynomial and can be integrated analytically.

If now we insert the expansion $\phi \approx \sum \phi_k N_k $ into eq.(\ref{eq:weak}), we get:
\begin{align}
    \sum_{m\in nodes} \iiint_{\mathcal{D}} \dfrac{1}{\rho^{ n+1 }} (\grad N_m \cdot \grad N_k) \phi_m \di V = \iiint_{\mathcal{D}} \vect{V}^* \cdot \grad N_k \di V - \iint_{\partial\mathcal{D}_N} N_k v_{bc} \di S
\end{align}
In the above eq. and below, we substitute $(i,j)$ into single index notation with $k,m,l$.

Since $\vect{V},\rho$ come from the FV discretization, they are constant within each cell $\Omega_{l}$ and we can compute:
\begin{subequations}\label{kggb}
\begin{align}
    \{K\}_{km} &\eqdef \sum_{ l \in \text{cells} } \dfrac{1}{\rho^{ n+1 }_l}\iiint_{\Omega_l} \grad N_k \cdot \grad N_m \di V \\
    \iiint \vect{V}^* \cdot \grad N_k \di V &= \sum_{l\in \text{cells} } \vect{V}^*_l \cdot\iiint_{\Omega_l} \grad N_k \di V \eqdef \sum_{l \in cells } ( G^x_{kl} u^*_l + G^{z}_{kl} w^*_l )  \\
    \iint_{\partial\mathcal{D}_N} N_k v_{bc} \di S &= \sum_{l \in \text{boundary faces} } ( v_{bc} )_l \iint_{(\partial \mathcal{D})_l} N_k \di S \eqdef \sum_{l \in \text{boundary faces} } B_{kl} (v_{bc})_l
\end{align}
\end{subequations}
Gathering the above relations, we arrive at the 
system in eq.(\ref{linpco}) and the update relations in eq.(\ref{eq:pco_discrete}), which are used to update the nonhydrostatic pressure field.

\section{Discrete Variance Decay}\label{app:dvd}

We present the derivation of the equation for the variance decay (see \cite{banerjee2024discrete}). Specifically, we consider the space-discrete advection scheme by ignoring the horizontal fluxes and considering a zero-order (piecewise constant) reconstruction scheme. By dropping the $i$ index for reasons of compactness and readability, the continuity and tracer equations become:\begin{align}
    & \Delta \xi_j \pdv{L_j}{t} + \vartheta_{j+1/2}-\vartheta_{j-1/2} = 0 \\
    & \Delta \xi_j \pdv*{(L_j\rho_j)}{t} + (\vartheta \rho)_{j+1/2}-(\vartheta\rho)_{j-1/2} = 0
\end{align}
where the fluxes due to the upwinding can be written as:
\begin{align}
    (\vartheta \rho)_{j+1/2} =  \vartheta_{j+1/2} \dfrac{\rho_{j+1}+\rho_j}{2} - |\vartheta_{j+1/2}|\dfrac{\rho_{j+1}-\rho_j}{2} \end{align}
Then, we can calculate the discrete variance decay:
\begin{align}
    & \frac{1}{2} \Delta \xi_j \pdv*{(L_j\rho_j^2)}{t} = {\Delta \xi_j} \left[ \rho_j \pdv*{(L_j\rho_j)}{t} - \dfrac{\rho_j^2}{2} \pdv{L_j}{t} \right] \then \nonumber \\
     & \dfrac{1}{2} \Delta \xi_j \pdv*{(L_j\rho_j^2)}{t} + \rho_j \left[(\vartheta \rho)_{j+1/2}-(\vartheta\rho)_{j-1/2} \right] - \dfrac{\rho_j^2}{2} \left[ \vartheta_{j+1/2}-\vartheta_{j-1/2} \right] = 0
\end{align}
Then, by substituting the fluxes and using the identity:
\begin{align}
    \rho_j (\rho_{j+1} - \rho_j)  = \dfrac{\rho_{j+1}+\rho_j}{2}(\rho_{j+1}-\rho_j) - \dfrac{1}{2}(\rho_{j+1}-\rho_j)^2
\end{align}
we arrive at the following evolution equation:
\begin{align}
    \dfrac{1}{2} \Delta \xi_j \pdv*{(L_j\rho_j^2)}{t} + F^{en}_{j+1/2} - F^{en}_{j-1/2} = - \dfrac{1}{4} |\vartheta_{j+1/2}| (\rho_{j+1}-\rho_j)^2- \dfrac{1}{4} |\vartheta_{j-1/2}| (\rho_{j}-\rho_{j-1})^2
\end{align}
with $F^{en}_{j+1/2} \eqdef \vartheta_{j+1/2}\dfrac{\rho_j\rho_{j+1}}{2} - \frac{1}{4} |\vartheta_{j+1/2}| (\rho_{j+1}+\rho_j)(\rho_{j+1}-\rho_j) $.

If now we restore the $i$ index, multiply by $\Delta x_i$ and sum over $i$ and $j$, we arrive at:
\begin{align}
     \odv{}{t} \left( \sum_{i=1}^{n_x} \sum_{j=1}^{n_l}
     \rho_{ij}^2 L_{ij} \Delta x_i \Delta \xi_j \right) =
     - \sum_{i=1}^{n_x}\sum_{j=1}^{n_l-1}
     \Delta x_i |\vartheta_{i,j+1/2}|
     (\rho_{i,j+1}-\rho_{ij})^2
\end{align}

\section{Variational formulation and functional minimizer}\label{app:var}

We aim to calculate the variational (Gâteaux) derivative of eq.(\ref{eq:functional}) with respect to $\vartheta$ in the direction $h$: 
\begin{align}
    \delta_\vartheta \mathcal{F} [h] \eqdef \lim_{s\to 0} \dfrac{ \mathcal{F}(\vartheta+hs) - \mathcal{F}(\vartheta) }{s} 
\end{align}

Through the classic result of the calculus of variations, the minimizer of $\mathcal{F}$ is such that $\delta_\vartheta \mathcal{F}[h] = 0$ for all admissible $h$ (see also \cite{gelfand2000calculus}). Following common practices, since the goal is to get the differential equation on the interior of the domain, we restrict $h=0$ at the boundary $\partial\mathcal{D}$ in order to bypass the so-called natural boundary conditions.
Bearing in mind that by eq.(\ref{eq:dzdt}) we have $z^{n+1} = z_{lag}^* - \Delta t \vartheta $ and thus $\delta_\vartheta z^{n+1} = -\Delta t \times h$, we calculate term by term the following:
\begin{align}
\delta_\vartheta \mathcal{F}[h]
= & 2 \iiint_{\mathcal{D}} \left[
    a_\vartheta T_{ref}\, \vartheta h
    - a_x {\Delta x^2} \pdv{}{x}(z^{n+1}-z_{\mathrm{ref}})\pdv{h}{x}
    \right. \nonumber \\
& \qquad \qquad \left.
    - a_\xi {\Delta \xi^2} \pdv{}{\xi}(z^{n+1}-z_{\mathrm{ref}})\pdv{h}{\xi}
    - a_M \Delta \xi^2 M^2 \pdv{z^{n+1}}{\xi}\pdv{h}{\xi}
  \right] \, \mathrm{d}x\,\mathrm{d}\xi
\end{align}

Thus, through integration by parts we conclude that:
\begin{align}
    & \delta_\vartheta \mathcal{F} [h] = 2 \iiint_{ \mathcal{D}} h \left[ a_\vartheta T_{ref} \vartheta + a_x {\Delta x^2} \pdv[order=2]{}{x} \left( z^{n+1}-z_{ref} \right)  \right.  \nonumber \\
    & \hspace{100pt} \left. + a_\xi {\Delta \xi^2} \pdv[order=2]{}{\xi} \left( z^{n+1}-z_{ref} \right) + a_M {\Delta \xi^2} \pdv*{ \left( M^2 \pdv{z^{n+1}}{\xi} \right) }{\xi}   \right] \di x \di \xi \nonumber \\
    & \hspace{100pt} + \iint_{\partial \mathcal{D}} h \times \left( \rm boundary \ terms \right) \di S
\end{align}
Substituting the expression $ z^{n+1} = z_{lag}^* - \Delta t \vartheta $ and using $h_{\partial\mathcal{D}}=0$, we get:
\begin{align}
    \delta_\vartheta \mathcal{F} [h] &= 2 \Delta t  \iiint h \left[ a_\vartheta \dfrac{ T_{ref} }{ \Delta t } \vartheta - a_x {\Delta x^2} \pdv[order=2]{}{x} \left\{ \vartheta - v_{lag} \right\} 
    \right. \nonumber \\
    & \hspace{80pt} \left. - a_\xi {\Delta \xi^2} \pdv[order=2]{}{\xi} \left\{ \vartheta - v_{lag} \right\} - a_M \Delta \xi^2 \pdv{}{\xi} \left\{ M^2 \pdv{}{\xi} \left( \vartheta - \dfrac{z^*_{lag}}{\Delta t} \right) \right\} \right] \di x \di \xi
\end{align}
where $v_{lag} \eqdef \dfrac{ z^*_{lag} - z_{ref} }{\Delta t}$. Since this holds for every $h$, the fundamental lemma of calculus of variations states that the minimizer of $\mathcal{F}$ satisfies eq.(\ref{eq:elliptic}).

\section{Convergence test details}\label{app:djl}

\begin{figure}[t]

    \centering
    \includegraphics[width=0.32\linewidth]{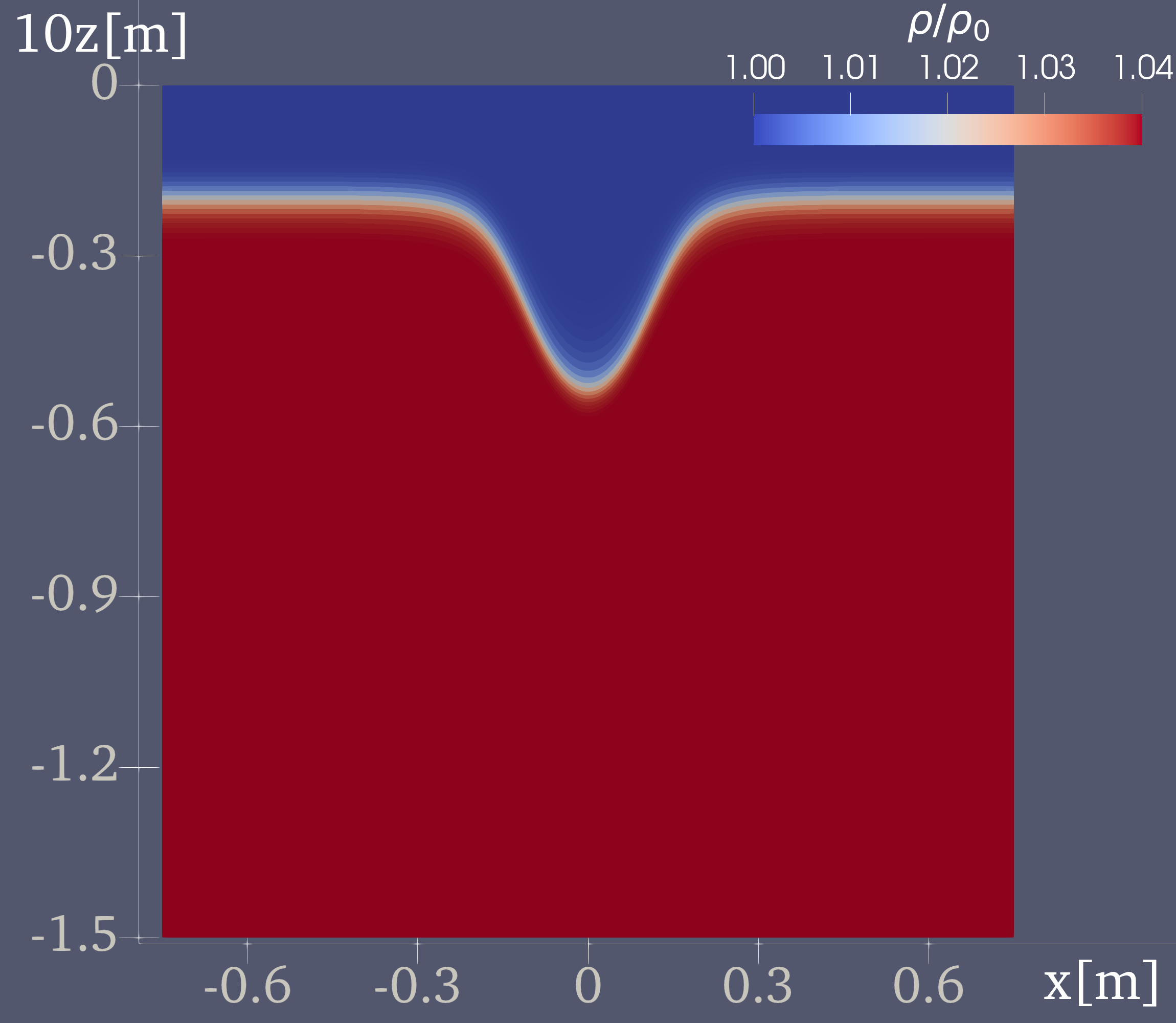}
    \includegraphics[width=0.32\linewidth]{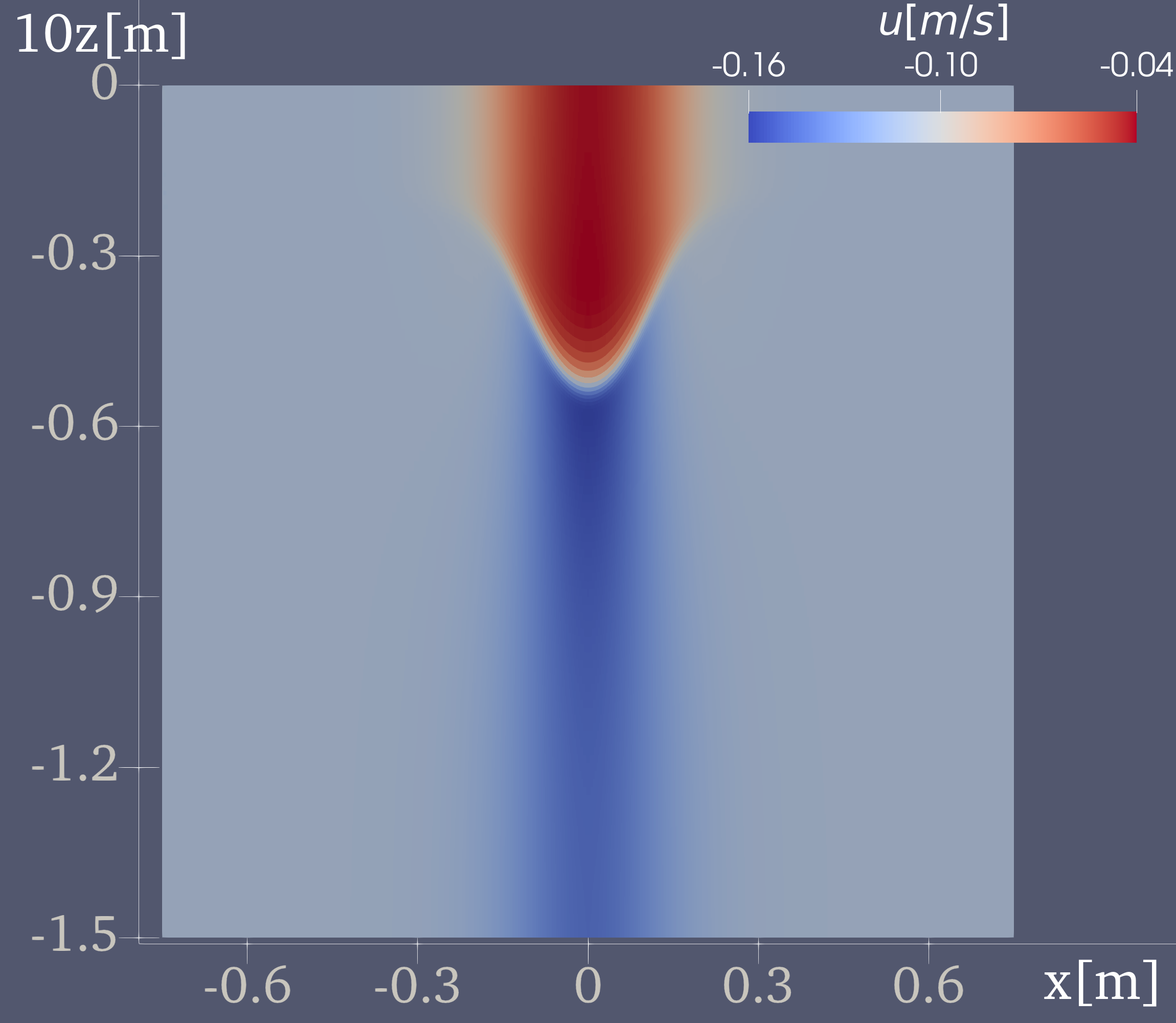}
    \includegraphics[width=0.32\linewidth]{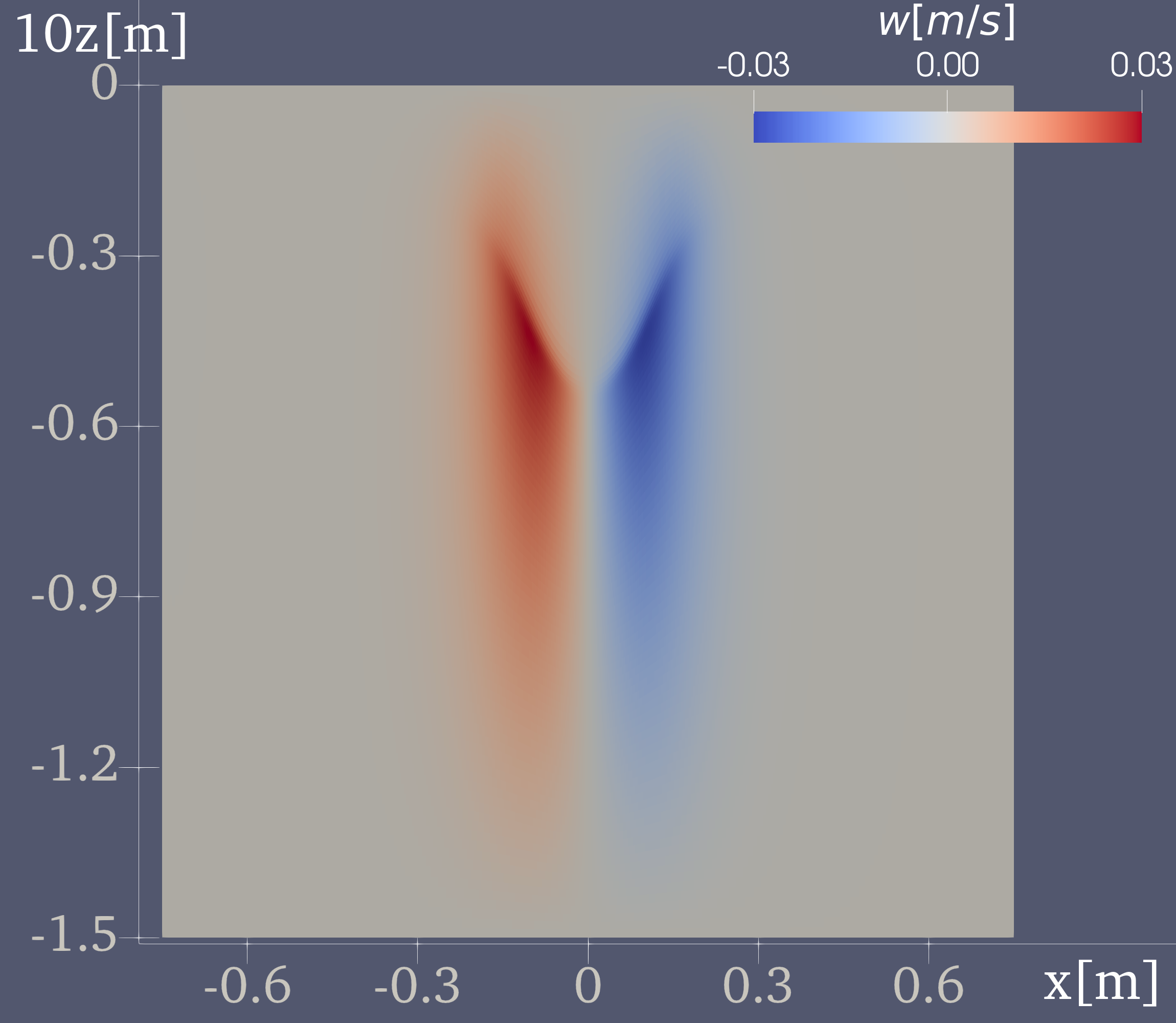}
    \caption{Snapshot of the soliton of \S\ref{sec:djl} (see Tab.\ref{tab:isw}) at $t=10s$ on a $1000 \times 100$ mesh. Contours of the normalized density, horizontal and vertical velocity are presented on a bounding box with the y axis being magnified by a factor of 10.}
    \label{fig:djl}
\end{figure}

\begin{figure}[t]

    \centering
    \includegraphics[width=0.6\linewidth]{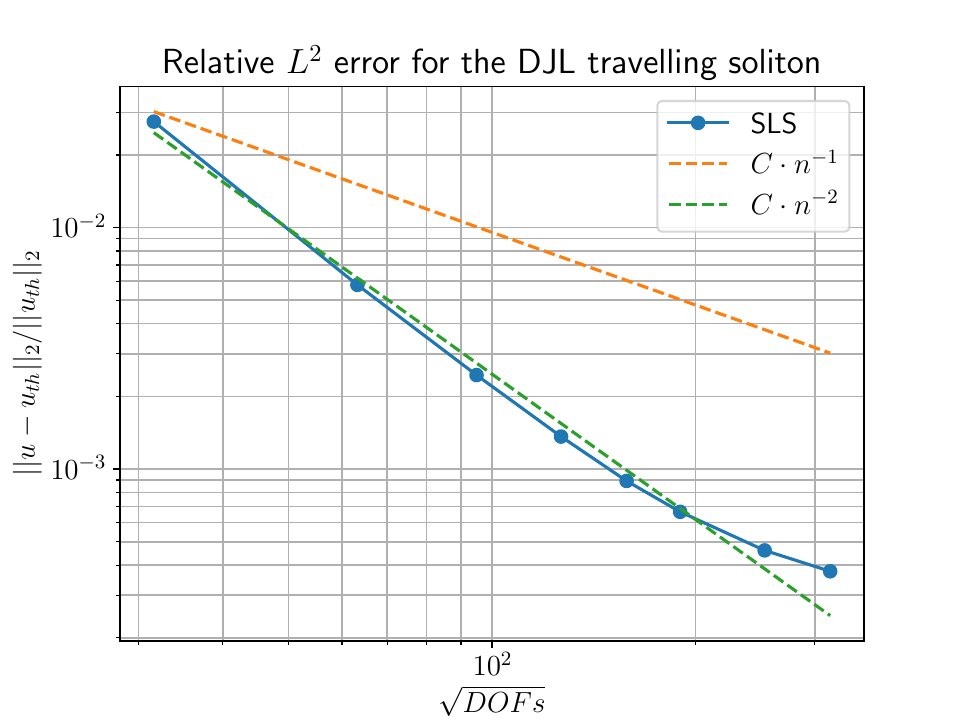}
    \caption{Convergence of the $L^2$ error norm of the velocity with respect to the square root of the total Degrees of Freedom (DOFs). The error is normalized using the norm of the initial velocity field. First and second-order ideal curves are presented in dashed lines. Both axes are logarithmic.}
    \label{fig:djl_error}
\end{figure}

In order to keep the soliton centered with respect to the computational domain, we subtract its traveling speed of $c=0.11279 m/s$ from the initial velocity field. By doing this, the soliton should remain fixed in the center of the frame and thus any deviation from the initial conditions is solely due to the numerical scheme. The solver is configured in isopycnal mode with $\vartheta = 0$ and the simulation is terminated at $t=10s$. To assess accuracy, the discrete $L^2$ error norms of the velocity field are calculated: $\norm{\vect{u}^n - \vect{u}^0 }_2= \sqrt{ \sum_{ij} ( {u}^n_{ij} - {u}_{ij}^{0} )^2 } $ with respect to the initial one $\vect{u}^0$. Various mesh sizes with $n_x / n_l = 10$ are considered.

The flow field solution for a total $ 1000 \times 100 $ degrees of freedom (DOFs) is presented in Fig.\ref{fig:djl}, while in Fig.\ref{fig:djl_error} the relative $L^2$ error is plotted over DOFs. While the error curve starts from a second-order trend, this is reduced to first-order at about $\sqrt{DOF} = 200$. This indicates that a first-order error term dominates at higher resolutions. That being said, the error curve is monotonic and no serious instabilities or spurious modes seem to pollute the convergence.

\end{document}